\documentclass[aps,twocolumn,showpacs,amsmath,amssymb,amsfonts,floatfix,superscriptaddress]{revtex4}
\usepackage{graphicx}
\usepackage{dcolumn}
\usepackage{bm}
\usepackage{latexsym}
\usepackage{float}
\usepackage{supertabular}
\usepackage{longtable}
\usepackage{hyperref}
\usepackage{amsmath}
\usepackage{wasysym}
\usepackage{color}
\usepackage{verbatim}
\usepackage{comment,xcolor}

\begin{document}
\title{Population Dynamics in a Changing Environment:  Random versus Periodic Switching} %
\author{Ami Taitelbaum\footnote[4]{Equally contributed to this work.}}
\affiliation{Racah Institute of Physics, Hebrew University of Jerusalem, Jerusalem 91904, Israel}
\author{Robert West\footnotemark[4]}
\affiliation{Department of Applied Mathematics, School of Mathematics, University of Leeds, Leeds LS2 9JT, U.K.} %
\author{Michael Assaf}
\email{michael.assaf@mail.huji.ac.il}
\affiliation{Racah Institute of Physics, Hebrew University of Jerusalem, Jerusalem 91904, Israel}
\author{Mauro Mobilia}
\email{M.Mobilia@leeds.ac.uk}
\affiliation{Department of Applied Mathematics, School of Mathematics, University of Leeds, Leeds LS2 9JT, U.K.}
\begin{abstract}
Environmental changes greatly influence the evolution of populations. Here, we study the dynamics of a  population  of two strains, one growing slightly faster than the other, competing for resources in a time-varying binary environment modeled by a carrying capacity switching either {\it randomly} or {\it periodically}
between states of abundance and scarcity. The population dynamics is characterized by demographic noise (birth and death events) coupled to a varying environment. We elucidate the similarities and differences of the evolution  subject to a stochastically- and periodically-varying environment. Importantly, the population size distribution is generally found to be broader under intermediate and fast random switching than under periodic variations, which results in markedly different asymptotic behaviors between the fixation probability of random and periodic switching.
We also determine the detailed conditions under which the fixation probability of the slow strain
is maximal.
\end{abstract}

\maketitle

The evolution of natural populations is influenced by varying environmental conditions:
the abundance of nutrients, toxins, or external factors like temperature
are subject to random and seasonal variations, and have an important impact on population dynamics~\cite{Morley83,Fux05,Caporaso11}.

Several models of a population response to a changing environment assume that external conditions vary either periodically or stochastically in time~\cite{Chesson81,Kussell05b,Assaf08,Assaf09,Loreau08,Beaumont09,Visco10,May73,Karlin74,He10,Tauber13,Assaf13,AMR13,Chisholm14,Kessler14,Kalyuzhny15,Assaf15,Melbinger15,Xue17,Assaf17,Assaf18,Dobramysl18,Marrec20}.
These external variations are often modeled by taking a binary environment that switches
between two states~\cite{Otto97,Thattai04,Kussell05,Acar08,Gaal10,Wienand11,Yurtsev2013,Patra2013,Ashcroft14,Patra2015,Hufton16,Hidalgo17,KEM1,KEM2,WMR18,Danino18,Hufton18,Su19,WM19,Shnerb19,Marrec20}.
In finite populations, demographic noise (DN) is another form of randomness that can lead to fixation (one species takes over the population~\cite{Kimura,Ewens}).
DN is strong in small populations and  negligible in large ones. Importantly,
the evolution of a population composition is often coupled with the dynamics
of its size~\cite{Roughgarden79,Leibler09,Melbinger2010,Cremer2011,Cremer2012,Melbinger2015a,Gokhale16}.
This can lead to coupling between DN and environmental variability (EV), with
external factors affecting the population size, which in turn modulates the DN strength.
The interplay between EV and DN plays a key role in microbial communities~\cite{Wahl02,Rainey03,Patwas09,Wienand15,Brockhurst07a,Brockhurst07b,Coates18,Cremer19}:
the variations of their composition and size are vital
to understand the mechanisms of antimicrobial resistance~\cite{Coates18,Marrec20}, and
may lead to \textit{population bottlenecks},
where new colonies consisting of few individuals are prone to fluctuations~\cite{Wahl02,Patwas09,Brockhurst07a,Brockhurst07b,Cremer19}. Interactions between microbial communities and environment have also been found to influence  cooperative behavior in  {\it Pseudomonas fluorescens} biofilms~\cite{Rainey03,Brockhurst07a,Brockhurst07b}. EV and DN are also important in ecology, \textit{e.g.}, in modeling tropical forests~\cite{Chisholm14,Kessler14,Kalyuzhny15}, and in
 gene regulatory networks~\cite{Assaf13,Assaf15}.

In most studies, there is no interdependence between the fluctuations stemming from DN and EV,
with growth rates often assumed to vary independently of the population size~\cite{May73,Karlin74,Thattai04,Kussell05,Acar08,Gaal10,Assaf08,Assaf09,He10,Tauber13,Assaf13,AMR13,Assaf15,Ashcroft14,Kussell05b,Melbinger15,Hufton16,Danino18,WMR18,Hufton18,Assaf18}.
Hence, there is as yet no systematic comparison of the dynamics under random and periodic switching: some works report that they
lead to similar evolutionary processes while others find differences, see {\it e.g.}, Refs.~\cite{Thattai04,Shnerb19}.
Here, we systematically study the {\it coupled influence} of EV and DN on the dynamics of a  population, where slow- and fast-growing strains compete for resources
subject to a randomly- and periodically-switching carrying capacity.

\begin{figure}[h!]
\includegraphics[width=0.7\linewidth]{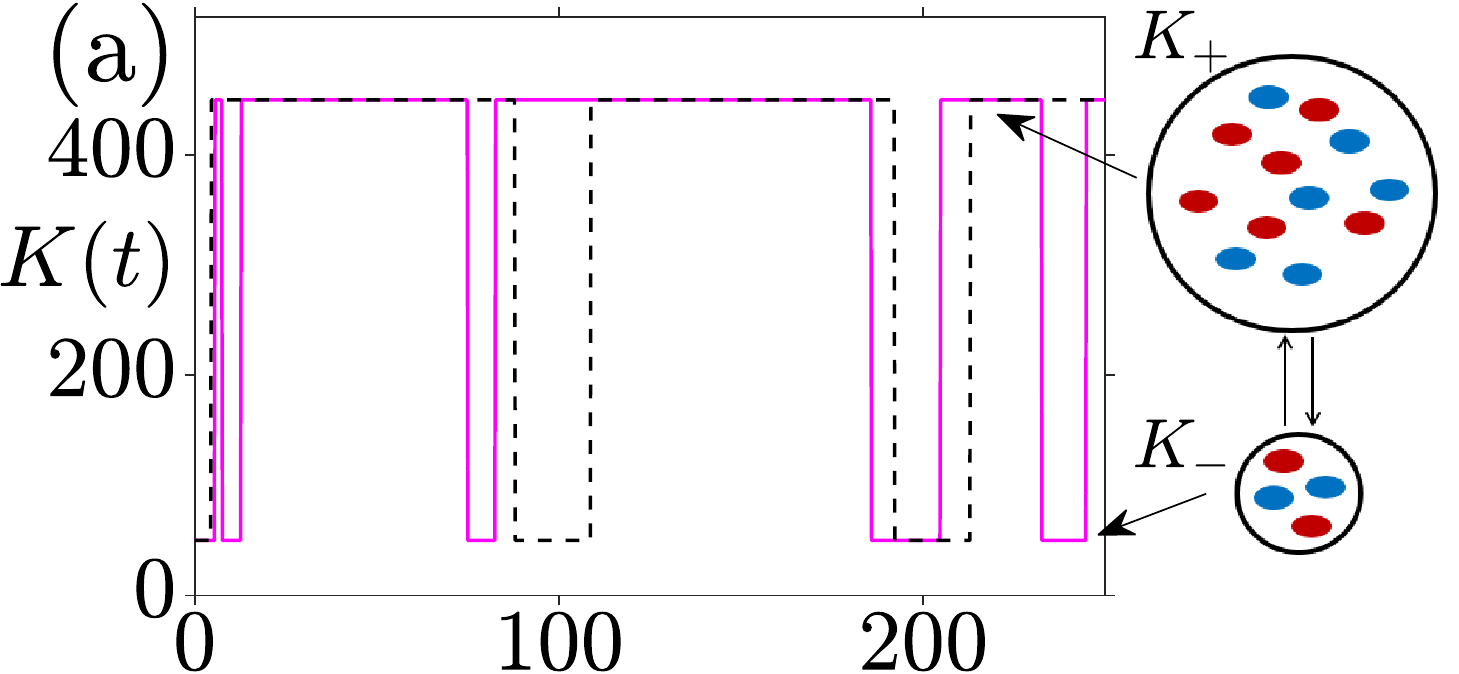}
\includegraphics[width=0.45\linewidth]{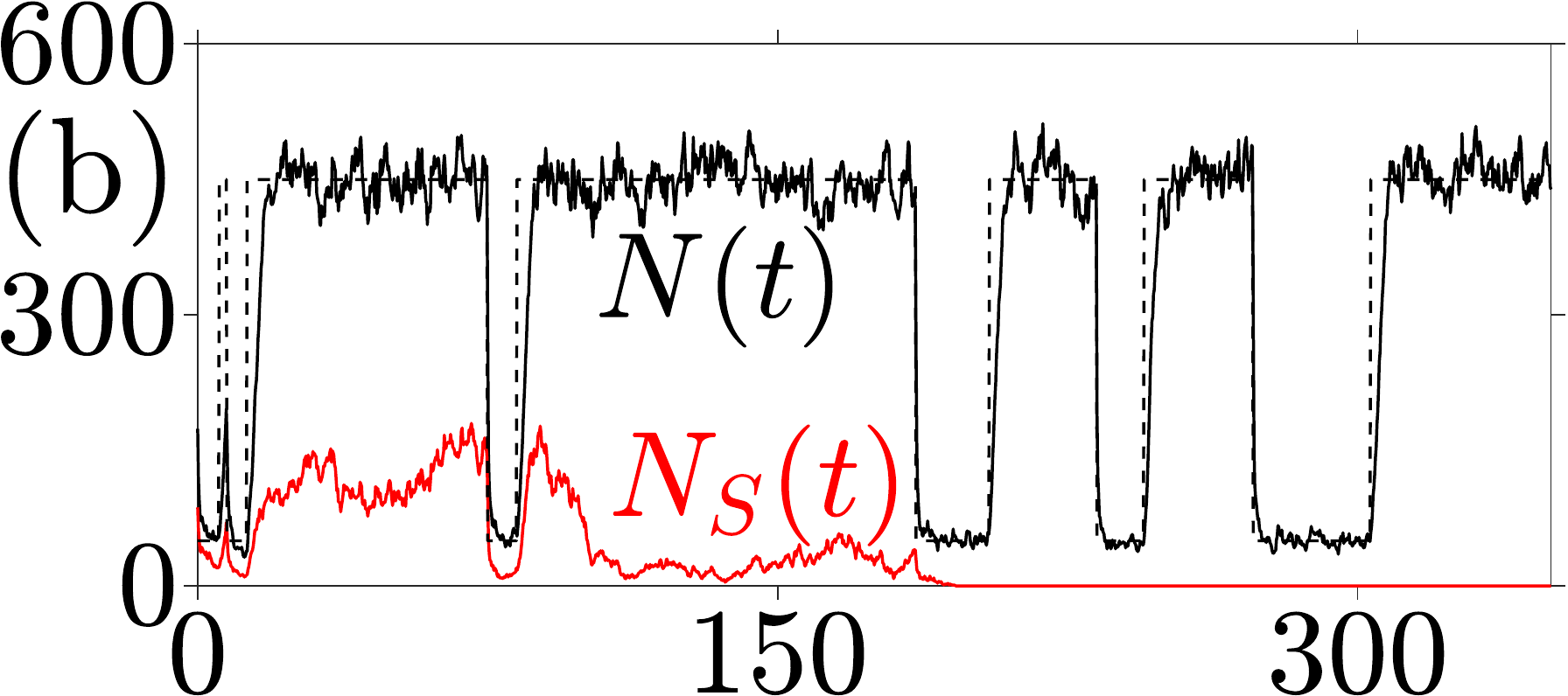}
\includegraphics[width=0.45\linewidth]{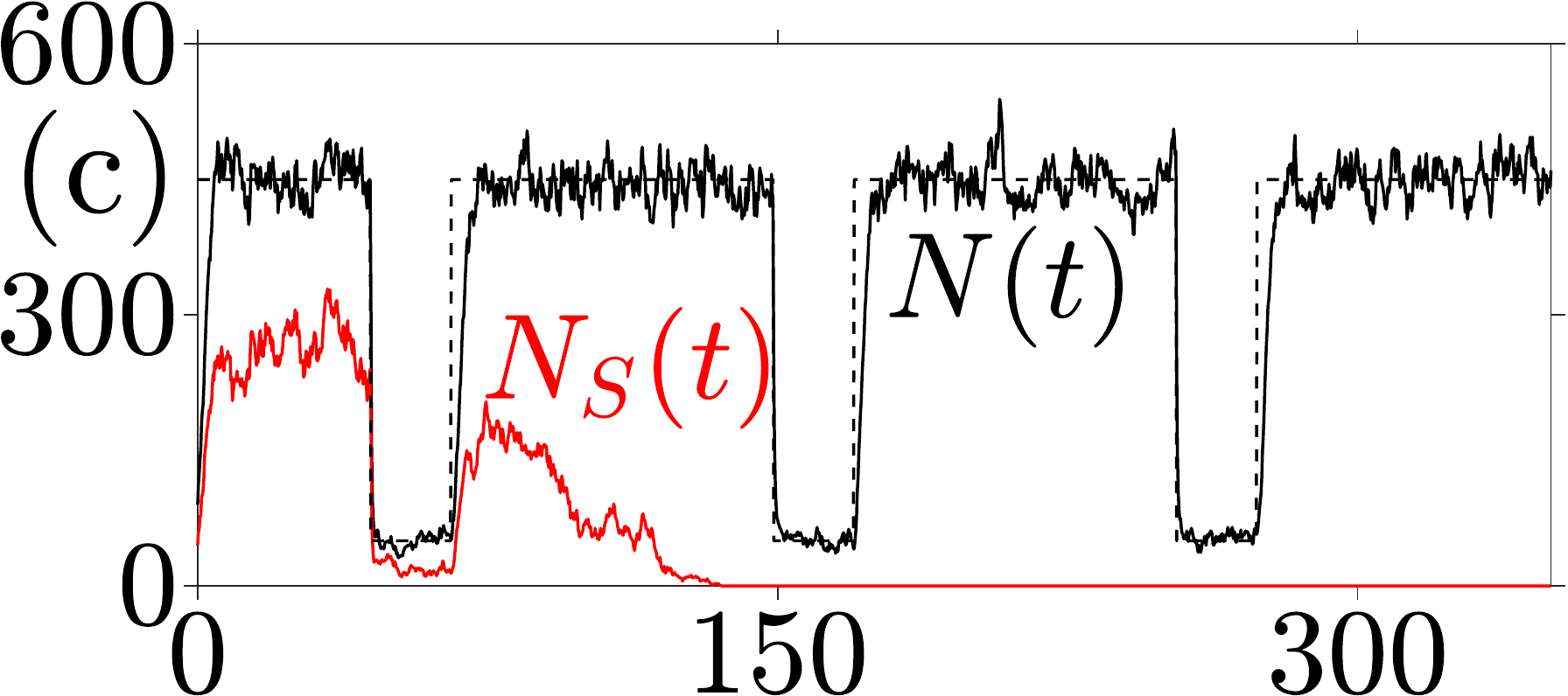}
\caption{
(a) $K$ vs. time $t$: asymmetric random (pink/light gray) and periodic (black dashed) switching between
$K_+$  and $K_-$ yield fluctuating population composition and size
(large/small circles), see text.
(b, c) Typical realizations of $N$ (black), $N_S$ (red/gray) and $K$ (black dashed) vs. $t$ under random (b) and periodic (c) switching: composition changes until fixation occurs.
Here $(s, K_0 , \nu, \gamma, \delta, x_0) = (0.02, 250, 0.03, 0.8, 0.6, 0.5)$.
}
\label{fig:Fig1}
\end{figure}

A distinctive feature of this model is that it accounts for the  stochastic  or periodic
depletion and recovery of resources via a binary environment, varying
with a finite correlation time or  period, and the {\it  DN and EV coupling}, see Fig.~\ref{fig:Fig1}. This setting is simple enough to enable  us to scrutinize  whether environmental  perturbations of different nature lead to the same  dynamics, and includes many features  (switching environment, varying population size)
that can be tested in controlled microbial experiments~\cite{Acar08,Leibler09,Cremer2012,Wienand15,Cremer19}.

To address the fundamental question of evolution under stochastic  and deterministic variations, we consider random and periodic environmental switching. This allows us to elucidate the influence of EV
 on the population size distribution (PSD) and the fixation properties.
We analytically show that
the PSD is generally broader under intermediate and fast random switching
than under periodic variations, leading to markedly different
fixation probabilities.
 We also determine the switching conditions for which the slow strain's fixation
 probability is maximized.

We consider a well-mixed population of time-fluctuating size $N(t)=N_{S}(t) + N_{F}(t)$
consisting of two strains. At time $t$,  $N_S(t)$ individuals are of a slow-growing strain $S$,
corresponding to a fraction $x=N_S/N$ of the population, and $N_F$ are of a fast-growing species $F$. The respective per-capita growth rates of
$S$ and $F$ are $(1-s)/\bar
f$ and $1/\bar f$, which sets the model's time scale~\cite{SM}. Here,
$\bar f=(1-s)x+1-x=1-sx$ is
the population average fitness and $0<s\ll1$ denotes the small selective growth
advantage of $F$ over $S$~\cite{Melbinger2010,Cremer2011,KEM1,KEM2}. Growth is limited by a logistic death rate $N/K$, where $K\gg 1$ is the carrying capacity.
Population dynamics is often idealized by assuming a static environment (constant $K$) yielding a constant or logistically-varying $N$~\cite{Moran62,Ewens,Blythe07,Antal,Nowak}.
Here, we instead consider a population of fluctuating size subject to a time-varying environment, and obeying the  birth-death process~\cite{KEM2,SM}: $N_{S/F}  \xrightarrow{T_{S/F}^{+}}  N_{S/F}+ 1 \quad \text{and} \quad
N_{S/F}  \xrightarrow{T_{S/F}^{-}}  N_{S/F}- 1$,
with transition rates $T_{S}^{+}= (1-s)N_S/\bar{f}$,  $T_{F}^{+}= N_F/\bar{f}$ and
$T_{S/F}^{-}= (N/K(t))N_{S/F}$.
 We model EV  via a switching carrying capacity
\begin{eqnarray}
\label{eq:K}
K(t) =K_0[1+\gamma\xi_{\alpha}(t)], \;\;\;\;\xi_\alpha(t)\in \{-1,+1\},
\end{eqnarray}
where $K_0 \equiv (K_+ + K_-)/2$ and $\gamma \equiv (K_+ - K_-)/(2K_0)$,
while $\alpha\in \{r,p\}$ and $\gamma={\cal O}(1)$.
Here, resources  vary either randomly ($\alpha=r$)
or periodically ($\alpha=p$),
between states of scarcity, $K=K_-$ ($\xi_\alpha=-1$), and abundance, $K=K_+$ ($\xi_\alpha=+1$),
where $K_+>K_- \gg 1$,
causing fluctuations of population size and composition,
see Fig.~\ref{fig:Fig1}. This specific choice of birth-death process coupled to a time-varying binary environment is arguably the simplest biologically-relevant model
to study population dynamics under the joint influence of EV and DN, see Sec.~S1.1 in \cite{SM}.

When $K(t)$ switches {\it randomly},
 $\xi_{r}$ is a colored {\it asymmetric  dichotomous (telegraph) Markov noise} (ADN)~\cite{Bena06,HL06}, with the transition
$\xi_{r} \to -\xi_{r}$ occurring at rate $\nu_\pm$ when $\xi_{r}=\pm 1$. The (average) switching rate is $\nu=(\nu_+ + \nu_-)/2$ while  $\delta=(\nu_- - \nu_+)/(2\nu)$ measures the switching asymmetry ($|\delta|<1$, with $\delta=0$ for symmetric switching). In this model, the ADN  is a stationary noise of mean  $\langle \xi_{r}(t)  \rangle= \delta$ and  autocorrelation function $\langle \xi_{r}  (t) \xi_{r}  (t')\rangle-\langle \xi_{r}  (t)\rangle \langle\xi_{r}  (t')\rangle = (1-\delta^2)~e^{-2\nu|t-t'|}$
($\langle \cdot \rangle$ denotes ensemble averaging).
When $K(t)$ switches {\it periodically},
 $\xi_p$ is a {\it rectangular wave} defined by the rectangular function, ${\rm rect}(\cdot)$~\cite{rect},
 of period $T=(1/\nu_+) +(1/\nu_-)=2/[(1-\delta^2)\nu]$:
\begin{eqnarray*}
 \label{eq:alphap}
\xi_p(t)\!=\!\sum_{j=-\infty}^{\infty}\!\left[{\rm rect}\left(\frac{t\!+\!\frac{1}{2\nu_+}\!+\!jT}{1/\nu_+}\right)\!-\!
{\rm rect}\left(\frac{t\!-\!\frac{1}{2\nu_-}\!+\!jT}{1/\nu_-}\right)\right],
\end{eqnarray*}
which becomes
 the square wave $\xi_p(t)=-{\rm sign}\left\{\sin{(\pi \nu t)}\right\}$ when $\delta=0$.  In our simulations, $\xi_p(t)$ averaged over a period $T$ has the  same mean and variance as $\xi_{r}(t)$. Hence, the mean and variance of $K(t)$
are the same for $\alpha\in \{r,p\}$:
$\langle K(t) \rangle= K_0(1+ \gamma \delta)$ and ${\rm var}(K)=(\gamma K_0)^2 (1-\delta^2)$~\cite{average}.

The model considered here gives rise to a long-lived {\it population size distribution} (PSD) followed by an eventual extinction of the entire population which occurs after a very long time (practically unobservable when $K_0\gg 1$~\cite{KEM1}~\cite{meta}). Below, we focus on intermediate times $t={\cal O}(s^{-1})$,  a timescale on which one species is likely to have gone extinct and the other fixated the population that is in its long-lived PSD~\cite{SM}.
We show that the {\it fixation probabilities} strongly depend on the PSD which is encoded in the underlying master equation~\cite{Assaf10,Redner,KEM2,WM19}, see \cite{SM} for details.

Insight into the dynamics is gained by ignoring fluctuations and considering the mean-field picture of a very large population with  constant  $K=K_0$. Here, $N$ and $x$ evolve according to $dN/dt\equiv \dot{N}=N(1-N/K_0)$ and $\dot{x}=-sx(1-x)/(1-sx)$~\cite{Melbinger2010,Cremer2011,SM}, with
 $x$ decaying on a timescale $t\sim s^{-1}\gg 1$ and $N(t)= {\cal O}(K_0)$ after $t= {\cal O}(1)$~\cite{IC}. Thus, a timescale separation occurs:  the relaxation of $x$ is much slower than that of $N$.

However, when dealing with a finite population, DN (random birth/death events) must be taken into account, yielding the  fixation of one of the species. The $S$ fixation probability, given a fixed population size $N$, and an initial fraction $x_0=N_S(0)/N(0)$ of $S$ individuals, is~\cite{Ewens,Antal,Redner}
\begin{eqnarray}
 \label{eq:phi1}
\phi(x_0)|_N &= \left[e^{-Nx_0\ln(1-s)}-1\right]/\left[e^{-N\ln(1-s)}-1\right],
\end{eqnarray}
which exponentially decreases with $N$. For $s\ll N^{-1/2}\ll 1$ (``diffusion approximation''), this  simplifies to $\phi(x_0)|_N \simeq  (e^{-Ns(1-x_0)}-e^{-Ns})/(1-e^{-Ns})$~\cite{Blythe07,KEM1,KEM2}.
While Eq.~(\ref{eq:phi1}) provides a good approximation for the fixation probability also when $N$ fluctuates about constant $K=K_0$, this picture changes drastically when, in addition to DN, the population is subject to a time-varying $K(t)$, see Fig.~\ref{fig:Fig1}. Below we study the joint influence of EV and DN  on the
PSD and fixation properties.
\begin{figure}[t!]
\includegraphics[width=0.303\linewidth]{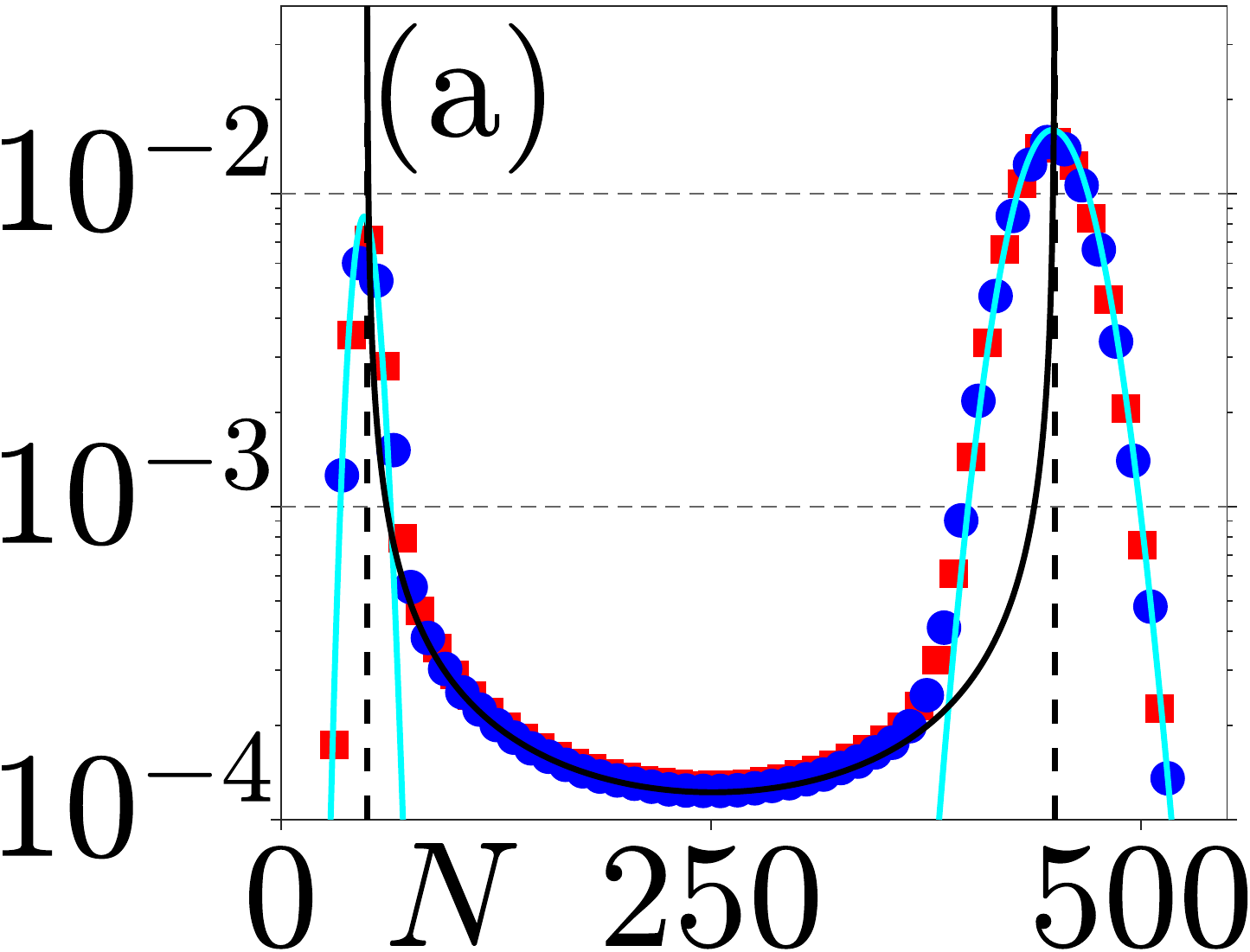}
\includegraphics[width=0.18\linewidth]{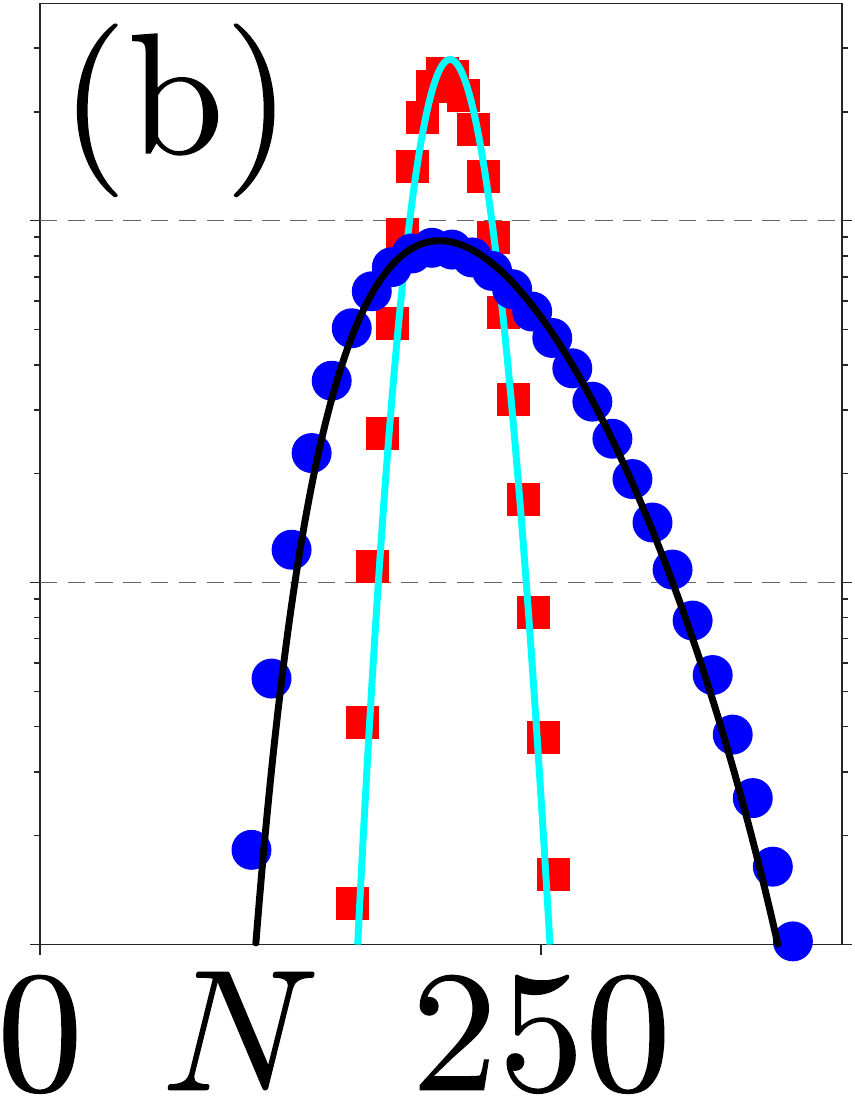}
\includegraphics[width=0.245\linewidth]{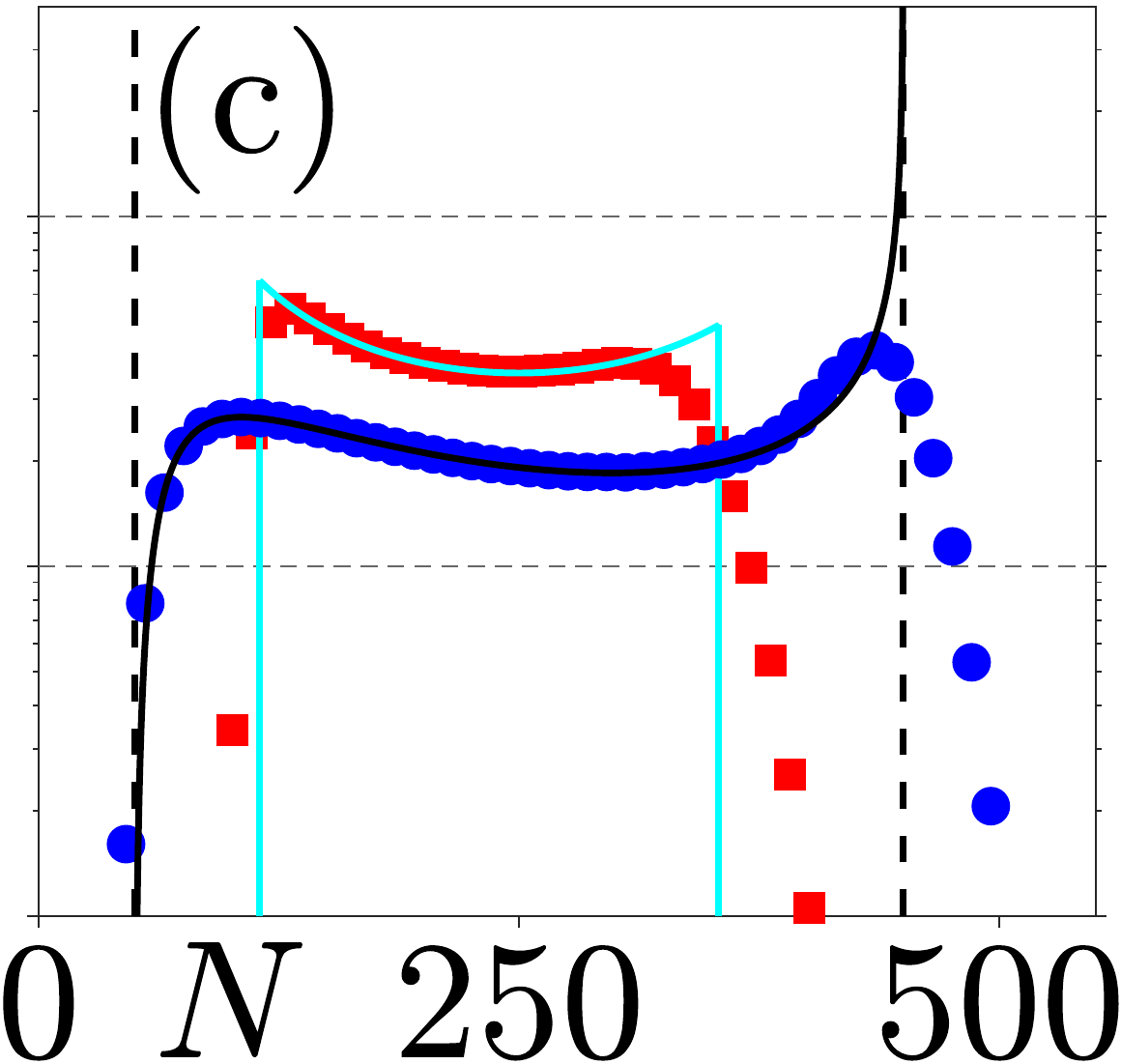}
\includegraphics[width=0.15\linewidth]{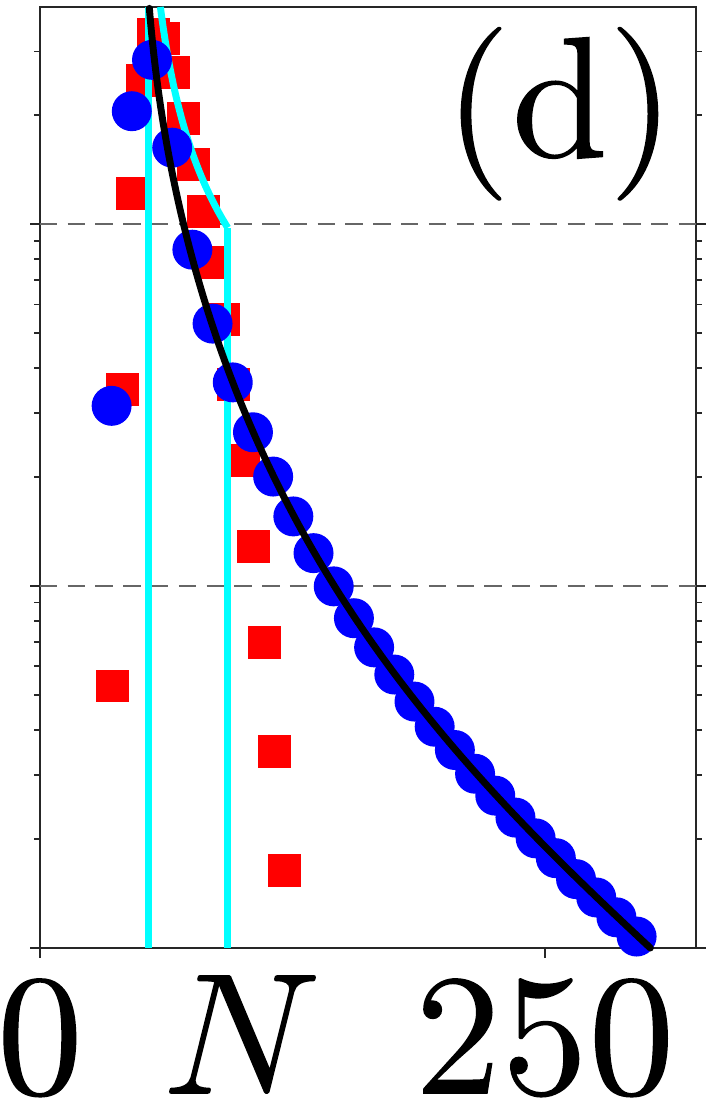}
\caption{$P_{\nu}^{(r)}(N)$ (blue/dark gray)
and $P_{\nu}^{(p)}(N)$ (red/gray) for different $\nu$: (a) $\nu=0.05$, (b) $\nu=17.5$, (c) $\nu=1.4$, (d) $\nu=1$. Symbols are from simulations;
solid black lines in  (a)-(d) are from $P_{\nu}^{{\rm PDMP}}$, those in
cyan/light gray
are from $P_0(N)$ in (a),
$P_{\nu}^{{\rm Kap}}$
in (b), and $P_{\nu}^{{\rm PPP}}$ in (c,d); vertical
 lines show $N=K_{\pm}$ (dashed) in (a,c)  and $N=N_{{\rm min}/{\rm max}}$ (cyan/light gray) in (c,d), see text and Sec.~S2.3 in \cite{SM}; horizontal dashed lines are eyeguides.
Here $(s, K_0, \gamma, x_0)=(0.05,250,0.8,0.6)$,
$\delta=0.7$ in (a)-(c)
and $\delta=-0.5$ in (d).
}
\label{fig:Fig2}
\end{figure}

{\it Population size distribution.} Simulations show that the marginal quasi-stationary PSD,
$P_\nu^{(\alpha)}(N)$ (unconditioned of $\xi_{\alpha}$), is characterized by different regimes depending on the switching rate $\nu$,
with markedly different features in the case of random and periodic variations when $\nu={\cal O}(1)$ and $\nu\gg 1$, see Fig.~\ref{fig:Fig2}.

The case of random switching can be treated as in \cite{KEM1,KEM2}
for $\delta=0$. Upon ignoring DN,  $N(t)$ is therefore
subject only to ADN according to the
piecewise-deterministic Markov process (PDMP)~\cite{PDMP1,Davis84,SM} defined by the stochastic differential equation
$\dot{N}=N\left[1-(N/{\cal K})(1-\gamma \xi_r)/(1-\gamma\delta)\right]$, where
${\cal K}\equiv  K_0(1-\gamma^2)/(1-\gamma\delta)$. When $\nu \to \infty$,
the ADN self-averages,
$\xi \xrightarrow{\nu \to \infty} \langle \xi\rangle=\delta$, and $N  \xrightarrow{\nu \to \infty} {\cal K}$. The marginal
PSD of this PDMP has support $[K_{-},K_{+}]$ and
can be computed explicitly~\cite{HL06,KEM2}: its expression $P_{\nu}^{{\rm PDMP}}(N)$ is given by Eq.~(S22) of \cite{SM}.
Although $P_{\nu}^{{\rm PDMP}}$  only accounts for EV,  when $K_0\gg 1$ and $\gamma={\cal O}(1)$, it
captures the  peaks of  $P^{(r)}_{\nu}$  and the average population size, see Figs.~\ref{fig:Fig2} and S3(b) in \cite{SM}.
However, $P_{\nu}^{{\rm PDMP}}$ ignores DN and cannot capture the width of $P^{(r)}_{\nu}$ about its peaks, see Fig.~\ref{fig:Fig2}(a,c,d). Yet, this can be remedied, by a linear noise approximation, see \cite{KEM2} and  Sec.~S3.2 in \cite{SM}.
 We can also obtain a PDMP-like approximation (ignoring DN)~\cite{Doering85,Bena06}  of the periodic PSD  by solving the  mean-field equation for $N(t)$ with periodic $K(t)$. By inverting  $N(t)$ we then obtain the piecewise periodic process (PPP) approximation
$P_{\nu}^{{\rm PPP}}$ of  $P^{(p)}_{\nu}$, given by (S19) in Sec.~S2.3 of \cite{SM}, which is valid over a broad range of switching rates, see  Fig.~\ref{fig:Fig2}(c,d) and below.

Furthermore, for periodic switching, the full $P^{(p)}_{\nu}$ can be found analytically in the limits of very slow ($\nu \to 0$) and fast ($\nu \gg 1$) variations.
For $\nu \to 0$ the carrying capacity is initially
randomly allocated and almost constant, \textit{i.e.} $K(t)\simeq K(0)$. The PSD is thus the same
for periodic and random switching:  $P^{(p)}_{0}= P^{(r)}_{0}\equiv
 P_{0}$, and can be computed from the master equation. Assuming $K_0\gg 1$ and $\gamma={\cal O}(1)$, the PSD is bimodal  with peaks about $N=K_{\pm}$, whose intensity depends on $\delta$~\cite{SM}:
$P_{0}(N)\simeq [(1+\delta)~K_+^{N+1} e^{-K_+}+(1-\delta)~K_-^{N+1} e^{-K_-}]/[2N\cdot N!]$. This result excellently agrees with simulations, see Fig.~\ref{fig:Fig2}(a).
Under fast periodic switching,
$P^{(p)}_{\nu}$ differs markedly from its random counterpart, see Fig.~\ref{fig:Fig2}(b).
An approximate expression of $P^{(p)}_{\nu}$ to leading order in $1/\nu$, here denoted by $P_{\nu}^{{\rm Kap}}$, and peaked at $N={\cal K}$ when  $\nu\to\infty$ is given by Eq.~(S15) in \cite{SM}.  $P_{\nu}^{{\rm Kap}}$ is obtained from the master equation by using the WKB approximation~\cite{Elgart04} and the Kapitza method~\cite{Landau76,Assaf08,Assaf18}, \textit{i.e.} separating the dynamics into fast and slow variables, see Sec.~2.2 of \cite{SM}.
In  Fig.~\ref{fig:Fig2}(b), we notice that both $P^{(p)}_{\nu}\simeq P_{\nu}^{{\rm Kap}}$ and $P^{(r)}_{\nu}\simeq P^{{\rm PDMP}}_{\nu}$ are unimodal and peaked about $N\approx {\cal K}$ when $\nu \gg 1$, but
$P_{\nu}^{{\rm Kap}}$ is much sharper and narrower than $P^{{\rm PDMP}}_{\nu}$. In fact, the variance of $P^{{\rm PDMP}}_{\nu}$ scales as $K_0^2/\nu$
 when $1 \ll \nu\ll K_0$, and is much larger than that of $P_{\nu}^{{\rm Kap}}$, see Sec.~S4.3 in \cite{SM}.

Note that while $P_0$ and $P_{\nu}^{{\rm Kap}}$ account for DN {\it and} EV,
$P_{\nu}^{{\rm PDMP}}$ and $P_{\nu}^{{\rm PPP}}$  only account for  EV.
Yet, DN is negligible compared to EV when $1\lesssim \nu\ll K_0$
and $1\lesssim \nu\ll \sqrt{K_0}$ in the random and periodic cases, respectively~\cite{SM}. $P_{\nu}^{{\rm PDMP}}$ and $P_{\nu}^{{\rm PPP}}$ are therefore suitable approximations of $P^{(\alpha)}_{\nu}$ in those regimes.

In particular, $P_{\nu}^{{\rm PDMP}}$ and $P_{\nu}^{{\rm PPP}}$ allow us to characterize interesting phenomena arising  in the intermediate {\it asymmetric} switching regime where $\nu\gtrsim 1$ with
$\nu_->1$ and $\nu_+<1$,
or $\nu_-<1$ and $\nu_+>1$, \textit{i.e.} when $1/(1+|\delta|)<\nu<1/(1-|\delta|)$.
In the former case ($\delta>0$),
$P^{(r)}_{\nu}$ has a peak at $N\approx K_+$ and, under sufficiently strong  EV,  exhibits also a peak $N^*$ between $K_-$ and $K_+$
(\textit{i.e.} $K_-<N^*<K_+$), whose position is aptly captured by $P_{\nu}^{{\rm PDMP}}$, see Fig.~\ref{fig:Fig2}(c) and Sec.~S3.1 in \cite{SM}. In Fig.~\ref{fig:Fig2}(c),
$P_{\nu}^{(p)}$ is less broad than $P_{\nu}^{(r)}$ and has also two peaks well reproduced by $P_{\nu}^{{\rm PPP}}$ whose 
support is narrower than that of $P_{\nu}^{{\rm PDMP}}$~\cite{SM}.
When $\nu\gtrsim 1$, with $\nu_-<1$ and $\nu_+>1$ ($\delta<0$), $P^{(r)}_{\nu}$ and $P^{(p)}_{\nu}$
exhibit a single peak at $N\approx K_-$, well predicted by
$P_{\nu}^{{\rm PDMP}}$ and $P_{\nu}^{{\rm PPP}}$,
 with the latter being narrower than the former in Fig.~\ref{fig:Fig2}(d).
In fact, Figs.~\ref{fig:Fig2}(b) and \ref{fig:Fig2}(c) show that the transition from
 bimodal to unimodal PSD
 (slow to fast switching) is generally more abrupt under periodic than under random switching.

\begin{figure}[t!]
\includegraphics[width=0.42\linewidth]{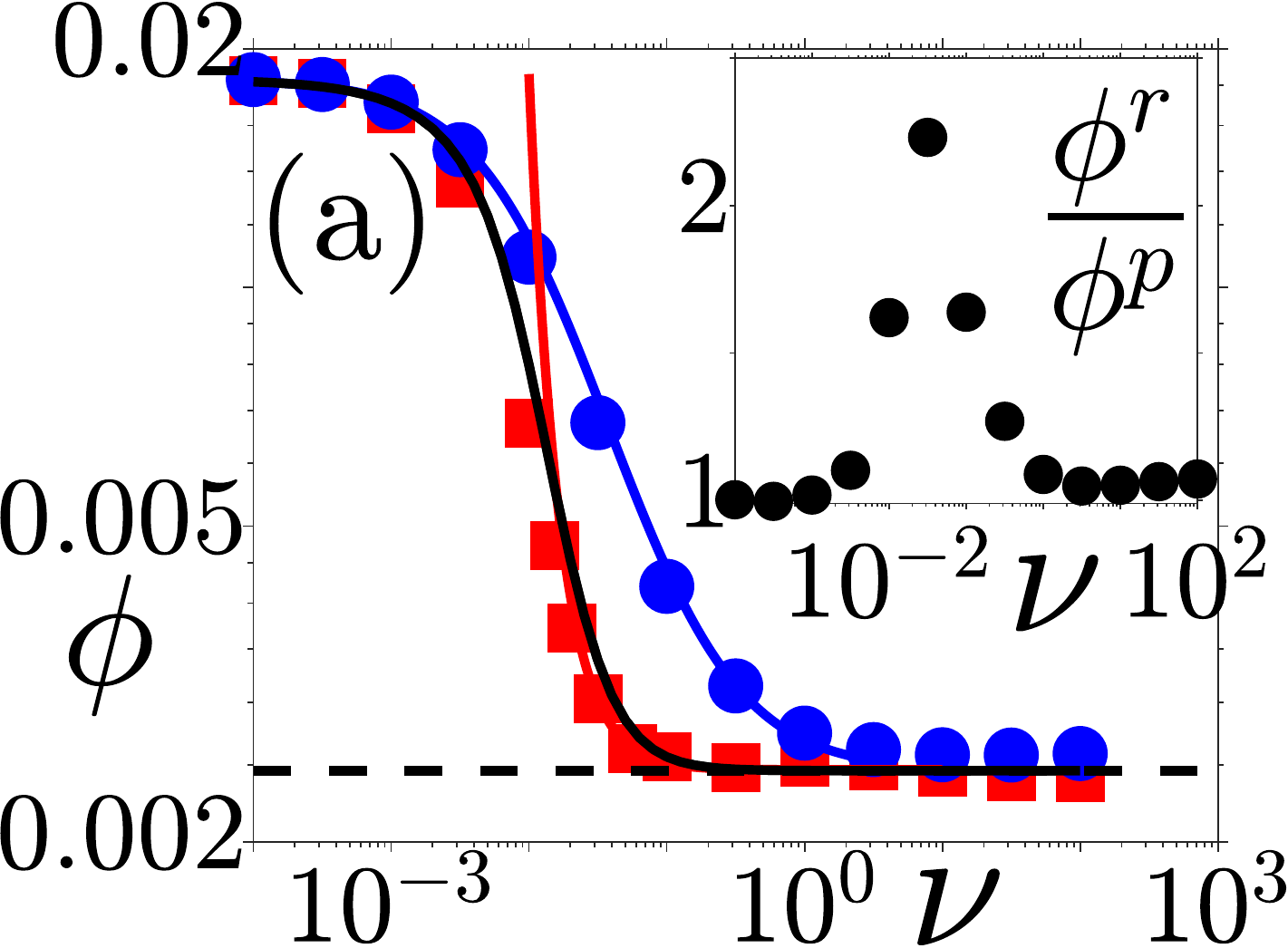}
\includegraphics[width=0.285\linewidth]{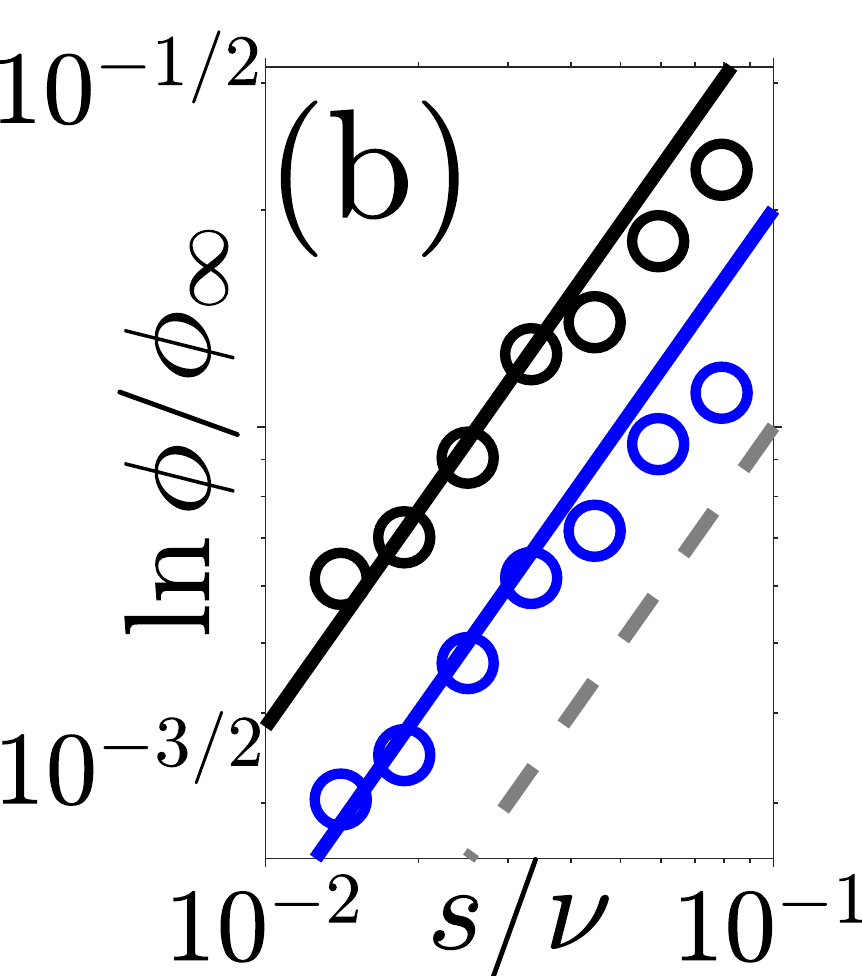}
\includegraphics[width=0.263\linewidth]{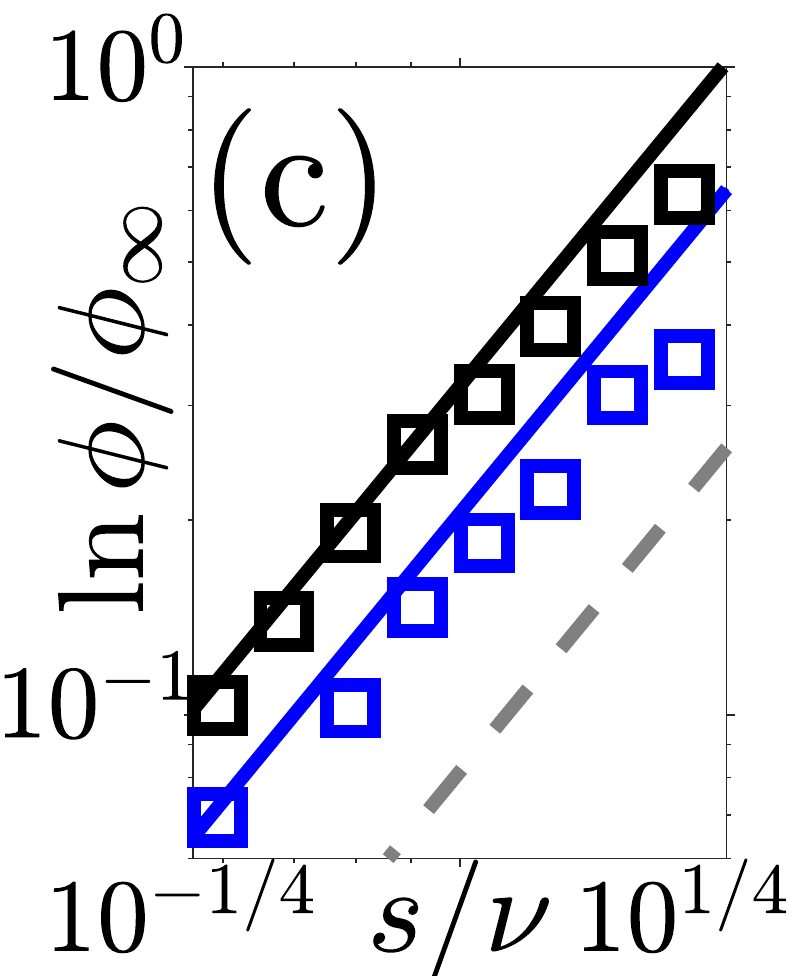}
\includegraphics[width=0.46\linewidth]{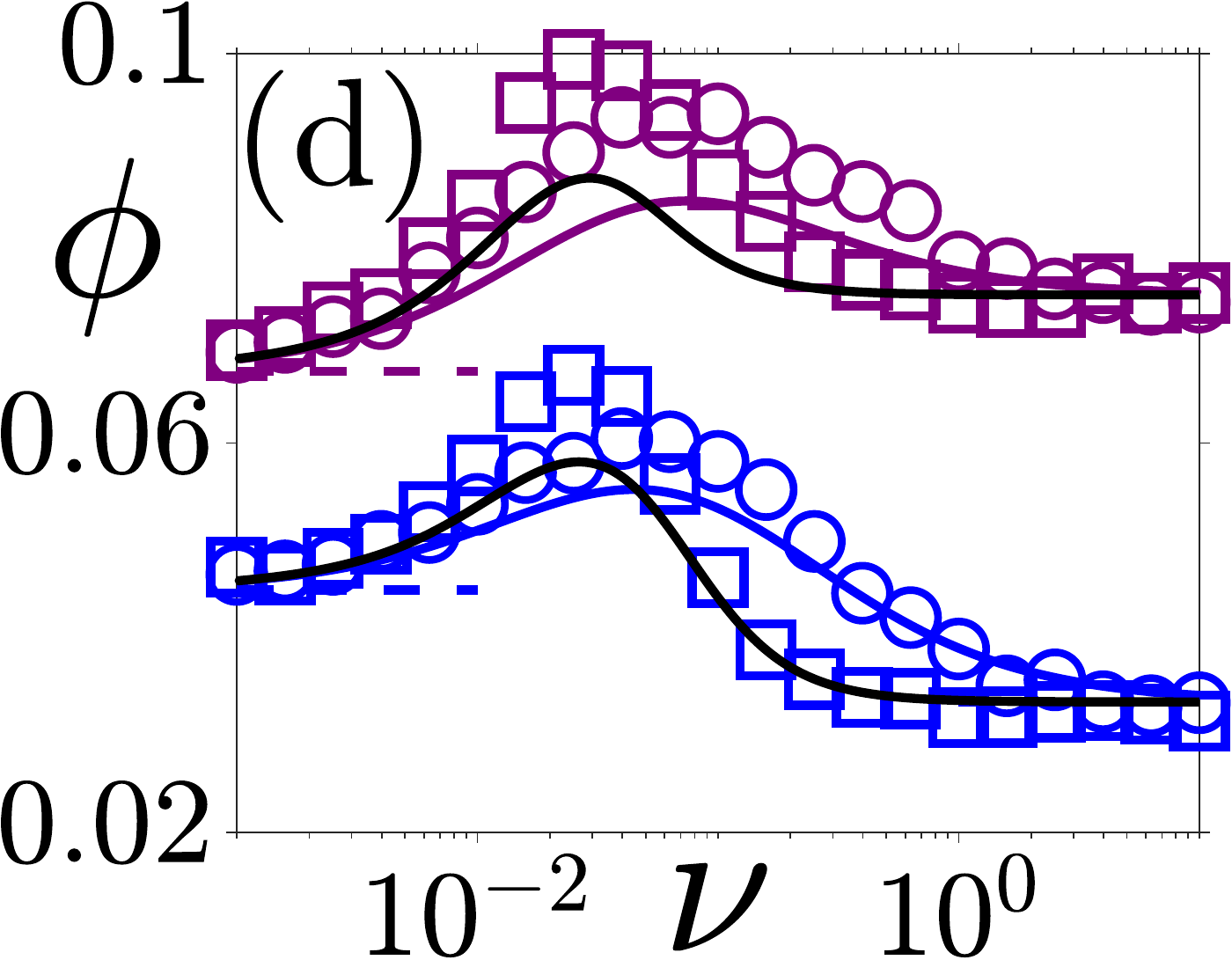}
\includegraphics[width=0.49\linewidth]{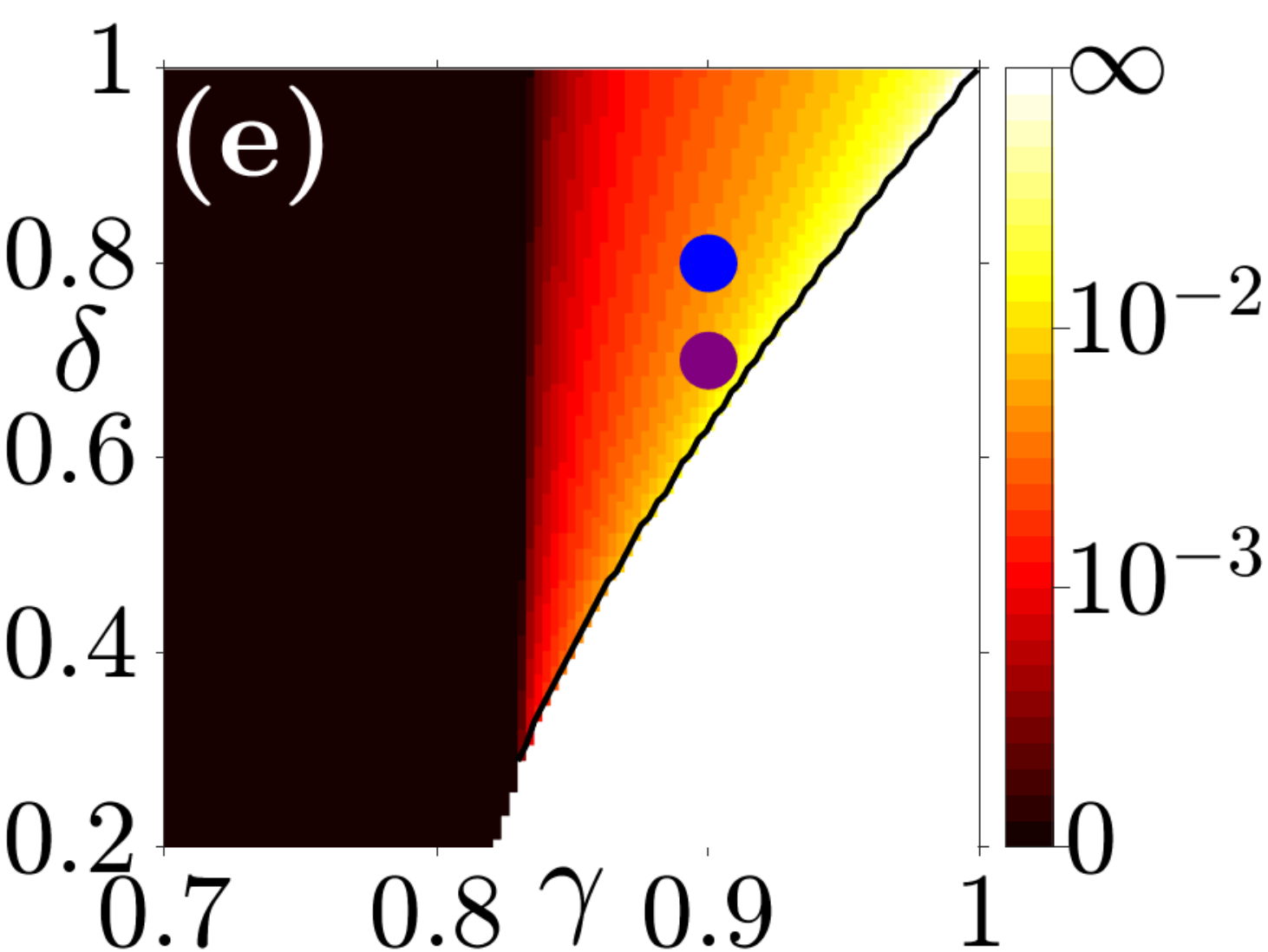}
\caption{(a)-(d)
fixation probability for random/periodic switching (circles/squares): symbols are from simulations. In (a) solid lines are  from (S38) (blue/dark gray, $\alpha=r$),
 (\ref{eq:saddle_ap}) (red/gray, $\alpha=p$)
 and  (S39) (black, $\alpha=p$) of \cite{SM}.
(a) $\phi_{\alpha}$  versus $\nu$
with $\delta=0.2$; dashed  line shows  $\phi^{(\infty)}$. Inset: $\phi_{r}/\phi_{p}$ versus $\nu$ with $\delta=0.2$.
 (b,c) $\ln{(\phi_{\alpha}/\phi^{(\infty)})}$
 versus $s/\nu$ for random (b) and periodic (c) switching
 with $\delta=0.2$ (black) and  $\delta=0$ (blue/dark gray).
 Dashed gray lines are eyeguides $\propto s/\nu$ in (b) and  $\propto (s/\nu)^2$ in (c).
 (d) Non-monotonic $\phi_{\alpha}(\nu)$  with $\delta=0.7$ (purple/ gray) and  $\delta=0.8$ (blue/dark gray). Solid lines are from
 (S38) (purple/gray and blue/dark gray) and from
 (S39) (black) of \cite{SM};
 dashed lines show  $\phi^{(0,\infty)}$.
 $\phi_r(\nu)$ and $\phi_p(\nu)$ are maximal
 at
 $\nu=\nu_{r}^*\approx 0.1$ and  $\nu=\nu_{p}^*\approx 0.07$, see text.
 (e)  Heatmap of  $\nu^*_{r}$ (see Sec.~S5.1 in \cite{SM} for details
 and heatmap of $\nu^*_{p}$):
$\nu^*_{r} \to 0,\infty$ in the black
and white areas, respectively;  $\phi_r(\nu)$ is non-monotonic in the red-yellow/gray area, with
$\nu_{r}^*\approx 0.01$ (red/dark gray) - $\nu_{r}^*\approx 0.1$ (yellow/light gray), see vertical bar.
Symbols are for  $\delta=0.7$ (purple/dark gray) and  $\delta=0.8$ (blue/black). Here $(s,K_0,\gamma,x_0)=(0.025,800,0.7,0.5)$
in (a)-(c) and $(0.05,250,0.9,0.6)$ in (d,e).}
\label{fig:Fig3}
\end{figure}
{\it Fixation probability.}
We denote by $\phi_{\alpha}$ the slow ($S$) species fixation probability  subject to $\alpha$-switching ($\alpha\in\{r,p\}$). As aforementioned, when $s\ll 1$ and $t\gtrsim{\cal O}(1)$, the system has settled in its long-lived PSD.
Thus, given $x_0$, $\phi_{\alpha}$ can be approximated by averaging $\phi(x_0)|_{N}$
 over  $P_{\nu/s}^{(\alpha)}(N)$, upon rescaling $\nu \to \nu/s$~\cite{KEM1,KEM2}
\begin{eqnarray}
 \label{eq:formula}
 \phi_{\alpha}(\nu)\simeq
 \int_0^{\infty}~P_{\nu/s}^{(\alpha)}(N)~\phi(x_0)|_{N}~dN, \quad \alpha\in\{r,p\}.
\end{eqnarray}
This result is  valid under weak selection,  $1/K_0\ll s\ll 1$,
when there are  ${\cal O}(\nu/s)$
 switches prior to fixation~\cite{KEM1,KEM2,SM}. The difference between $\phi_{r}$ and $\phi_{p}$
stems from the different $\nu$-dependence
of $P_{\nu}^{(r)}$ and   $P_{\nu}^{(p)}$,  see  Fig.~\ref{fig:Fig2}. Approximations of $\phi_r$ and  $\phi_p$ are obtained by respectively substituting $P_{\nu/s}^{(\alpha)}$ by $P_{\nu/s}^{{\rm PDMP}}$ and  $P_{\nu/s}^{{\rm PPP}}$ into Eq.~(\ref{eq:formula}). This yields expressions (S38) and (S39) of \cite{SM} which are valid over a broad range of $\nu$~\cite{KEM1,SM}, see Fig.~\ref{fig:Fig3} and S2(c,d) of \cite{SM}. Notably, when $\nu/s\gg 1$, $\phi_{p}$ is better approximated by substituting  $P_{\nu/s}^{(p)}$ by $P_{\nu/s}^{{\rm Kap}}$ in Eq.~(\ref{eq:formula}), see below and \cite{SM}.

When $\nu \to 0$ (slow switching), on average
there are  almost no switches prior to fixation and $P_{\nu/s}^{(\alpha)}$
is peaked at $N=K_{\pm}$. Hence, with Eq.~(\ref{eq:formula}),
$\lim_{\nu \to 0}\phi_{\alpha}(\nu) \simeq \phi^{(0)}
=[(1-\delta)\phi(x_0)|_{K_-}+(1+\delta)\phi(x_0)|_{K_+}]/2$.
 Fig.~\ref{fig:Fig3}(d)  confirms that $\phi_{r}$ and $\phi_{p}$ approach  $\phi^{(0)}$ when $\nu/s\ll 1$.

When $\nu/s\gg 1$ (fast switching),
$P_{\nu/s}^{(\alpha)}$ is sharply peaked at $N\simeq {\cal K}$, see
Fig.~\ref{fig:Fig2}(b), and  to leading order
$\lim_{\nu \to \infty}\phi_{\alpha}(\nu)\simeq  \phi^{(\infty)}=\phi(x_0)|_{{\cal K}}$~\cite{KEM1,KEM2}. Simulation results of Fig.~\ref{fig:Fig3}
confirm that at $\nu\gg s$, $\phi_{r}(\nu)$  and  $\phi_{p}(\nu)$ converge to  $\phi^{(\infty)}$.
Thus,  the fixation probability under fast random/periodic switching is the same to lowest order in $1/\nu$. Yet, the \textit{rate} of convergence differs, see Fig.~\ref{fig:Fig3}(a).
 This is explained by computing the  next-to-leading order of $\phi_{\alpha}$ in $\nu/s\gg 1$. For this, we use Eq.~(\ref{eq:formula}) with Eq.~(\ref{eq:phi1})
 and  $P_{\nu/s}^{{\rm PDMP}}$  and $P_{\nu/s}^{{\rm Kap}}$    for random and periodic switching, respectively. A saddle-point calculation,
  with $1/K_0\ll s\ll 1$, yields (see Sec.~S4 in \cite{SM})
  \begin{eqnarray}
   \label{eq:saddle_ap}
   \ln{\left(\frac{\phi_{\alpha}(\nu)}{\phi^{(\infty)}}\right)}\simeq
\begin{cases}
{\cal A}_r (s/\nu) &\quad (\alpha=r)\\
{\cal A}_p (s/\nu)^{2} &\quad (\alpha=p).
\end{cases}
  \end{eqnarray}
Here $\phi^{(\infty)}=e^{m/2}$, $m\equiv 2{\cal K}(1-x_0)\ln{(1-s)}$,  and
  ${\cal A}_r=m(4+m)(1-\delta^2)(\gamma/(1-\gamma\delta))^2/16$
while  ${\cal A}_p={\cal K}(1-(1+m/{\cal K})^3)(\gamma/(1-\gamma\delta))^2/72$.
 Thus, when $K_0s \gg 1$,  $\phi_{\alpha}(\nu)$ converges to $\phi^{(\infty)}$ much faster in the periodic
  than in the random case, see Fig.~\ref{fig:Fig3}(a)-(c).
The different asymptotic behavior can be understood by noting that $P_{\nu}^{(r)}$ is generally broader than $P_{\nu}^{(p)}$, with respective variances scaling as $\nu^{-1}$ and $\nu^{-2}$.  $N$ can thus attain smaller values  under random than periodic switching, which
 enhances   $\phi_{r}$ with respect to $\phi_{p}$~\cite{monotone}. When $\nu/s\gg 1$, $\phi_{r,p}$ is determined by the mean $\langle N\rangle\simeq{\cal K}$ of $P_{\nu}^{(\alpha)}$, and the different rate of convergence to $\phi^{(\infty)}$ stems from
 the deviations of $\langle N\rangle$ from ${\cal K}$, which decrease as
 $\nu^{-1}$ when $\alpha=r$ and $\nu^{-2}$ when $\alpha=p$, see Sec.~S4.3 in \cite{SM}.
  Another signature of the different asymptotic behavior is  the
  sharp peak of the  ratio $\phi_{r}/\phi_{p}$  at a nontrivial $\nu$, see Fig.~\ref{fig:Fig3}(a, inset).

 Under intermediate (rescaled) switching, $\phi_{\alpha}$  exhibits a rich behavior, see Fig.~\ref{fig:Fig3}(d).
When the switching asymmetry is sufficiently large,
$\phi_{\alpha}$ is a non-monotonic function of $\nu$
in a nontrivial region  $\gamma>\gamma_c(s)$, $\delta>\delta_c(\gamma,s)$ of the parameter space that
can be found from Eq.~(\ref{eq:formula}), see Fig.~\ref{fig:Fig3}(d,e) and Sec.~S5.1 in \cite{SM}.
The PDMP- and PPP-based
approximations [Eqs.~(S38) and (S39) in \cite{SM}] adequately capture  the $\nu$-dependence of
$\phi_{\alpha}$ in this regime, and its maximum at $\nu_{\alpha}^*\sim s$.
This optimal switching rate, which maximizes the $S$ species fixation probability at given ($\gamma$, $\delta$, $s$), corresponds to ${\cal O}(1)$ switches prior to fixation. The relative increase in $\phi_{\alpha}(\nu)$, given by
$\phi_{\alpha}(\nu_{\alpha}^*)/{\rm max}(\phi^{(0)},\phi^{(\infty)})-1$  reaches up to $30\%$, see Fig.~\ref{fig:Fig3}(d,e). In agreement with the PDMP- and PPP-based approximations, we find that $\nu_p^*\lesssim \nu_r^*$,
and $\phi_{p}(\nu_{p}^*)$ is narrower around the peak than $\phi_{r}(\nu_{r}^*)$, see Figs. ~\ref{fig:Fig3}(d,e) and ~S2(e)  of \cite{SM}.
When  the asymmetry is not too large  ($|\delta|<\delta_c$), $\phi_{\alpha}(\nu)$ is a monotonic function: it  increases/decreases with $\nu$ below/above a critical selection intensity $s_c$ (with $\gamma, \delta$ fixed), see Sec.~S5.2 and Fig.~S2(d) in \cite{SM}. Remarkably, transitions between monotonic and non-monotonic behavior of $\phi_{\alpha}(\nu)$ are also found when  $S$ produces public goods benefiting the entire population, see Sec.~S7 in~\cite{SM}.

Inspired by the evolution of microbial communities in fluctuating environments,
we have studied the dynamics of a  population  of two strains competing for resources  subject to a binary carrying capacity, switching {\it randomly} or {\it periodically} in time. We have analyzed how the coupling of demographic noise and environmental variability
affects the population size and  fixation properties.
We have shown that the population size distribution is generally broader under random variations than under periodic changes in the intermediate/fast switching regime, which lead to markedly different asymptotic behaviors of the fixation probabilities.
 We have also determined the conditions under which the probability that the slow species prevails is maximal.
Our work sheds light on the similarities and differences of evolution in  stochastically- versus deterministically-varying environments, and is thus relevant to  microbial communities, often subject to frequent and extreme environmental changes.

 We are grateful to E. Frey, A.~M. Rucklidge, and K. Wienand for useful  discussions.  AT and MA acknowledge support from the Israel Science Foundation grant No. 300/14 and the United States-Israel Binational Science Foundation grant No. 2016-655. The support of an EPSRC Ph.D. studentship to RW (Grant No. EP/N509681/1) is also gratefully acknowledged.

\clearpage
\renewcommand{\thefigure}{S\arabic{figure}}
\renewcommand{\theequation}{S\arabic{equation}}
\renewcommand{\thesection}{S\arabic{section}}
\renewcommand{\thesubsection}{S\arabic{section}.\arabic{subsection}}
\renewcommand{\thesubsubsection}{S\arabic{section}.\arabic{subsection}.\arabic{subsubsection}}
\setcounter{figure}{0}
\setcounter{equation}{0}
\setcounter{section}{0}

\onecolumngrid



			\begin{center}
				{{\itshape\Large Appendix: Supplementary Material to} \\~\\ \Large Population Dynamics in a Changing Environment:  \\Random versus Periodic Switching}
			\end{center}

In this Supplemental Material, we provide some further technical details and supplementary information in support of the results discussed
in the main text. We also provide additional information
concerning the population's mean fixation time (MFT), and the generalization of the model
in a scenario where the slow strain is a public goods producer.

\vspace{0.25cm}
In what follows, unless stated otherwise, the notation is the same as in the main text and the
equations and
figures refer to those therein.
This document and additional supporting resources are available at the following URL: \href{https://doi.org/10.6084/m9.figshare.12613370}{https://doi.org/10.6084/m9.figshare.12613370}.

\section{Model description, master equation and simulation methods}
\label{app:Appendix1}
In this section, we describe in detail the model and discuss
our modelling choices. We then give the master equation (ME) of the birth-death process according to which the population evolves, and describe the methods used to simulate the population dynamics in the case of random and periodic switching.
\subsection{Model description}
As explained in the main text, the population evolves according to a multivariate birth-death process where
reproduction of $S/F$ individuals, $N_{S/F}\to N_{S/F}+ 1$, occurs at a  transition  rate $T_{S/F}^{+}$, and death $N_{S/F}\to N_{S/F}- 1$, occurs at a transition rate $T_{S/F}^{-}$,
with~\cite{KEM1,KEM2}

 \begin{eqnarray}
  \label{eq:rates}
 T_{S}^{+}=\frac{f_S}{\bar{f}}~N_{S},  \quad  T_{F}^{+}=\frac{f_F}{\bar{f}}~N_{F} \quad \text{and}  \quad
 T_{S}^{-}= \frac{N}{K(t)}N_{S},  \quad T_{F}^{-}= \frac{N}{K(t)}N_{F}.
 \end{eqnarray}
 In the main text we  explicitly consider $f_S=1-s$ and $f_F=1$, with $0<s\ll 1$,
 yielding the population's average (relative birth) fitness  $\bar{f}= (N_S f_S   + N_F f_F)/N=1-sx$, where $x\equiv N_S/N$ is the fraction of $S$ individuals (slow growers). In the transition rates (\ref{eq:rates}),   the carrying capacity $K(t)$ varies in time either randomly or periodically according to Eq.~(1) of the main text,
 and switches with rates $\nu_{\pm}$, see also below.
  It is worth noting that our choice of  $f_i, i\in \{S,F\}$
  sets the  typical time scale of the dynamics.
In a more general setting, the biological factors determining the per capita growth and death rates can be written as the
product of a global and relative terms:
$T_i^{+}=g(x, N)f_i(x) N_i/\bar{f}$
 and $T_i^{-}=d(x, N)w_i(x) N_i/\bar{w}$, where $\bar{w}= (N_S w_S   + N_F w_F)/N$. In this formulation,  $g(x, N)$
and $d(x,N)$ are respectively referred to as the global birth fitness
and global weakness and are species independent (acting similarly on both strains),
whereas $f_i(x)$ and $w_i(x)$ are the species-dependent
relative birth fitness and relative weakness, respectively~\cite{Melbinger2010,Cremer2011}.
In this general setting, $g$ and $f_i$ affect the strains' birth rates, while $d$ and $w_i$ determine their survival or viability. Within this framework,   various evolutionary scenarios can be investigated, see below and Refs.~\cite{KEM1,KEM2,Melbinger2010,Cremer2011,Cremer2012,Melbinger2015a,WM19}.

In this work, as in many applications, see {\it e.g.} Refs.~\cite{Melbinger2010,Cremer2011,Cremer2012,Melbinger2015a},  we have assumed that $S$ and $F$ (slow and fast growers) have equal survival
chances and are subject to a logistic growth, and hence we set  $w_S =
w_F = 1$, and $d(x,N) = N/K$ for the global weakness. For the sake of simplicity, we have assumed that the relative birth fitness
(referred to as ``fitness'' for brevity) is constant for each species, with $f_S=1-s$ and $f_F=1$, while
the global birth fitness is $g=1$  in the main text, where we focus on the ``pure resource competition scenario'' of Refs.~\cite{KEM1,KEM2}. In Sec.~7 of this Supplemental Material (SM),
we also  consider a ``public good scenario'' in which the slow growers (strain $S$)
are public good (PG) producers, and the global growth birth fitness (global growth rate) is  $g(x)=1+bx$ (with $b>0$), i.e., a global growth rate increasing linearly with the level of PG production represented by the fraction $x$ of $S$ individuals in the population. This choice corresponds to the ``balanced growth scenario'' considered in
Refs.~\cite{Melbinger2010,Cremer2011,Cremer2012,Melbinger2015a} with a constant carrying capacity. In such a scenario, birth and death events balance each other, and the population size fluctuates about its carrying capacity after a short transient. Interestingly, the ``balanced growth scenario'' (with PG production, $b>0$) has been used in Ref.~\cite{Cremer2012} to explain the
Simpson's paradox found in the microbial experiments of Ref.~\cite{Leibler09}. This framework also allows us to model the effect of bacteriostatic (biostatic) and bactericidal (biocidal) antimicrobials
on the time evolution of sensible microorganisms in communities of sensible and resistant cells:
bacteriostatic suppresses sensible cells
growth, and hence affects $f_i$ (but neither $d$ nor $w_i$), while bactericidal induces sensible cells death and thus affects $w_i$ (but neither $g$ nor $f_i$), see, {\it e.g.},~Refs.~\cite{Coates18,Marrec20}.

While different other model formulations are of course possible,  studying the birth-death process defined by Eqs.~(\ref{eq:rates}) and (1) of the main text, is arguably the simplest way to investigate analytically, in  a biologically simple and relevant  setting, the effect of demographic noise (random birth/death events)
{\it coupled} to environmental variability. Namely, this coupling is achieved
via the switching carrying capacity
that drives the dynamics of the population size. At this point, it is useful to summarize the main properties of the birth-death process defined by Eqs.~(\ref{eq:rates}) and (1):
\begin{enumerate}
 \item[-] As reported in the main text,
 at mean-field level (constant $K=K_0\gg 1$, large population),
the population size obeys the  logistic equation $dN/dt\equiv \dot{N}=\sum_i(T_i^+ -T_i^-)=N[1-(N/K)]$, while
the population composition evolves according to the  replicator-like equation~\cite{Nowak} $\dot{x}=(T_S^+ -T_S^-)/N - x(\dot{N}/N)=-x(1-x)[f_S -\bar{f}]/\bar{f}=-sx(1-x)/(1-sx)$~\cite{KEM1,KEM2}. This model, and its generalization (see Sec.~7 of this SM), therefore have a sound eco-evolutionary dynamics.
 \item[-] When the population size is constant ($N=K_0$) and there
 is no environmental variability (only demographic noise), the
 dynamics can be mapped onto that of the well-known fitness-dependent Moran model \cite{Nowak,Ewens,Blythe07} defined by the reactions $SF \to SS$
 and $SF \to FF$, respectively occurring at rates  $\widetilde{T}_S^+=T_S^+T_F^-/N=(1-s)x(1-x)N/(1-sx)$   and  $\widetilde{T}_S^-=T_S^-T_F^+/N=x(1-x)N/(1-sx)$,
see Ref.~\cite{KEM2}. This allows us to obtain Eq.~(2) in the main text, used
in Eq.~(3) to compute the fixation probability when $K$ varies in time.
 \item[-] The model studied here is conceptual, but many of its features are {\it biologically relevant}.  With modern bioengineering techniques, it is in fact possible  to perform controlled microbial experiments in settings allowing to test the theoretical predictions of models featuring
 switching
environment, time-varying population size, PG production, cooperation dilemma, see, {\it e.g.}, Refs.~\cite{Leibler09,Acar08,Wienand15,Cremer19}.
 \item[-] The birth-death process underpinning this model can
 generalized in different ways.
 In addition to the scenario with PG production, see above and Sec.~7 of this SM, a possible generalization is
  the ``dormancy scenario'' of Ref.~\cite{Cremer2011} where $g(x,N)=1+x-(N/K_0), d(x,N)=0$ and same $f_i,w_i$ as here. The above general framework can also accommodate more realistic and complex processes in which  $g(x,N)$ and $f_i(x)$, and/or  $w_i(x)$,  depend on $\xi_{{\alpha}}$, with $\alpha\in\{r,p\}$ and hence, also vary with the environment along with $d=N/K$.

\end{enumerate}

\subsection{Master equation of the underlying birth-death process}
\label{app:Appendix1.1}

Using $\vec{N}=(N_S,N_F)$ and $\pm$ as a shorthand notation for $\xi_r=\pm 1$,
the ME for the birth-death process defined by (\ref{eq:rates}), where
 the carrying capacity $K(t)$ varies {\it randomly}
by switching according to $K_+ \rightarrow K_-$ with rate $\nu_+$ and  $
K_- \rightarrow K_+$ with rate $\nu_-$ [see Eq.~(1) in the main text, with $\alpha=r$],
 reads
 \begin{subequations}
  \label{eq:ME}
 \begin{align}
  \label{eq:ME1}
  \frac{d P_{\nu}^{(r)}({\vec N},\!+\!,t)}{dt}&=(\mathbb{E}^{-}_{S}-1)[T_{S}^{+} P_{\nu}^{(r)}({\vec N},\!+\!,t)] +
  (\mathbb{E}^{-}_{F}-1)[T_{F}^{+} P_{\nu}^{(r)}({\vec N},\!+\!,t)]\\
  &+
  (\mathbb{E}^{+}_{S}-1)[T_{S}^{-} P_{\nu}^{(r)}({\vec N},\!+\!,t)] +
  (\mathbb{E}^{+}_{F}-1)[T_{F}^{-} P_{\nu}^{(r)}({\vec N},\!+\!,t)]+  \nu_- P_{\nu}^{(r)}({\vec N},\!-\!,t)- \nu_+ P_{\nu}^{(r)}({\vec N},\!+\!,t),\nonumber\\
   \label{eq:ME2}
   \frac{d P_{\nu}^{(r)}({\vec N},\!-\!,t)}{dt}&=(\mathbb{E}^{-}_{S}-1)[T_{S}^{+} P_{\nu}^{(r)}({\vec N},\!-\!,t)] +
  (\mathbb{E}^{-}_{F}-1)[T_{F}^{+} P_{\nu}^{(r)}({\vec N},\!-\!,t)]\\
  &+
  (\mathbb{E}^{+}_{S}-1)[T_{S}^{-} P_{\nu}^{(r)}({\vec N},\!-\!,t)] +
  (\mathbb{E}^{+}_{F}-1)[T_{F}^{-} P_{\nu}^{(r)}({\vec N},\!-\!,t)]+  \nu_+ P_{\nu}^{(r)}({\vec N},\!+\!,t)-\nu_- P_{\nu}^{(r)}({\vec N},\!-\!,t), \nonumber
  \end{align}
 \end{subequations}
 where $\mathbb{E}^{\pm}_{S/F}$ are shift operators such that
 $\mathbb{E}^{\pm}_{S}f(N_S,N_F,\xi,t)=
 f(N_S\pm 1,N_F,\xi,t)$ and similarly for $\mathbb{E}^{\pm}_{F}$.
 Clearly,   Eqs.~(\ref{eq:ME1}) and~(\ref{eq:ME2})
 are coupled and the terms on the 2nd lines' right-hand-side account for environmental switching.

For periodic switching, the carrying capacity $K(t)=K_0[1+\gamma\xi_{p}(t)],$ varies deterministically
with $\xi_p(t)\equiv \xi_p(t+T)$, where the shape of $\xi_p(t)$ is taken to be a rectangular wave of period $T=(1/\nu_+) + (1/\nu_-)$~\cite{footnote7}.
 In this case, the ME of the birth-death processs (\ref{eq:rates}) with periodically switching $K(t)$ reads
\begin{eqnarray}
\label{eq:ME3}
\frac{d P_{\nu}^{(p)}({\vec N},t)}{dt}&=&(\mathbb{E}^{-}_{S}-1)[T_{S}^{+}
P_{\nu}^{(p)}({\vec N},t)]
+
(\mathbb{E}^{-}_{F}-1)[T_{F}^{+} P_{\nu}^{(p)}({\vec N},t)]\nonumber\\
&+&
(\mathbb{E}^{+}_{S}-1)[T_{S}^{-}(\xi_p) P_{\nu}^{(p)}({\vec N},t)] ]+
(\mathbb{E}^{+}_{F}-1)[T_{F}^{-}(\xi_p) P_{\nu}^{(p)}({\vec N},t)],
\end{eqnarray}
where  $T_{S/F}^{-}(\xi_p)$ are now the time-dependent transition rates
given by (\ref{eq:rates})
that vary periodically with
$\xi_p$.
Note, that in both MEs (\ref{eq:ME})-(\ref{eq:ME3}), $P_{\nu}^{(\alpha)}({\vec N},t)=0$ whenever  $N_S<0$ or $N_F<0$.

\subsection{Simulation methods}
\label{app:Appendix1.2}
While the MEs (\ref{eq:ME}) and (\ref{eq:ME3}) fully describe the population dynamics
in the case of random and periodic  switching, respectively, in general, they cannot be solved analytically. However,  to gain insight into to the stochastic dynamics, one can employ efficient numerical simulations. In the case of random switching,  process (\ref{eq:ME}) defined by the birth-death (\ref{eq:rates}) and switching  $\xi_r\rightarrow-\xi_r$ reactions, can be exactly simulated using the standard Gillespie algorithm \cite{Gillespie76}.  In the case of periodic switching, it is convenient to simulate the birth-death process (\ref{eq:ME3}) with time-dependent (periodic) transition rates (\ref{eq:rates}) using the simulation  method outlined below.

\subsubsection{Simulation of the periodic switching case with the
modified next reaction method}
In the periodic case we used the modified next reaction method \cite{Anderson07}, which is a suitable algorithm for systems with explicit time dependent rates. Unlike the classic Gillespie Algorithm, this version considers all possible birth/death processes as independent reactions. We can calculate the time step $\Delta t_{i}$ in which
the next reaction occurs by generating a random number from a uniform
distribution $r_{i}\in U\left(0,1\right)$ for the probability that reaction $i$ did \textit{not} occur after time interval
$\Delta t_i$. Here, we have four stochastic reactions $i\in \{1,\dots, 4\}$ (birth/death of $S$ and $F$)
each with a propensity function
$a_i\in\{T_S^+, T_S^-(\xi_p(t)), T_F^+,  T_F^-(\xi_p(t))\}$, and thus we have
$r_{i}=\exp\left[-\int_{t}^{t+\Delta t_{i}}a_{i}\left(t'\right)dt'\right]$.

We start the simulation at time $t=0$, and for each reaction we
set the ``internal time" $T_{i}=0$ and the quantity $P_{i}=\ln\left(1/r_{i}\right)$. We also set the initial number of each species, the environmental state (with probability determined by the duty cycle), and the initial time to the next switch $\Delta t_\text{switch}$.
Here, the time step $\Delta t_{i}$ is found by computing $\int_{t}^{t+\Delta t_{i}}a_{i}\left(t'\right)dt'=P_{i}-T_{i}$,
which can be easily solved, since $K$ is discrete and thus in each iteration it is constant. At this point we find the reaction that has the minimal time step
$\Delta t_{\mu}=\min_{_{i}}\left\{ \Delta t_{i}\right\} $, propagate time $t\rightarrow t+\text{\ensuremath{\Delta t_{\mu}}}$, and update
the population size, the internal times $T_{i}\rightarrow T_{i}+\int_{t}^{t+\Delta t_{\mu}}a_{i}\left(t'\right)dt'$,
and $P_{i}\rightarrow P_{i}+\delta_{i,\mu}\ln\left(1/r_{i}\right)$.
Then we recalculate the rates $a_{i}$, generate another  random number $r_i \in U\left(0,1\right)$, and repeat these steps iteratively
until one of the species has undergone extinction.
We treat the deterministic switches $\xi\rightarrow-\xi$, that occurred during a period of $1/\nu_{\pm}$,
as follows: if $\Delta t_{\text{switch}} <\Delta t_{\mu}$, we switch $\xi\rightarrow-\xi$ and propagate the time $t\rightarrow t+\text{\ensuremath{\Delta t_\text{switch}}}$.

\section{Approximations of the quasi-stationary population density: periodic switching}
\label{AppendixD}
In this section we compute the quasi-stationary population size distribution (PSD), $P_N$,
in the slow switching regime, as well as  under fast and intermediate {\it periodic} switching. This is done by first computing the PSD in the case of constant carrying capacity, assuming a static environment $\xi_{\alpha}(t)=\xi$ and carrying capacity $K(t)=K$. To do so, we start with the ME for  $P(N)|_K$ -- the probability that the total population size is $N$ given a carrying capacity $K$
\begin{equation}\label{eq:}
\frac{dP(N)|_K}{dt}=\left(N-1\right)P\left(N-1\right)|_K+\frac{\left(N+1\right)^{2}}{K}P\left(N+1\right)|_K-\left(N+\frac{N^{2}}{K}\right)P(N)|_K.
\end{equation}
The PSD can be found by putting $\dot{P}(N)|_K=0$ and demanding a reflecting boundary condition at $N=1$. The latter assumes that the probability flux to the extinction state $P(N=0)$ is negligibly small, which is legitimate since the mean time to extinction is assumed to be much larger than the time scales we are interested in here, see main text. The normalized solution of the resulting recursion equation reads
\begin{equation}
P(N)|_K=\frac{1}{\text{Ei}\left(K\right)-\gamma-
\text{\ensuremath{\ln}}\left(K\right)}\frac{K^{N}}{NN!}\simeq \frac{K}{N}\frac{K^{N}e^{-K}}{N!},
\label{eq:24-1}
\end{equation}
where $\text{Ei}\left(x\right)=-\int_{-x}^{\infty}dte^{-t}/t$
 is the exponential integral function, $\gamma=0.577...$
is the Euler--Mascheroni constant, and the last approximation holds when $K\gg1$.

\subsection{Quasi-stationary PSD under slow switching}
When  $\nu\to 0$, on average there are no switches prior to fixation, and the population  evolves in a static environment
$\xi_{\alpha}=\pm 1$, with $\xi_{\alpha}$
that is  distributed with a probability $p(\xi_{\alpha})=(1\pm \delta)/2$. Namely, if $\xi_{\alpha}=\pm 1$, the population
is subject to a constant carrying capacity $K=K_{\pm}$. Hence, using Eq.~(\ref{eq:24-1}), the PSD under slow switching reads
\begin{equation}
P_{0}(N)=
\sum_{\xi_{\alpha}=\pm 1} P(N|\xi_{\alpha})~p(\xi_{\alpha})\simeq
\left(\frac{1+\delta}{2}\right) \frac{K_+}{N}\frac{K_+^{N} e^{-K_+}}{N!}+
\left(\frac{1-\delta}{2}\right) \frac{K_-}{N}\frac{K_-^{N} e^{-K_-}}{N!}.
\end{equation}
As explained in the main text, this result is valid both for periodic and random switching.

\subsection{Quasi-stationary PSD under fast periodic switching: Kapitza method}
\label{Kapitza}
In the opposite limit $\nu\gg 1$, the carrying capacity $K$ rapidly oscillates
around $K_{0}$. To find the PSD in the case of fast {\it periodic} switching, we employ the Kapitza method~\cite{Assaf08}, valid for a general periodic $\xi_{p}(t)$, which involves separating the dynamics into fast and slow variables, and averaging the fast variables over the period of variation.

Our starting point is ME (\ref{eq:}), but now with $K=K_0[1+\gamma\xi_{p}(t)]$, i.e., explicitly time-dependent rates. To treat Eq.~(\ref{eq:}) semi-classically, we define the probability generating function $G\left(p,t\right)=\sum_{m=0}^{\infty}P\left(m,t\right)p^{m},$ where $p$ is an auxiliary variable. Conservation of probability yields $G\left(1,t\right)=1$. The definition of $G$ is useful
since
\begin{equation}
P\left(N,t\right)=\frac{1}{N!}\frac{\partial^{N}G\left(p,t\right)}{\partial p^{N}}|_{p=0}.\label{eq:prob}
\end{equation}
Multiplying Eq.~(\ref{eq:}) by $p^{N}$ and summing over all $N$'s,
we obtain a second-order partial differential equation for G
\begin{equation}
\frac{\partial G}{\partial t}=\left(1-p\right)\left(-p\frac{\partial G}{\partial p}+\frac{1}{K\left(t\right)}\frac{\partial G}{\partial p}+\frac{p}{K\left(t\right)}\frac{\partial^{2}G}{\partial p^{2}}\right).\label{eq:2}
\end{equation}
This  equation cannot be solved in general. An approximate solution can be found by using the fact that the typical carrying capacity is large, $K_0\gg 1$, and employing the WKB ansatz $G=G_{0}\exp\left[-K_{0}S\left(p,t\right)\right]$
in Eq.~(\ref{eq:2})~\cite{Assaf17}. Keeping leading- and subleading-order terms with respect to ${\cal O}(K)$, we arrive at the following Hamilton-Jacobi equation
\begin{equation}
-\frac{\partial S}{\partial t}=q\left(1-p\right)\left(-p+\frac{1}{K\left(t\right)}+\frac{K_{0}}{K\left(t\right)}pq\right)\equiv H\left(p,q\right),\label{eq:3-1}
\end{equation}
where $H$ is the Hamiltonian, $S$ is the action associated with the Hamiltonian, and we have defined $q=-\frac{\partial S}{\partial p}$ as the coordinate conjugate to the variable $p$, see~\cite{Dykman}.

Let us separate the fast and slow time scales by denoting $q\left(t\right)=X\left(t\right)+\zeta\left(t\right)$
 and $p\left(t\right)=Y\left(t\right)+\eta\left(t\right)$. Here $X$
and $Y$ are slow variables, while $\zeta$ and $\eta$ are small corrections (to be verified a-posteriori)
that rapidly oscillate around $0$~\cite{Assaf08}. Expanding the Hamiltonian (\ref{eq:3-1}) up to second order around $q=X$ and $p=Y$ we find
\begin{align}
H\left(q,p,t\right) \simeq H\left(X,Y,t\right)+\zeta\frac{\partial H}{\partial X}+\eta\frac{\partial H}{\partial Y} +\frac{\zeta^{2}}{2}\frac{\partial^{2}H}{\partial X^{2}}+\frac{\eta^{2}}{2}\frac{\partial^{2}H}{\partial Y^{2}}+\zeta\eta\frac{\partial^{2}H}{\partial X\partial Y}\equiv\tilde{H}\left(X,Y,t\right)\nonumber.\label{eq:Htayolr}
\end{align}
Using the Hamilton equations $\dot{q}=\dot{X}+\dot{\zeta}\simeq \partial_Y\tilde{H}(X,Y,t)$ and $\dot{p}=\dot{Y}+\dot{\eta}\simeq-\partial_X\tilde{H}(X,Y,t)$, and equating the rapidly oscillating terms yields in the leading order in
$K_{0}\gg1$: $\zeta \simeq\left(X^{2}-2X^{2}Y\right)(B/\nu)$ and $\eta\simeq-\left(2XY-2XY^{2}\right)(B/\nu)$. Here $B(t)={\cal O}(1)$ is defined in Eq.~(\ref{eq:C}), and in the calculation $X$ and $Y$ were considered as constants during the period of rapid oscillations. In addition, we have neglected terms of order $\zeta$ and $\eta$, but kept their time derivatives (proportional to $\nu\gg 1$).

Following this result, we define a canonical transformation from the old $\left(q,p\right)$
to the new $\left(X,Y\right)$ variable
\begin{equation}
q  \simeq X+X^{2}\left(1-2Y\right)\frac{B}{\nu}+2X^{3}\left(1-2Y\right)^{2}\frac{B^{2}}{\nu^{2}},\;\;\;\;
p\simeq Y-2\left(Y-Y^{2}\right)X\frac{B}{\nu}-2X^{2}\left(Y-3Y^{2}+2Y^{3}\right)\frac{B^{2}}{\nu^{2}},\label{eq:canonical}
\end{equation}
which can be obtained using the generating function $F_{2}\left(q,Y,t\right)=qY-q^{2}\left(Y-Y^{2}\right)(B/\nu)$. This transformation is canonical up to second order in the small parameter
$1/\nu$, as the Poisson brackets satisfy $\left\{ q,p\right\} _{\left(X,Y\right)}=1+O\left(\frac{1}{\nu^{3}}\right).$
Using Eqs.~(\ref{eq:3-1}) and (\ref{eq:canonical}) and defining $H'=H+\frac{\partial F_{2}}{\partial t}$, by averaging
over a period of a rapid oscillation, we find
\begin{align}
\overline{H}(X,Y)=XY\left(1-Y\right)\left[AX-1+X^{2}C\left(2-4Y+4Y^{2}-XA\right)\frac{1}{\nu^{2}}\right],\label{eq:12}
\end{align}
where we have defined the following $O\left(1\right)$ variables
\begin{equation}\label{eq:C}
A\left(\gamma,\delta\right)\!=\!\overline{\frac{K_{0}}{K(t)}}\!=\!\frac{1}{T}\int_{t_{0}}^{t_{0}+T}\frac{K_{0}}{K\left(t\right)}dt;\;\;\;\;
C\left(\gamma,\delta\right)\!=\!\overline{B^{2}}\!=\!\frac{1}{T}\int_{t_{0}}^{t_{0}+T}B^{2}\left(t\right)dt;\;\;\;\;
B\left(t\right) \!=\!K_{0}\nu\intop dt\left[\frac{1}{K(t)}-\overline{\frac{1}{K(t)}}\,\right],
\end{equation}
and used the fact that $B\left(t\right)$ is periodic. It can be shown that the ${\cal O}(\nu^0)$ terms in  Hamiltonian~(\ref{eq:12}) yield the PSD in the constant environment case [Eq.~(\ref{eq:24-1})].

Having found the time-\textit{independent} Hamiltonian (\ref{eq:12}), which effectively takes into account the rapid
environmental oscillations, we can compute the PSD by finding the nontrivial  zero-energy trajectory of $\overline{H}\left(X,Y\right)$. Up to second order in $\frac{1}{\nu}$, this trajectory is given by $X\left(Y\right)=1/A-C/(A^{3}\nu^{2})\left(2Y-1\right)^{2}+O\left(\nu^{-3}\right)$.
Thus, recalling that $q=-\frac{\partial S}{\partial p}$, and using the fact
that the transformation $\left(q,p\right)\rightarrow\left(X,Y\right)$
is canonical, we find $S(Y)=-\int XdY=-Y/A+C/(6A^{3}\nu^{2})(2Y-1)^{3}$. As a result, the generating function becomes
\begin{equation}
G(Y)\simeq G_{0}\exp\left[-K_{0}S(Y)\right]=G_{0}\exp\left[K_{0}\frac{Y}{A}-K_{0}\frac{C}{6A^{3}\nu^{2}}\left(2Y-1\right)^{3}\right],\nonumber
\end{equation}
where $G_0$ is a constant, see below. Therefore, the PSD can be found by employing the Cauchy theorem to
Eq.~(\ref{eq:prob}):
\begin{equation}
P(N)=\frac{1}{2\pi i}\varoint\frac{G(Y)}{Y^{N+1}}dY=\frac{G_{0}}{2\pi i}\varoint\frac{1}{Y}\exp\left[K_{0}g\left(Y\right)\right]dY,\nonumber
\end{equation}
where the integration has to be performed over a closed contour in
the complex $Y$ plane around the singular point $Y=0$, and we have
defined $g\left(Y\right)=-S\left(Y\right)-\frac{N}{K_{0}}\ln Y$.
This integral can be calculated using the saddle point approximation
\cite{Bender}. The saddle point, up to second order in $1/\nu$, is found at $Y^{*}=AN/K_0+[CN/(AK_0)]\left(2AN/K_0-1\right)^{2}\nu^{-2}$. Furthermore, since $g''\left(Y^{*}\right)>0$ the integration contour in the vicinity
of the saddle point must be chosen perpendicular to the real axis.
As a result, the Gaussian
integration yields $P(N)\cong\left[G_{0}/\left(Y^{*}\sqrt{2\pi K_{0}\left|g\prime\prime\left(Y^{*}\right)\right|}\right)\right]e^{K_{0}g\left(Y^{*}\right)}$.
Note, however, that only the leading-order result can be taken into account here; accounting for the prefactor would be an excess of accuracy since we have ignored the $p$-dependent prefactors
in both $G$ and in $P$.
Putting it all together, we finally obtain
\begin{equation}
P(N)\simeq {\cal C}\exp \left[N-N\ln\left(AN/K_0\right)-K_{0}\frac{C}{6A^{3}\nu^{2}}\left(2AN/K_{0}-1\right)^{3}\right] ,\label{eq:QSD-TD}
\end{equation}
where ${\cal C}$ is a normalization constant which can be found by demanding $\int P(N)dN=1$.

\subsubsection{Rectangular wave}
\label{Rectangular and square wave}
Our derivation above has been carried out for
a general periodic function $\xi_p\left(t\right)$. We now
compute the PSD in the particular case of a rectangular
wave. Using the expression of $\xi_p(t)$ given in the main text, we find
\begin{align}
B\left(t\right) & =\frac{\gamma}{1-\gamma^{2}}\times\begin{cases}
(\delta-1)\nu t-1/2 & - \frac{1}{\nu_+}\leq t\leq0\\
(\delta+1)\nu t-1/2 & 0\leq t\leq\frac{1}{\nu_-}
\end{cases},\label{eq:B_cal_sw}
\end{align}
where the constant of integration was determined by the demand
that $\overline{B}=0$. Plugging this into Eq.~(\ref{eq:C}) yields
$\overline{K(t)^{-1}}=A/K_0=(1-\gamma\delta)/[K_{0}\left(1-\gamma^{2}\right)]$ and
$C=(1/12)\gamma^{2}/\left(1-\gamma^{2}\right)^{2}$. Using these results, Eq.~(\ref{eq:QSD-TD}) becomes
\begin{equation}
\label{eq:PKapitzaSM}
P_{\nu}^{{\rm Kap}}\simeq {\cal P}(N){\rm exp}\left[-\frac{K_0}{72\nu^2}~\left(\frac{\gamma}{1\!-\!\gamma^2}\right)^2\left(\frac{2N\!-\!{\cal K}}{K_0}\right)^3
\right]\!,
\end{equation}
which is the expression of $P_{\nu}^{{\rm Kap}}$ used in the main text, with
$\lim_{\nu \to \infty} P_{\nu}^{{\rm Kap}} = {\cal P}(N)\propto {\rm exp}[N(1-\ln{(N/{\cal K})})]$, peaked at ${\cal K}=  K_0(1-\gamma^2)/(1-\gamma\delta)$. Hence, $P_{\nu}^{{\rm Kap}}(N)$ is unimodal and peaked about $N\approx {\cal K}$ when $\nu \gg 1$, see Fig.~2(b).

\subsection{Quasi-stationary PSD under intermediate periodic switching}\label{app:PPP}
We now consider the quasi-stationary PSD in the regime of intermediate periodic switching where $\nu={\cal O}(1)$.
In this regime, progress can be made upon neglecting demographic noise, and by considering only the environmental periodic modulation for $N(t)$. This leads to an approximation of $P_{\nu}^{(p)}$,
here referred to as ``piecewise periodic process''
and denoted by $P_{\nu}^{{\rm PPP}}$, that is the periodic counterpart of the PDMP approximation, see Eq.~(\ref{eq:PpdmpSM}) and below. This approach is similar in nature to that of Refs.~\cite{Bena06,Doering85} (whose focus was on {\it symmetric} switching).

Our starting point is the mean-field rate equation for the total population size, in the case of periodic switching, upon ignoring demographic noise. Using the definition of $\xi_{p}\left(t\right)$ from the main text, the equation reads
\begin{equation}
\dot{N}=N\left[1-\frac{N}{K\left(t\right)}\right]=N\left\{1-\frac{N}{K_{0}\left[1+\gamma\xi_{p}\left(t\right)\right]}\right\},
\end{equation}
At $t\rightarrow\infty$, after the transient has decayed, the periodic solution reads
\begin{equation}
N=f\left(t\right)=\begin{cases}
K_{0}\left(1-\gamma^{2}\right)\left[1-\gamma+2\gamma e^{-\tilde{t}}\frac{1-e^{-1/\nu_{-}}}{1-e^{-T}}\right]^{-1} \quad & 0<\tilde{t}<\frac{1}{\nu_{+}}\left(\xi=1\right)\\
K_{0}\left(1-\gamma^{2}\right)\left[1+\gamma-2\gamma e^{-\tilde{t}}\frac{e^{1/\nu_{+}}-1}{1-e^{-T}}\right]^{-1}\quad & \frac{1}{\nu_{+}}<\tilde{t}< T\left(\xi=-1\right)
\end{cases},
\end{equation}
where $\tilde{t}=t-1/\nu_{-}-\lfloor\frac{t-1/\nu_{-}}{T}\rfloor T$, such that $0\leq\tilde{t}\leq T=(1/\nu_{+})+(1/\nu_{-})$. As a result, for each segment of the solution, one can express $\tilde{t}$ as function
of $N$: $\tilde{t}=g_{\pm}\left(N\right)=\ln\left(B_{\pm}\right)+\ln\left(N\right)-\ln\left(1-N/K_{\pm}\right)$. Here, $B_{\pm}$ is a cumbersome expression independent on $N$ and hence irrelevant for our purposes, whereas the subscripts $+$ and $-$ stand for the first and second segment in each period, respectively. Therefore, we can approximate the PSD, $P_{\nu}^{{\rm PPP}}$, as
\begin{equation}\label{PSD_PPP}
P_{\nu}^{{\rm PPP}}(N)=\int_{0}^{T}d\tilde{t}'P_{\nu}^{{\rm PPP}}\left(N,\tilde{t}'\right),
\end{equation}
where $P_{\nu}^{{\rm PPP}}\left(N,\tilde{t}'\right)\sim \delta\left[N-f\left(\tilde{t}\right)\right]$ is the probability that the population
size at time $\tilde{t}$ is $N$, and we have omitted the normalization constant. Here we have neglected demographic noise by assuming that the instantaneous total population size $N$ is sharply peaked around its deterministic solution.
Performing the integral in Eq.~(\ref{PSD_PPP}) , we find
\begin{equation}\label{PPPapprox}
P_{\nu}^{{\rm PPP}}(N)={\cal C}\left[\left|\frac{dg_{+}}{dN}\right|+\left|\frac{dg_{-}}{dN}\right|\right]={\cal C}\left[\frac{1}{K_{+}-N}+\frac{1}{N-K_{-}}\right],
\end{equation}
where ${\cal C}$ is a normalization constant.
This expression is valid for $N_{\text{min}}\leq N\leq N_{\text{max}}$,
where the boundaries $N_{\text{min}}=N(\tilde{t}=0)$ and $N_{\text{max}}=N(\tilde{t}=1/\nu_{+})$ satisfy
\begin{equation}
N_{\text{min}}=K_{0}\left(1\!-\!\gamma^{2}\right)\left[1\!-\!\gamma\!+\!2\gamma\frac{1\!-\!e^{-1/\nu_{-}}}{1\!-\!e^{-T}}\right]^{-1},\quad\quad
N_{\text{max}}=K_{0}\left(1\!-\!\gamma^{2}\right)\left[1\!-\!\gamma\!+\!2\gamma\frac{e^{-1/\nu_{+}}\!-\!e^{-T}}{1\!-\!e^{-T}}\right]^{-1},
\end{equation}
while the normalization constant is given by
\begin{equation}
\label{C}
{\cal C}^{-1}=\ln\left[\frac{\left(K_{+}-N_{\text{min}}\right)\left(N_{\text{max}}-K_{-}\right)}{\left(K_{+}-N_{\text{max}}\right)\left(N_{\text{min}}-K_{-}\right)}\right].
\end{equation}
The PPP approximation $P_{\nu}^{{\rm PPP}}$ of the periodic PSD is
shown in Fig.~2(c,d) of the main text, where it is found to agree well with the simulation results and to reproduce the main features of $P_{\nu}^{(p)}$ in the intermediate switching regime. It also accurately captures the average population size, as shown in Fig.~\ref{fig:FigS3}(b).

\section{Quasi-stationary PSD for random switching: the PDMP approximation}
When demographic noise is neglected, by assuming that the fluctuating population size is always large,
and the only source of noise stems from the randomly switching carrying capacity, we have seen that the
PSD, $P_{\nu}^{(r)}(N)$,  can be described in terms of the
marginal stationary probability density of the underlying piecewise-deterministic Markov process (PDMP). Upon omitting the normalization constant, this PSD reads~\cite{KEM2,HL06}
\begin{eqnarray}
\label{eq:PpdmpSM}
P^{{\rm PDMP}}_{\nu}(N) \propto \frac{1}{N^2} \left[\left(\frac{K_+}{N}-1\right)^{\nu_+ -1}
\left(1-\frac{ K_-}{N}\right)^{\nu_- - 1}\right],
\end{eqnarray}
where  the dependence on $\gamma, \delta$
and $\nu$ is given by $K_{\pm} =(1\pm \gamma)K_0$ and $\nu_{\pm}=(1\mp \delta)\nu$.
Clearly, $P^{{\rm PDMP}}_{\nu}$ has support $[K_{-},K_{+}]$ and accounts for environmental noise, but ignores all demographic fluctuations.
The expression of $P^{{\rm PDMP}}_{\nu}$ gives a suitable description of $P_{\nu}^{(r)}$ in the intermediate switching regime where interesting phenomena arise (see Sec.~S4.3 below for a detailed discussion of the validity of the PDMP-like approximations).
\begin{figure}[h!]
	\centering
	\includegraphics[width=0.8\linewidth]{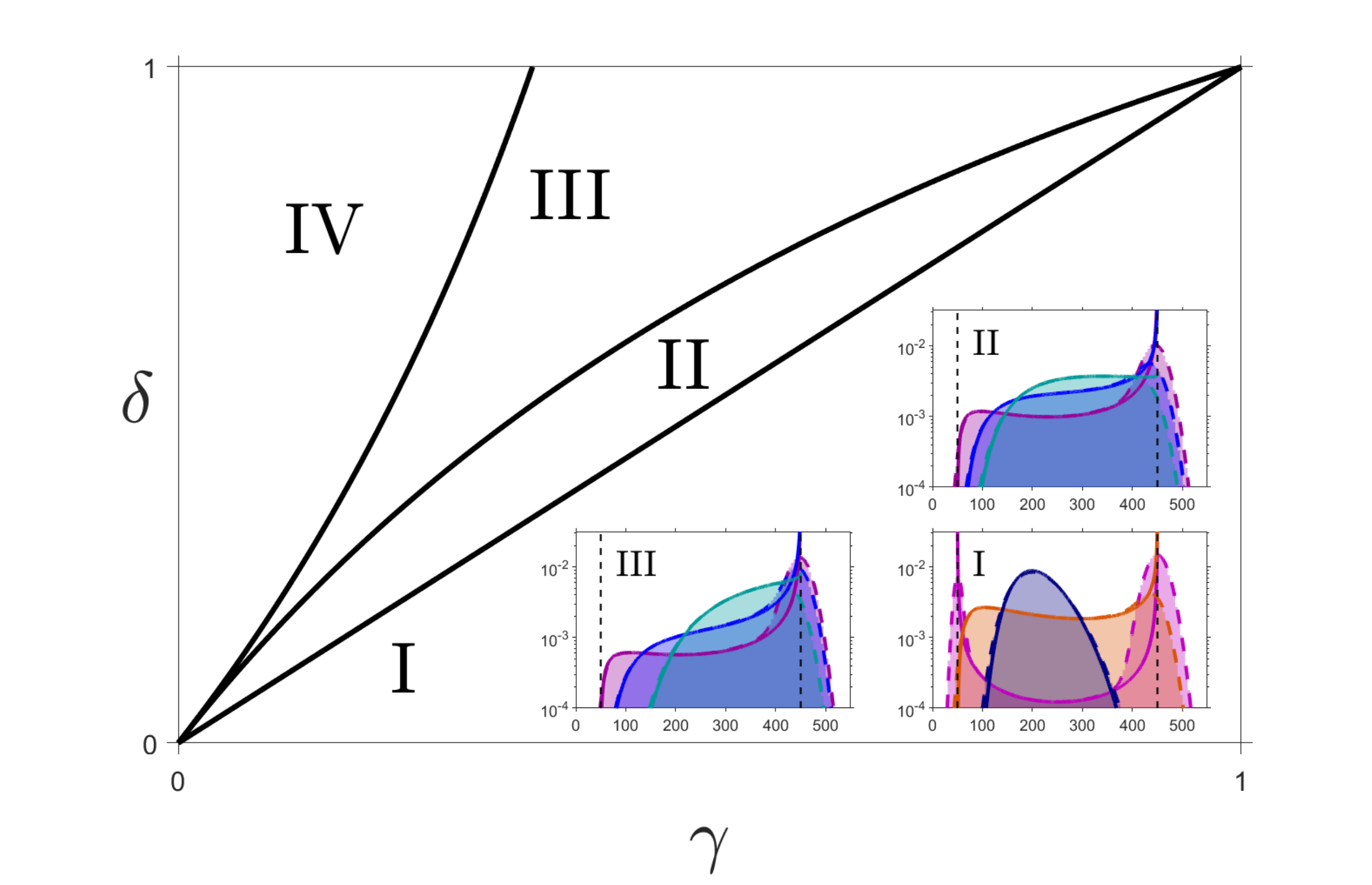}
	\caption{Phase diagram for the PSD, $P_{\nu}^{(r)}(N)$,
	and its approximations
	$P_{\nu}^{{\rm PDMP}}$ and
	$P_{\nu}^{{\rm LNA}}$ (insets), see Eq.~(\ref{eq:PLNA}).
	We distinguish
	 four regions described in the text: In addition to a peak about $K_+$, the PSD has always a local maximum $K_-<N^*<K_+$ in the	intermediate switching regime in I; in regime II and III, the PSD and  $P_{\nu}^{{\rm PDMP}}$
	have  a peak about $K_+$ and, depending on $\nu$, possibly another peak
	at some values $K_-<N^*<K_+$, see insets;
	the PSD and  $P_{\nu}^{{\rm PDMP}}$ have one single peak about $K_+$ in IV.
		 Insets illustrate the form of $P_{\nu}^{(r)}(N)$,
		 $P_{\nu}^{{\rm PDMP}}$ and
	$P_{\nu}^{{\rm LNA}}$  in regions I-III. In the insets,
	 solid lines are from the $P_{\nu}^{{\rm PDMP}}$, given by Eq.~(\ref{eq:PpdmpSM}),
	 dashed lines are from $P_{\nu}^{{\rm LNA}}$, given by Eq.~(\ref{eq:PLNA}),  solid areas are from computer simulations, and the vertical dashed lines are eyeguides showing $N=K_{\pm}$.
	Parameters
	are: $(K_0,\gamma,s,x_0) = (250,0.8,0.05,0.5)$  and (inset I)
	$\delta= 0.7$, $\nu = (0.05, 1.4,17.5)$ (pink, orange, blue);
	(inset II) $\delta= 0.85$, $\nu = (1, 3,6.5)$ (purple, blue,
	green); (inset III)
	$\delta= 0.92$, $\nu = (1, 3,12)$ (purple, blue, green).
	In inset I, $N^*$ is in the intermediate regime for $\nu=1.4$ (orange).
	In inset II, $N^*$ is  in the intermediate regime for $\nu=1$ (purple)
	and $\nu=6.5$ (green). In inset III, $N^*$ is in the intermediate regime for $\nu=1$ (purple).
	We notice that the LNA  excellently agrees with simulation results for the PSD:
	 $P_{\nu}^{(r)}(N)$ and $P_{\nu}^{{\rm LNA}}$ are almost indistinguishable
	 in each inset.}
	\label{fig:FigS4}
\end{figure}
\subsection{PSD dependence on $\gamma$ and $\delta$ in the  intermediate switching regime}
The PSD, $P_{\nu}^{(r)}(N)$, and its PDMP approximation,
$P^{{\rm PDMP}}_{\nu}$, are bimodal, with peaks about $K_{\pm}$, when $\nu < 1$,
 and unimodal when  $\nu >1$ with a peak $N^*$ that is the smaller solution to
\begin{eqnarray}\label{Nstareq}
N^2 - \left(\nu(1-\gamma\delta) + 1\right)K_0 N +(1-\gamma^2)K_0^2 \nu =0,
\end{eqnarray}
with $N^* \to \mathcal{K}$ as $\nu \to \infty$~\cite{KEM1,KEM2,WM19}. In addition, two other regimes can arise under \emph{asymmetric} switching at \emph{intermediate} rate when
$1/(1+|\delta|)<\nu<1/(1-|\delta|)$. Here, the PSD
has a different form not found when $\delta=0$:
When $\delta<0$ and $1/(1-\delta)<\nu<1/(1+\delta)$, $P^{{\rm PDMP}}_{\nu}$ and $P_{\nu}^{(r)}$
have a peak at $N\simeq K_-$. When $\delta>0$ and $1/(1+\delta)<\nu<1/(1-\delta)$,
 $P^{{\rm PDMP}}_{\nu}$ and $P_{\nu}^{(r)}$ have a peak at $N\simeq K_+$
and, depending on $\delta, \gamma$ and $\nu$, also a peak at $N^*$.
The condition for the existence of such a peak at  $K_-<N^*<K_+$
can be inferred from the PDMP approximation~(\ref{eq:PpdmpSM})
by noting
 that \eqref{Nstareq} has real roots when
\begin{equation}\label{pdf_eq_help}
(1-\gamma\delta)^2 \nu^2 - 2(1+\gamma(\delta-2\gamma))\nu +1 >0.
\end{equation}
We thus distinguish four regions, I-IV, in the $(\delta,\gamma)$ - space, see Fig.~\ref{fig:FigS4}:
\begin{itemize}
	\item[I:] $\delta < \gamma$, where $N^*$ exists for all intermediate $\nu$.
	\item[II:] $\gamma< \delta < \frac{2\gamma}{1+\gamma}$, where  $N^*$ exists for all intermediate
	$\nu$ that lie outside the interval between the two solutions of \eqref{pdf_eq_help}, here
	denoted by $\nu_{1,2}$ (with $\nu_2\geq \nu_1)$.
	\item[III:] $\frac{2\gamma}{1+\gamma} < \delta <  \frac{2\gamma}{1-\gamma}$,
	where $N^*$ only exists if $\frac{1}{1+\delta} <\nu< \nu_1$ .
	\item[IV:] $\delta> \frac{2\gamma}{1-\gamma}$, where $N^*$ does not exist.
\end{itemize}
Simulation results of Figs.~2 and~\ref{fig:FigS4} confirm that the above analysis
correctly reflects the properties of $P_{\nu}^{(r)}$, see the videos of the Figshare resources~\cite{SM}.

As  shown by Fig.~2 of the main text, the PSD under intermediate \textit{periodic} switching
is qualitatively characterized by the same features as $P_{\nu}^{(r)}$, with some generic quantitative differences: $P_{\nu}^{(p)}$ is generally narrower and has sharper peak than $P_{\nu}^{(r)}$. All these features are well captured by the  PPP approximation (\ref{PPPapprox})-(\ref{C}) of $P_{\nu}^{(p)}$. In particular, $P_{\nu}^{\rm PPP}$ has a narrower support $[N_{\text{min}}, N_{\text{max}}]$ than the support $[K_{-}, K_{+}]$ of $P^{{\rm PDMP}}_{\nu}$, since $N_{\text{min}}>K_-$ and $N_{\text{max}}<K_+$, see Fig.~2 (c,d) and the Figshare resources of \cite{SM}.

\subsection{Linear noise approximation about the PDMP solution}
\label{app:LNA}
While $P^{{\rm PDMP}}_{\nu}$ (\ref{eq:PpdmpSM})
 captures well the position of the peaks of the PSD and some of its main features, the  PDMP approximation fails to capture the width of $P_{\nu}^{(r)}(N)$. In order to account for the demographic
 noise responsible for the shape of $P_{\nu}^{(r)}(N)$ near its peaks,
 we can perform a linear noise approximation (LNA) about the PDMP~\cite{KEM2}
 \begin{eqnarray}
 \label{eq:PDMP}
 \frac{d}{dt}N=N\left[1-\frac{N}{\cal K}\left(\frac{1-\gamma \xi_r}{1-\gamma\delta}\right)\right],
\end{eqnarray}
whose probability density   $P_{\nu}^{{\rm PDMP}}(N,\xi_r)$
 in the environmental state $\xi_r=\pm 1$, is given by~\cite{HL06}
\begin{eqnarray*}
P_{\nu}^{{\rm PDMP}}(N,\xi_r) &\propto &
\begin{cases}
\frac{1+\delta}{N^{2}}
\left[\frac{K_+}{N}-1\right]^{\nu_+-1}~ \left[1-\frac{K_-}{N}\right]^{\nu_-}, \quad & (\xi_r=+1)\\
\frac{1-\delta}{N^{2}} \left[\frac{K_+}{N}-1\right]^{\nu_+}~ \left[1-\frac{K_-}{N}\right]^{\nu_--1}
, \quad & (\xi_r=-1),
\end{cases}
\end{eqnarray*}
where
$K_{\pm} =(1\pm \gamma)K_0$ and $\nu_{\pm}=(1\mp \delta)\nu$.
As in Refs.~\cite{KEM2,Hufton16}, we also make the simplifying assumption that demographic noise is approximately the same in each environmental state,
yielding the Gaussian distribution $\propto{\rm exp}\left(-(N-\widetilde{N})^2/(2\widetilde{N})\right)/\sqrt{\widetilde{N}}$ for the demographic fluctuations $N-\widetilde{N}$ about the  PDMP (\ref{eq:PDMP}). Proceeding as in the case of symmetric switching ($\delta=0$), see Ref.~\cite{KEM2} where full details are provided, and omitting the normalization constant, we obtain the LNA of the marginal
stationary probability density about the PDMP (\ref{eq:PDMP})
\begin{eqnarray}
\label{eq:PLNA}
\hspace{-2mm}
P_{\nu}^{{\rm LNA}}(N) \propto \int_{K_-}^{K_+}\frac{e^{\frac{-\left(N-\widetilde{N}\right)^2}
{ 2\widetilde{N}} }}{\widetilde{N}^{5/2}}\!\left\{(1\!+\!\delta) \left[\frac{K_+ }{\widetilde{N}}\!-\!1\right]^{\nu_+-1}
\!\left[1\!-\!\frac{K_-}{\widetilde{N}}\right]^{\nu_-}\!\!+\!(1\!-\!\delta)\left[\frac{K_+ }{\widetilde{N}}\!-\!1\right]^{\nu_+} \!\left[1\!-\!\frac{K_-}{\widetilde{N}}\right]^{\nu_--1}\right\}\! d\widetilde{N}\!.
\end{eqnarray}

The results shown in the insets of Fig.~\ref{fig:FigS4} illustrate that $P_{\nu}^{{\rm LNA}}(N)$ is an excellent approximation of
the PSD: it accurately predicts all the details of the PSD $P_{\nu}^{(r)}(N)$ obtained
from stochastic  simulations.
However, while $P_{\nu}^{{\rm LNA}}$ significantly improves over $P_{\nu}^{{\rm PDMP}}$ to describe the PSD, we have verified
that computing $\phi_r$ in the realm of the PDMP-based approximation [i.e. with Eq.~(\ref{eq:Ppdmp-basedSM})] or by averaging $\phi(x_0)|_N$ over
$P_{\nu}^{{\rm LNA}}$, as an approximation of $P_{\nu}^{(r)}$, according to Eq.~(3), yields essentially the same results: As shown in  Fig.~\ref{fig:FigS1}(c), the fixation probability calculated using $P_{\nu}^{{\rm LNA}}$ gives only a minute improvement over the results obtained with $P_{\nu}^{{\rm PDMP}}$. The LNA approximation  (\ref{eq:PLNA})
is thus useful to describe the PSD, but the PDMP approximation is sufficient to compute the fixation probability.

Note, that while we have not carried it out explicitly, a similar LNA treatment can be done in the periodic case. This would allow us to accurately reproduce the PSD in the low and intermediate periodic switching regime.

\section{Fixation probability under fast switching: saddle-point calculations}
\label{sec:Saddle}

In this section we perform a saddle-point approximation to find the fixation probability, $\phi_{\alpha}$, in the fast switching regime $\nu/s\gg 1$, and then discuss the validity of the PDMP-like (PDMP and PPP) approximations.

To perform a saddle-point calculation of $\phi_{\alpha}$ under fast switching, we rewrite Eq.~(3) of the main text in terms of the total population density $y = N / K_0$. Accounting for the normalization of the probability distribution, the fixation probability can be written as
\begin{equation}
\phi_{\alpha}(\nu)=\frac{\intop_{0}^{\infty}P_{\nu/s}^{(\alpha)}\left(y\right)\exp\left[K_{0}\left(1-x_{0}\right)\ln\left(1-s\right)y\right]dy}{\intop_{0}^{\infty}P_{\nu/s}^{(\alpha)}\left(y\right)dy}\equiv \frac{\intop_{0}^{\infty}\exp \left[f_{\text{num}}^{(\alpha)}(y)\right]dy}{\intop_{0}^{\infty}\exp \left[f_{\text{den}}^{(\alpha)}(y)\right]dy},\label{eq:avg_fix-1}
\end{equation}
where we have defined  $f_{\text{den}}^{(\alpha)}(y)=\ln P_{\nu/s}^{(\alpha)}(y)$, and $f_{\text{num}}^{(\alpha)}(y)=f_{\text{den}}^{(\alpha)}(y)+K_{0}\left(1-x_{0}\right)\ln\left(1-s\right)y$, and $\alpha$ denotes either $r$ (random) or $p$ (periodic). Evaluating both integrals separately via the saddle point
approximation, we obtain
\begin{equation}
\label{eq:phi-saddle}
\phi_{\alpha} (\nu)\simeq  \sqrt{\kappa^{(\alpha)}_1/\kappa^{(\alpha)}_2}\; e^{f_{\text{num}}^{(\alpha)}\left(y^{(\alpha)}_2\right) - f_{\text{den}}^{(\alpha)}\left(y^{(\alpha)}_1\right)} .
\end{equation}
Here $y^{(\alpha)}_1$ and $y^{(\alpha)}_2$ are the positions of the saddle points of the denominator and numerator, respectively, and satisfy $(d/dy)f_{\text{den}}^{(\alpha)}\left(y^{(\alpha)}_1\right) = 0$
and $(d/dy)f_{\text{num}}^{(\alpha)}\left(y^{(\alpha)}_2\right) = 0$. In addition, $\kappa^{(\alpha)}_1=(d^2/dy^2)f_{\text{den}}^{(\alpha)}\left(y^{(\alpha)}_1\right)$ and $\kappa^{(\alpha)}_2=(d^2/dy^2)f_{\text{num}}^{(\alpha)}\left(y^{(\alpha)}_2\right)$
represent the curvatures at the saddle point of the denominator and numerator, respectively.

\subsection{Fast random switching}
\label{app:fast-rand}

Here we compute Eq.~(\ref{eq:phi-saddle}) in the case of randomly switching environment in the realm of the PDMP approximation,
with $P_{\nu/s}^{(r)}\simeq P_{\nu/s}^{{\rm PDMP}}$. To compute the denominator of Eq.~(\ref{eq:avg_fix-1}), with Eq.~(\ref{eq:PpdmpSM}), we define
\begin{equation}
\label{eq:gr}
f_{\text{den}}^{(r)}(y)=\ln P_{\nu/s}^{{\rm PDMP}}(y) = -2 {\frac{\nu}{s}} \ln y + \left[\left(1-\delta\right)\frac{\nu}{s} -1\right] \ln
\left(1+\gamma -y\right) + \left[\left(1+\delta\right) {\frac{\nu}{s}} -1\right] \ln \left(y-1+\gamma\right).
\end{equation}
Thus, the saddle point is found at
\begin{equation}
y^{(r)}_1\simeq \frac{\left(1-\gamma^{2}\right)}{\left(1-\delta\gamma\right)}\left[1+\frac{\gamma\left(\delta-\gamma\right)}{\left(1-\delta\gamma\right)^2\nu/s}\left(1+\frac{\left(1-2\gamma^{2}+\delta\gamma\right)}{\left(1-\delta\gamma\right)^2\nu/s}\right)\right].\nonumber
\end{equation}
As a result, we find
\begin{eqnarray}
\hspace{-5mm}f_{\text{den}}^{(r)}\left(y^{(r)}_1\right) & \simeq& \left(\nu/s\right)\left\{\left(1+\delta\right)\ln\left[\frac{\gamma\left(1+\delta\right)\left(1-\gamma\right)}{\left(1-\delta\gamma\right)}\right]+\left(1-\delta\right)\ln\left[\frac{\gamma\left(1-\delta\right)\left(1+\gamma\right)}{\left(1-\delta\gamma\right)}\right]-2\ln\left[\frac{1-\gamma^2}{1-\delta\gamma}\right]\right\}\nonumber\\
&+& \ln\left[\frac{\left(1-\delta\gamma\right)^2}{\gamma^2 \left(1-\delta^2\right)\left(1-\gamma^2\right)}\right]  +\frac{\left(\delta-\gamma\right)^2}{\left(1-\delta^2\right)\left(1-\delta\gamma\right)^2\nu/s},\nonumber\\ \kappa^{(r)}_1&=&\frac{d^2}{dy^2}f_{\text{den}}^{(r)}\left(y^{(r)}_1\right)\simeq \frac{-2 \left(1\!-\!\delta\gamma\right)^4 \nu/s}{\left(1\!-\!\delta^2\right)\gamma^2\left(1\!-\!\gamma^2\right)^2} + \frac{2\left(1\!-\!\delta\gamma\right)^2 \left(1 \!+\!6\delta\gamma\!-\!2\delta^3\gamma\!-\!5\gamma^2\!-\!3\delta^2(1\!-\!\gamma^2)\right)}{\left(1-\delta^2\right)^2 \left(1-\gamma^2\right)^2\gamma^2}\!.
\end{eqnarray}
To compute the numerator of~(\ref{eq:phi-saddle}) we define $f_{\text{num}}^{(r)}(y)=f_{\text{den}}^{(r)}(y)+K_{0}\ln(1\!-\!s)\left(1\!-\!x_{0}\right)y$, and find the saddle point at
\begin{eqnarray}
y^{(r)}_2&\simeq& \frac{\left(1-\gamma^{2}\right)}{\left(1-\delta\gamma\right)}\Bigg\{ 1+\frac{\gamma \left[2\left(1-\delta\gamma\right)\left(\delta-\gamma\right) +b\gamma \left(1-\gamma^2\right)\left(1-\delta^2\right) \right]}{2\left(1-\delta\gamma\right)^3 \nu/s}\bigg[1+\nonumber\\
&&\frac{2-2\gamma^2\left(2+\delta^2-2\delta\gamma\right) - b\gamma\left(1-\gamma\right)^2 \left(2\delta - 3\gamma +\delta^2 \gamma\right)}{2\left(1-\delta\gamma\right)^3\nu/s}\bigg]\Bigg\} ,\nonumber
\end{eqnarray}
where $b=K_{0}\left(1-x_{0}\right)\ln\left(1-s\right)$.
As a result, we find
\begin{eqnarray}
f_{\text{num}}^{(r)}\left(y^{(r)}_2\right) & \simeq& \left(\nu/s\right)\left\{\left(1+\delta\right)\ln\left[\frac{\gamma\left(1+\delta\right)\left(1-\gamma\right)}{\left(1-\delta\gamma\right)}\right]+\left(1-\delta\right)\ln\left[\frac{\gamma\left(1-\delta\right)\left(1+\gamma\right)}{\left(1-\delta\gamma\right)}\right]-2\ln\left[\frac{1-\gamma^2}{1-\delta\gamma}\right]\right\}\nonumber\\
&+& \ln\left[\frac{\left(1-\delta\gamma\right)^2}{\gamma^2 \left(1-\delta^2\right)\left(1-\gamma^2\right)}\right] + \frac{b\left(1-\gamma^2\right)}{1-\delta\gamma} +\frac{\left[2\left(\delta-\gamma\right)\left(1-\delta\gamma\right)+b\gamma\left(1-\gamma^2\right)\left(1-\delta^2\right)\right]^2}{4\left(1-\delta^2\right)\left(1-\delta\gamma\right)^4\nu/s},\nonumber\\
\kappa^{(r)}_2&=&\frac{d^2}{dy^2}f_{\text{num}}^{(r)}\left(y^{(r)}_2\right) \simeq -\frac{-2 \left(1-\delta\gamma\right)^4 \nu/s}{\left(1-\delta^2\right)\gamma^2\left(1-\gamma^2\right)^2} - \frac{2\left(1-\delta\gamma\right)}{\left(1-\delta^2\right)^2 \left(1-\gamma^2\right)^2\gamma^2}\bigg[\left(5-3b\right)\gamma^2 +3b\gamma^4 + \nonumber\\
& &\delta^2\left(3+\left(3+4b\right)\gamma^2 - 4b\gamma^4\right) -\delta^4\gamma^2\left(2+b\left(1-\gamma^2\right)\right) +\nonumber\\
&& \delta^3\gamma\left(-1+3\gamma^2-2b\left(1-\gamma^2\right)\right) +\delta\gamma\left(2b\left(1-\gamma^2\right)-5\left(1+\gamma^2\right)\right)-1\bigg].\label{eq:saddle-G-noise}
\end{eqnarray}

\subsection{Fast periodic switching}
\label{app:fast-per}
Here we compute Eq.~(\ref{eq:phi-saddle}) in the case of periodically switching environment using $P_{\nu/s}^{(p)}\simeq P_{\nu/s}^{{\rm Kap}}$ with Eq.~(\ref{eq:PKapitzaSM}). To compute the denominator of Eq.~(\ref{eq:phi-saddle}) we define
\begin{equation}
\label{eq:grper}
f_{\text{den}}^{(p)}(y)=\ln P_{\nu/s}^{{\rm Kap}}(y) = K_0 \left[y-y\ln\left(\frac{K_0}{\cal K}y\right) -
\frac{1}{72\nu^2/s^2}~\left(\frac{\gamma}{1-\gamma^2}\right)^2\left(2y-\frac{\cal K}{K_0}\right)^3\right].
\end{equation}
Thus, using Eq.~(\ref{eq:C}) the saddle point is found at $y^{(p)}_1\simeq A^{-1}\left[1-C/(A^{2}\nu^{2})\right]$, where $A$ and $C$ are given in Sec.~2.2.1.
As a result, we find
\begin{align}
f_{\text{den}}^{(p)}\left(y^{(p)}_1\right) & \simeq \frac{1}{A}\left(1-\frac{C}{6A^{2}\nu^{2}}\right),\;\;\;\; \kappa^{(p)}_1=\frac{d^2}{dy^2}f_{\text{den}}^{(p)}\left(y^{(p)}_1\right)\simeq -A-5\frac{C}{A\nu^{2}}.\label{eq:saddle-g-noiseper}
\end{align}
To compute the numerator of~(\ref{eq:phi-saddle}) we define $f_{\text{num}}^{(p)}(y)=f_{\text{den}}^{(p)}(y)+K_{0}\ln(1\!-\!s)\left(1\!-\!x_{0}\right)y$, and find the saddle point at
\begin{equation}
y^{(p)}_2\simeq A^{-1}\left(1-s\right)^{1-x_{0}}\left\{ 1-C/(A^{2}\nu^{2})\left[2\left(1-s\right)^{1-x_{0}}-1\right]^{2}\right\} .\nonumber\label{eq:saddle_n}
\end{equation}
As a result, we find
\begin{align}\label{eq:saddle-g-noise-1-1}
f_{\text{num}}^{(p)}(y^{(p)}_2) & \simeq \frac{\left(1-s\right)^{1-x_{0}}}{A}-\frac{C}{6A^{3}\nu^{2}}\left(2\left(1-s\right)^{1-x_{0}}-1\right)^{3},\nonumber\\
\kappa^{(p)}_2=\frac{d^2}{dy^2}f_{\text{num}}^{(p)}\left(y^{(p)}_2\right)& \simeq -\frac{A}{\left(1-s\right)^{1-x_{0}}}\left\{ 1+\frac{C}{A^{2}\nu^{2}}\left[2\left(1-s\right)^{1-x_{0}}-1\right]\left[6\left(1-s\right)^{1-x_{0}}-1\right]\right\}.
\end{align}

Thus, for both random and periodic switching \eqref{eq:phi-saddle} predicts the same
fixation probability $\phi_r= \phi_p \simeq \phi^{(\infty)}=
\phi(x_0)|_{\mathcal{K}}$,
for  $\nu \to \infty$. Yet, the asymptotic \textit{convergence} to  $\phi^{(\infty)}$ is markedly different
[see Eq.~(4) in the main text]:
\begin{equation*}
\ln \left(\frac{\phi_{\alpha}}{\phi^{(\infty)}}\right) = \begin{cases} {\cal A}_{r} (\nu/s)^{-1} &\mbox{for randomly switching environment} \\
{\cal A}_{p} (\nu/s)^{-2} & \mbox{for periodically switching environment, with} \end{cases}
\end{equation*}
\begin{eqnarray}\label{asymptoticPhi}
{\cal A}_{r}&=& (1-x_0)\ln\left(1-s\right) \mathcal{K} \frac{\left(1-\delta^2\right)\gamma^2}{2(1-\delta\gamma)^2}\left(1+\frac{(1-x_0)\ln\left(1-s\right)\mathcal{K}}{2}\right),\nonumber\\
{\cal A}_{p} &=& \frac{\cal K}{72} \left\{1-\left[1+2\left(1-x_0\right)\ln\left(1-s\right)\right]^3\right\}
\left(\frac{\gamma}{1-\gamma\delta}\right)^2.
\end{eqnarray}
These show that $\phi_{p} (\nu)$ approaches  $\phi^{(\infty)}$
much faster than $\phi_{r} (\nu)$ as $\nu$ increases: the convergence  towards the fast switching limit is attained
much quicker with periodic than random switching, see Figs.~3(a) and \ref{fig:FigS1}(c).

\subsection{Validity of the PPP and PDMP approximations in the intermediate/fast switching regime}\label{Sec:PDMP_var}

Simulation results show that $P_{\nu}^{{\rm PPP}}$ and $P_{\nu}^{{\rm PDMP}}$  are generally good approximations of $P_{\nu}^{(p)}$ and
 $P_{\nu}^{(r)}$ for a broad range of $\nu$, from slow to fast switching. We now combine the results of Sections \ref{app:PPP}, \ref{app:LNA} and \ref{app:fast-rand} of this SM
to assess the theoretical validity of the PPP and PDMP approximations, $P_{\nu}^{{\rm PPP}}(N)$ and $P_{\nu}^{{\rm PDMP}}(N)$, given by Eqs.~(\ref{PPPapprox}) and (\ref{eq:PpdmpSM}), under intermediate/fast switching.
This can be done by computing the variance of $P_{\nu}^{{\rm PPP}}$ and $P_{\nu}^{{\rm PDMP}}$, i.e., $\sigma^2_{{\rm PPP}}$ and
$\sigma^2_{{\rm PDMP}}$, and by comparing these results with
${\cal K}$, which is the variance of the PSD when, in the limit $\nu\to\infty$, it is solely
governed by demographic noise. Indeed, when $\nu\to\infty$, the PSD [both the Kapitza approximation given by Eq.~(\ref{eq:PKapitzaSM}) as well as the LNA given by Eq.~(\ref{eq:PLNA})] reduces to a Gaussian of mean
and variance ${\cal K}$, i.e., $P_{\nu\to \infty}(N) \propto e^{-(N-{\cal K})^2/(2{\cal K})}/\sqrt{{\cal K}}$.
 To compute the variances in the limit of $\nu\gg 1$, we perform a saddle-point calculation as in the previous section, and find to leading order in $1/\nu$ that
 \begin{eqnarray}
  \label{eq:varPdmp}
  \sigma^2_{{\rm PPP}}&=&\int_{N_{{\rm min}}}^{N_{{\rm max}}}\left(N-\langle N\rangle_{{\rm PPP}} \right)^{2}P_{\nu}^{{\rm PPP}}(N)dN=
\frac{1}{12}\left(\frac{\gamma}{1-\gamma\delta}\right)^2\frac{{\cal K}^2}{\nu^2},\nonumber\\
  \sigma^2_{{\rm PDMP}}&=&\int_{K_-}^{K_+} (N-\langle N \rangle_{{\rm PDMP}})^2P_{\nu}^{{\rm PDMP}}(N)dN=\frac{1}{2}\left(\frac{\gamma }{1-\gamma \delta}\right)^2\left(1-\delta^2\right)~\frac{{\cal K}^2}{\nu}.
 \end{eqnarray}
Here we have used Eqs.~(\ref{PPPapprox}) and (\ref{eq:PpdmpSM}), while
\begin{eqnarray}\label{means}
\langle N\rangle_{{\rm PPP}}&=&\int_{N_{{\rm min}}}^{N_{{\rm max}}}NP_{\nu}^{{\rm PPP}}(N)dN={\cal K}\left[1+\frac{\gamma^2}{12(1-\gamma\delta)^2\,\nu^2}\right],\nonumber\\
\langle N \rangle_{{\rm PDMP}}&=&\int_{K_-}^{K_+} N~P_{\nu}^{{\rm PDMP}}(N)dN
= {\cal K}\left[  1 + \frac{\gamma^2 (1-\delta^2)}{2  (1-\delta\gamma)^2\,\nu}\right],
\end{eqnarray}
to leading order in $1/\nu$.
Notably, from Eqs.~(\ref{eq:varPdmp}) one can see that when $\nu\gg 1$, $\sigma^2_{{\rm PPP}}\sim \nu^{-2}$ while $\sigma^2_{{\rm PDMP}}\sim \nu^{-1}$. This indicates that the PSD's width under random switching is significantly larger than in the periodic case, which allows the total population size to probe smaller values of $N$  in the random than periodic case. This ultimately leads to a larger fixation probability $\phi_r$ than $\phi_p$ (when $s>s_c$, see Sec.~S5.2). Importantly, since it is the PSD's mean that determines the fixation probability at high switching rates, the fact the PSD's mean here converges at a different rate to $\cal K$ for  periodic and random switching gives rise to the different asymptotic behavior of $\phi_r$  and $\phi_p$ when $\nu\to\infty$, yielding Eq.~(4) in the main text, see also Eq.~(\ref{asymptoticPhi}).
Notably, the convergence details $\phi_{\alpha}\xrightarrow{\nu\to \infty} \phi^{(\infty)}$ are expected to generally depend on the underlying periodic/random processes $\xi_{\alpha}(t)$.

What is the regime of applicability of these PDMP-like approximations?
According to Eqs.~(\ref{eq:varPdmp}), $\sigma^2_{{\rm PPP}}\gg {\cal K}$ for  $1 \ll \nu\ll \sqrt{K_0}$, while $\sigma^2_{{\rm PDMP}}\gg {\cal K}$
for $1 \ll \nu\ll K_0$; in these regimes the variance stemming from periodic/random switching is much larger than the variance caused by demographic fluctuations.
 Hence,  $P_{\nu}^{{\rm PPP}}$ and $P_{\nu}^{{\rm PDMP}}$ are  accurate approximations of $P_{\nu}^{(p)}$ and $P_{\nu}^{(r)}$ in the fast switching regime respectively when  $\nu\ll \sqrt{K_0}$ and  $\nu\ll K_0$. Remarkably, environmental noise also dominates over demographic fluctuations for slow/intermediate switching regime when $\nu\lesssim 1$, see Fig.~2. Therefore, these PDMP-like approximations neglecting demographic noise accurately describe of $P_{\nu}(N)$  over a broad range of $\nu$.
It is worth noting that for $\nu\gg 1$, the variance of  $P_{\nu}^{{\rm Kap}}$ to leading order in $1/\nu$ satisfies $\sigma^2_{{\rm Kap}}={\cal K}\left[1+{\cal O}\left({\cal K}/\nu^{2}\right)\right]$. Thus, as we have checked, the variances of $P_{\nu}^{{\rm PPP}}$ and $P_{\nu}^{{\rm Kap}}$
coincide in the leading order, for $1 \ll \nu\ll\sqrt{K_0}$. However, $\sigma^2_{{\rm Kap}}\gg \sigma^2_{{\rm PPP}}$ when $\nu\gtrsim\sqrt{K_0}$, which indicates that the Kapitza-based approximation is superior to that of the PPP in this regime. This also reiterates that, at very  high switching rates, one must take demographic noise into account as is done using the Kapitza method (see previous section).  Since the Kapitza-based approximation works well for any {\it arbitrary} switching rate $\nu\gg 1$, the calculation of Sec.~S4.2 leading to the  fast switching asymptotic behavior of $\phi_p(\nu)$ has been carried out using $P_{\nu/s}^{\rm Kap}$ instead of $P_{\nu/s}^{\rm PPP}$.

However, while the PPP and  PDMP approximations characterize well the PSD in the random and periodic cases, respectively when  $\nu\ll K_0$  and $\nu\ll K_0^{1/2}$, they can still be aptly used for the purpose of calculating the fixation probability, $\phi_{\alpha}$, at arbitrary high switching rates according to (\ref{eq:Pppp-basedSM}) and (\ref{eq:Ppdmp-basedSM}), see Figs.~3(a,d) and \ref{fig:FigS1}~(d). This is because  only the vicinity of the PSD's  maximum contributes to the leading-order
calculation of $\phi_{\alpha}$ at high $\nu$,  which is well captured, for any high switching rate, by these PDMP-like approximations.

\section{Further details about Figure 3(d,e) in the main text}
\label{AppendixA}
Here we  elaborate on our findings, see main text, that under certain conditions the fixation probability of the $S$ species, $\phi_{\alpha}$, at given $s,\gamma,\delta,K_0,x_0$ is optimal for a nontrivial switching rate $\nu_{\alpha}^*$,
see Fig.~3(d,e). We also discuss the critical selection intensity below/above which
$\phi_{\alpha}(\nu)$ is an increasing/decreasing function under weak switching asymmetry.
\subsection{Region of the parameter space in which the fixation probability is nonmonotonic }
Our starting point is Eq.~(3) of the main text which, when substituting $P_{\nu/s}^{(r)}$ by its PDMP approximation, reads
\begin{eqnarray}
\label{eq:Ppdmp-basedSM}
\phi_r(\nu)\simeq
\int_{K_-}^{K_+}~P_{\nu/s}^{{\rm PDMP}}(N)~\phi(x_0)|_{N}~dN.
\end{eqnarray}
\begin{figure}[h!]
\centering
\hspace*{0.93in}
\includegraphics[width=0.30\linewidth]{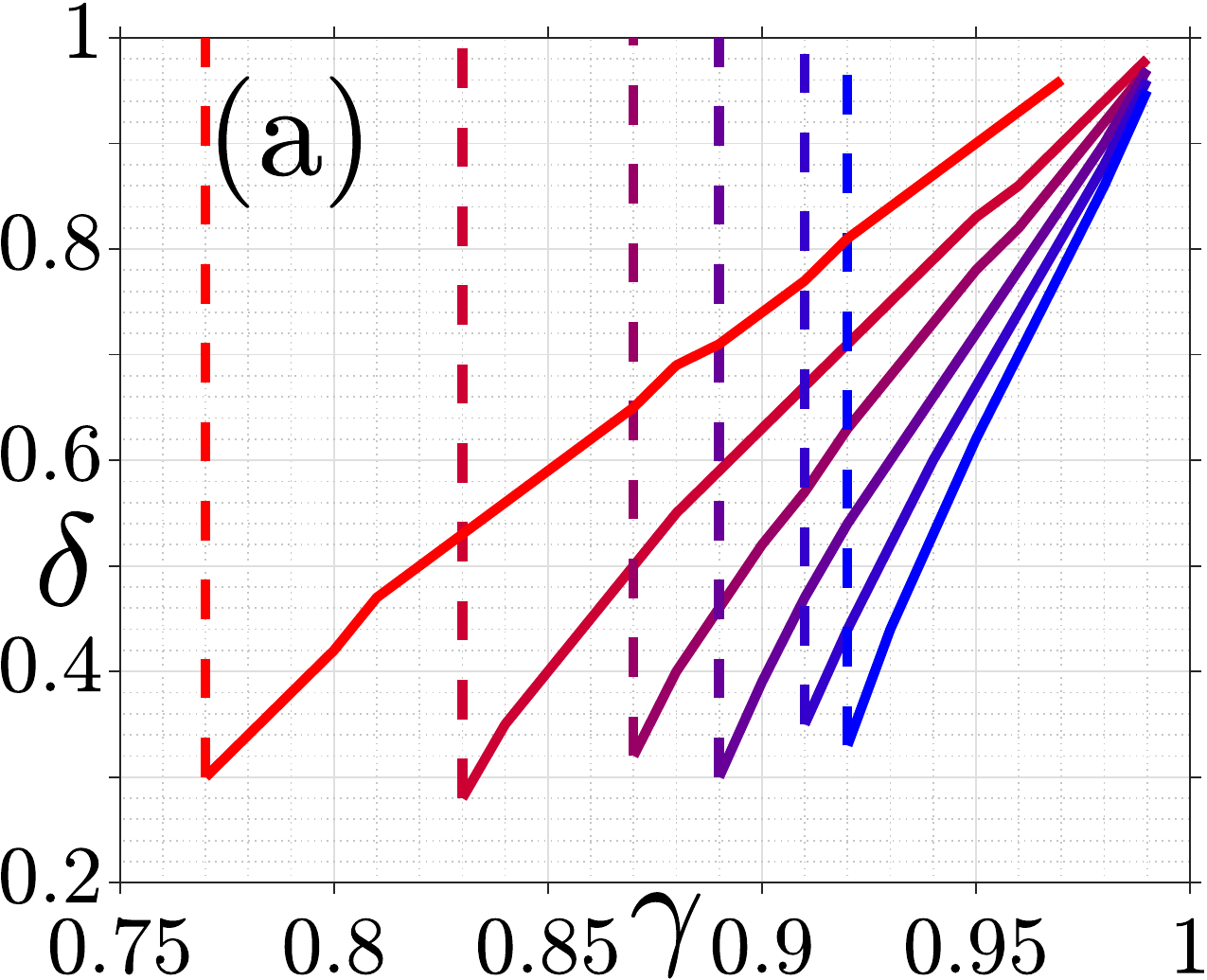} \hspace{1em} \includegraphics[width=0.31\linewidth]{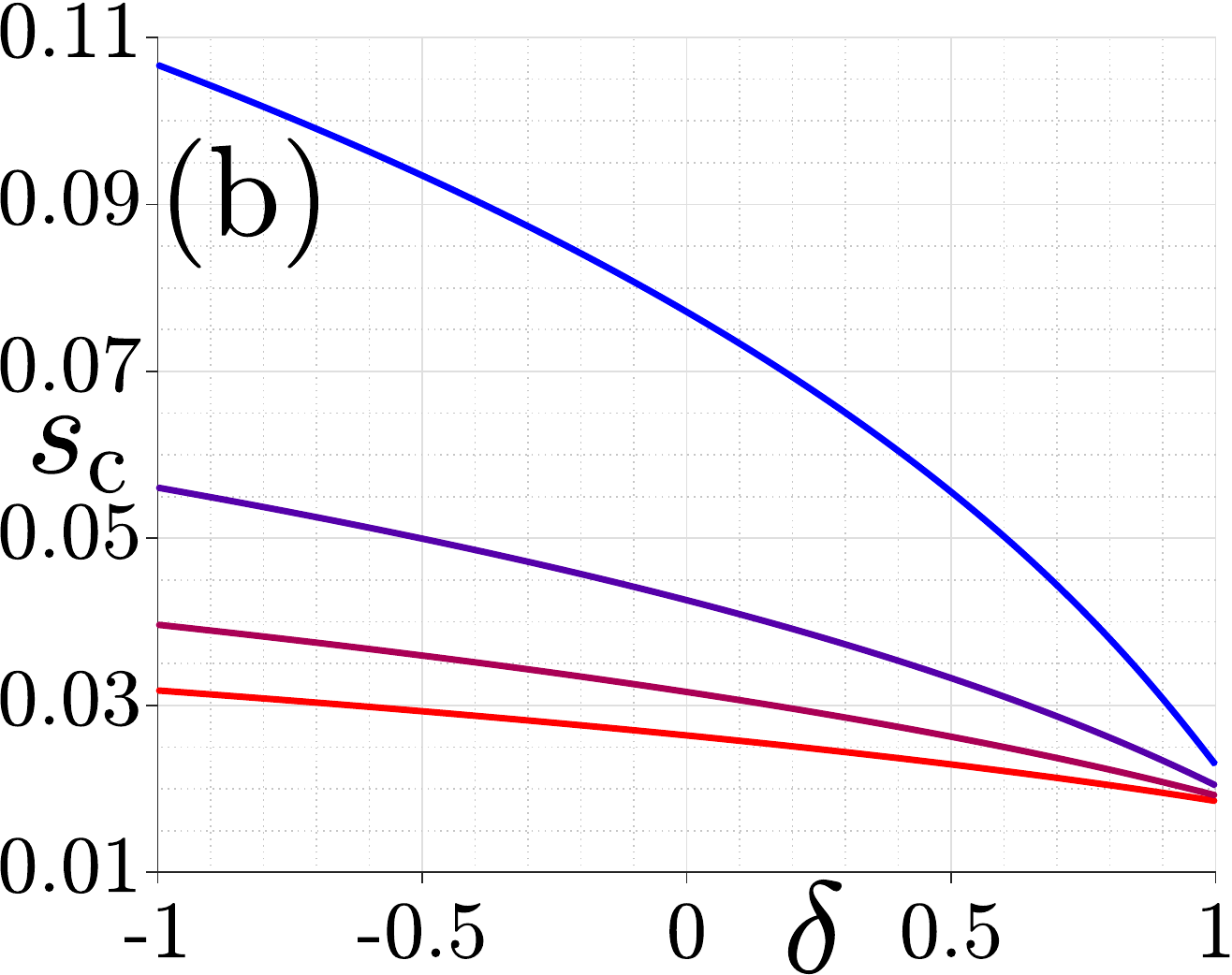}\\[0ex]
\hspace*{.75in}
\hspace*{0.02in}
\includegraphics[width=0.265\linewidth]{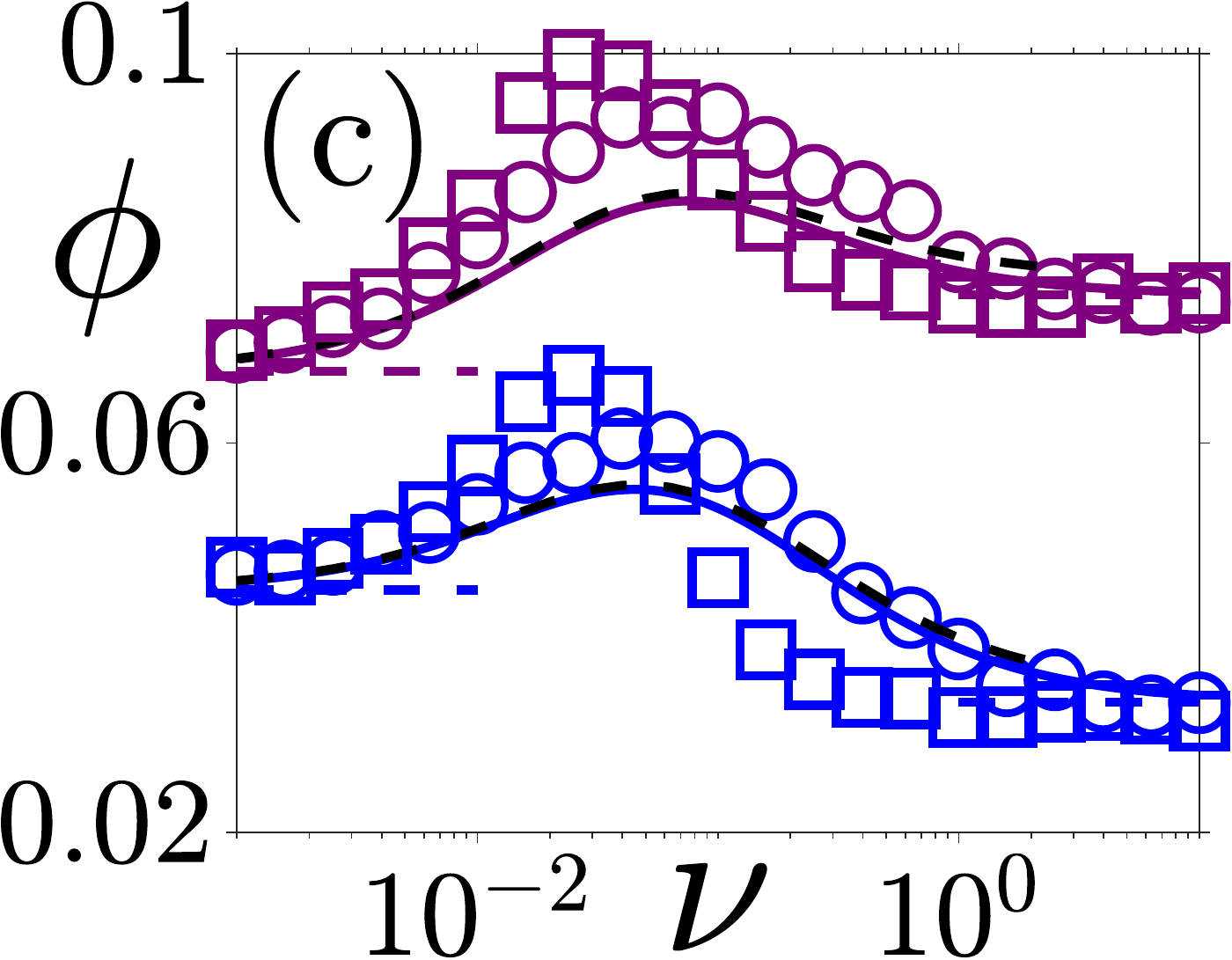}
\hspace*{0.06in}
\vspace{-0.04in}
\includegraphics[width=0.265\linewidth]{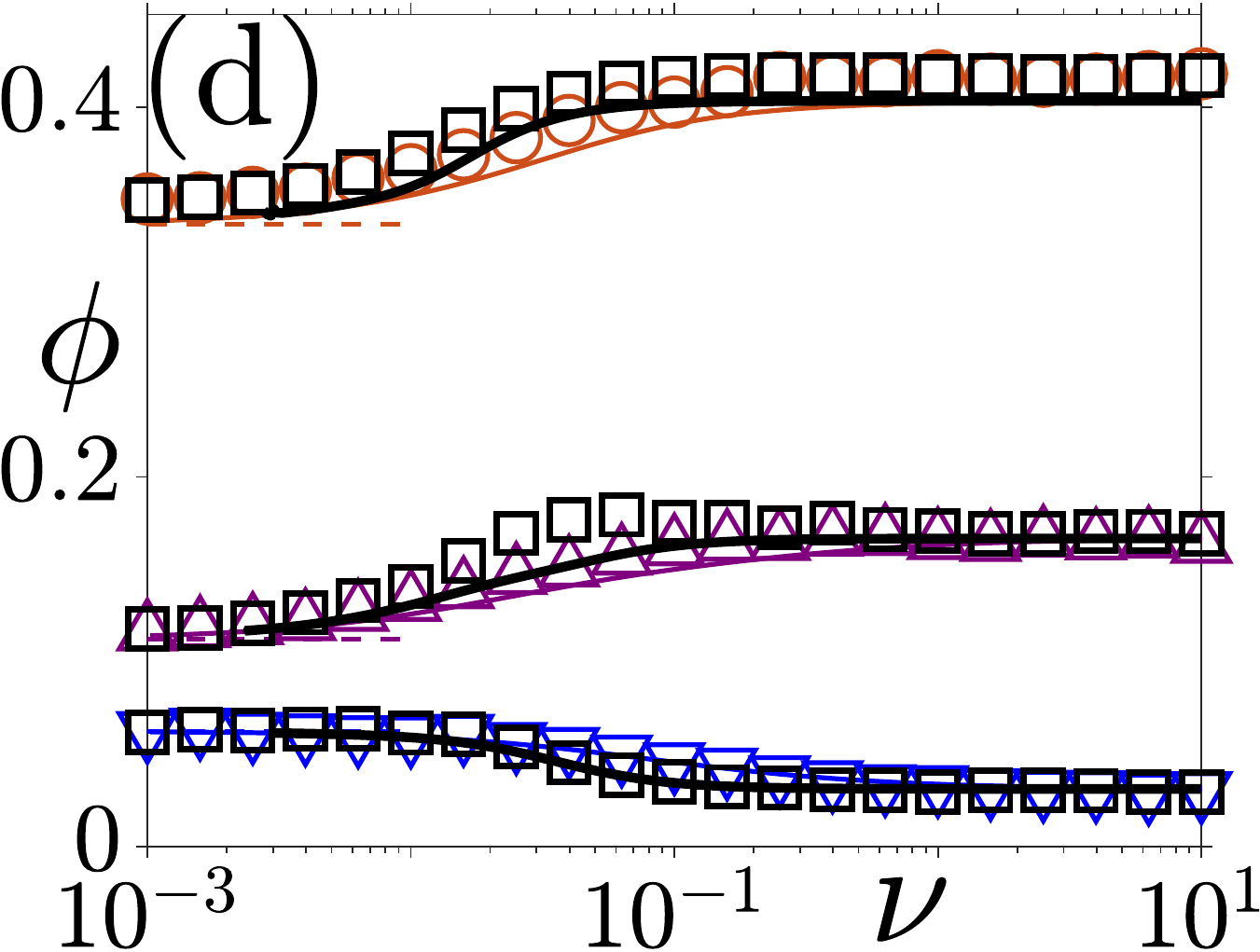}
\hspace*{0.06in}
\includegraphics[width=0.28\linewidth]{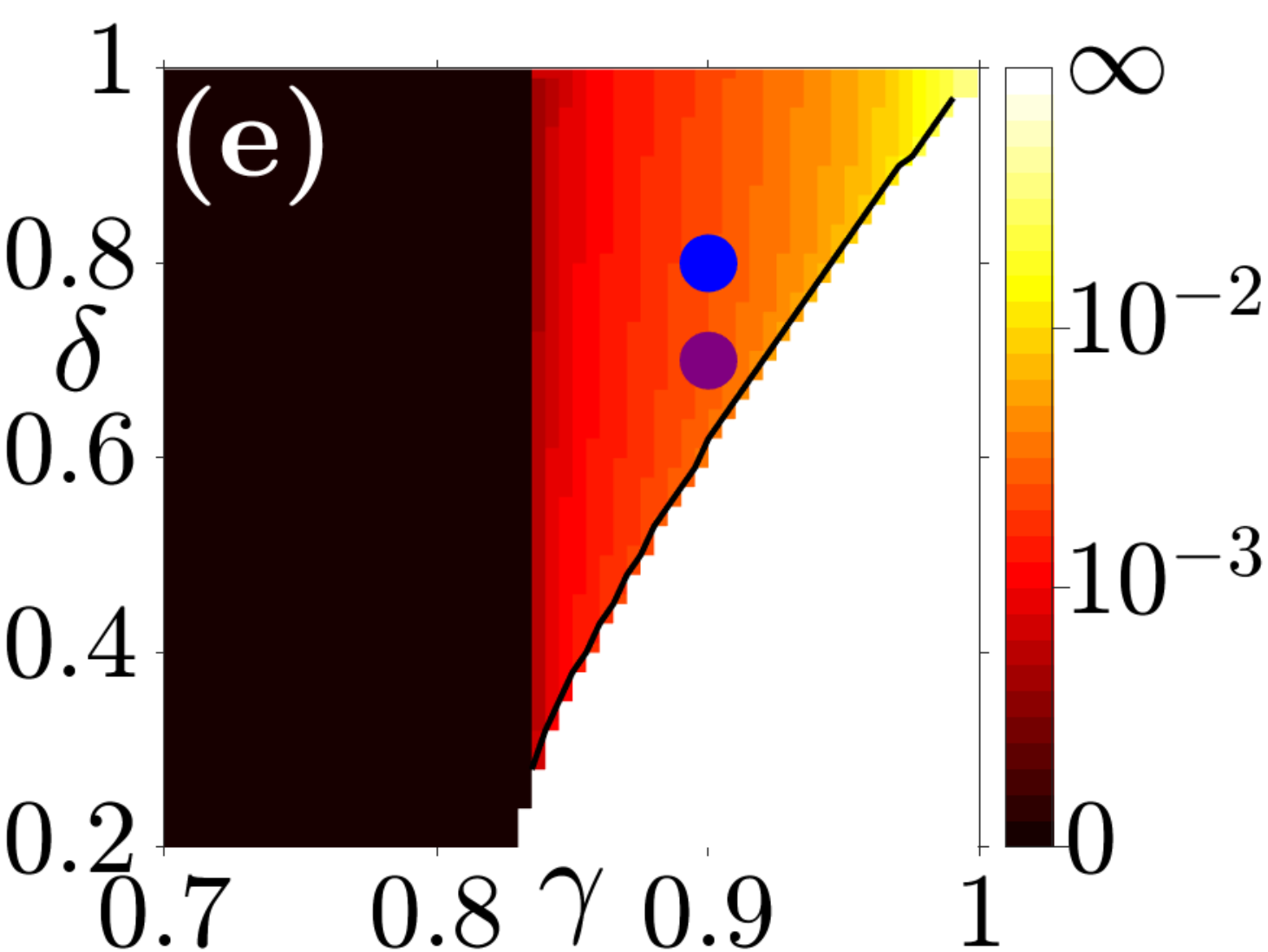}
\caption{ (a) Triangular-like region in the parameter space in which $\phi_{r}(\nu)$ has a nontrivial maximum
at $\nu=\nu_r^*$ for $s = 0.04, 0.05, 0.06, 0.07, 0.08, 0.09$
	(red to blue, left to right) obtained from Eq.~(\ref{eq:Ppdmp-basedSM}). This region, defined by
	$\gamma>\gamma_c$, $\delta>\delta_c(\gamma,s)$ is delimited by the solid and dashed lines, see text and compare with
	Fig.~3(e) in the main text.
	(b) Critical selection intensity $s_c$
	as a function of $\delta$ for $\gamma = 0.6, 0.7, 0.8, 0.9$ (red to blue, bottom to top)
	for $K_0 = 250$ and $x_0 = 0.6$, see text.
	(c) $\phi_r$ (circles) and $\phi_p$ (squares)  versus $\nu$ with $(s,K_0,\gamma, x_0) = (0.05, 250, 0.9, 0.6)$, $\delta=0.7$ (purple) and  $\delta=0.8$ (blue);
	symbols are from simulations;  colored solid lines are from Eq.~\eqref{eq:Ppdmp-basedSM}
	and  dashed lines are based on Eq.~(3) by averaging over Eq.~\eqref{eq:PLNA}, showing the minute improvement over the PDMP-based approximation \eqref{eq:Ppdmp-basedSM}
	achieved by using the $P_{\nu}^{{\rm LNA}}$ to approximate the PSD in Eq.~(3),	see text.
(d) Colored symbols show
$\phi_r$   versus $\nu$ for
$(s, K_0, x_0)=(0.05,250,0.6)$ and
$(\gamma,\delta)=(0.9, -0.5)$ (orange),
$(\gamma,\delta)=(0.9, 0.5)$ (purple),
$(\gamma,\delta)=~(0.8, 0.6)$~(blue). Solid colored lines are from (\ref{eq:Ppdmp-basedSM})
and
dashed lines show $\phi^{(0,\infty)}$  for $\nu\to 0,\infty$.
Results for
 $\phi_p$ and same values of $(\gamma,\delta)$
 are shown as black squares, with
 solid black lines from Eq.~(\ref{eq:Pppp-basedSM}).
(e) Heatmap of  $\nu^*_{p}$:
$\nu^*_{p} \to 0,\infty$ in the black
and white areas, respectively;  $\phi_p(\nu)$ is non-monotonic in the red-yellow area, with the values of
$\nu_{p}^*$ given in the vertical bar; parameters are $(s,K_0, x_0) = (0.05, 250,  0.6)$.
Symbols indicate $\nu_{p}^*$  for $\gamma=0.9$, $\delta=0.7$ (purple) and  $\delta=0.8$ (blue), for $\phi_p$ shown  in (c).
To be compared with Fig.~3(e) of the main text
showing the heatmap of $\nu^*_{r}$, see text.
}
\label{fig:FigS1}
\end{figure}
We now give further details on how to determine from this equation the region of the parameter space of Fig.~3(e)
in which $\phi_{\alpha}(\nu)$ is
non-monotonic, and how this region changes when  $s$ is increased.
Using the diffusion approximation $\phi(x_0)|_N \simeq  (e^{-Ns(1-x_0)}-e^{-Ns})/(1-e^{-Ns})$,  and Eq.~(\ref{eq:PpdmpSM}),  we compare the PDMP-based approximation
of $\phi_r(\nu)$ [Eq.~(\ref{eq:Ppdmp-basedSM})] for different switching rates (slow, intermediate and fast switching) for a given set $(K_0,s,\gamma, \delta,x_0)$,
and determine for which of these values $\phi_r$ is maximal.

If $\phi_r(\nu\ll s)>\phi_r(\nu\gg s), \phi_r(\nu\sim s)$, we say
that the optimal fixation probability is $\phi_r(0)=\phi^{(0)}$
 at slow switching, i.e., $\nu_r^*=0$. Similarly, if $\phi_r(\nu\gg s)>\phi_r(\nu\ll s), \phi_r(\nu\sim s)$,
the optimal fixation probability is $\phi_r(\infty)=\phi^{(\infty)}$, i.e., $\nu_r^*=\infty$, see Fig.~3(a). Otherwise, the $S$ fixation probability
  is maximal at a non-trivial switching rate $\nu_r^*\sim s$
that Eq.~(\ref{eq:Ppdmp-basedSM}) captures reasonably well. In this case, $\phi_r$ varies non-monotonically with $\nu$,  see Fig.~3(d,e).
We have performed extensive stochastic simulations
of the model's dynamics  and found that this behavior arises in a triangular-like region in the subset of the parameter space where
$\gamma$ and $\delta$ exceed some critical values $\gamma_c(s)$ and $\delta_c(\gamma, s)$,  see Fig.~3(e).
In order to determine the boundary $(\gamma_c(s),\delta_c(\gamma, s))$ of this triangular-like region at fixed $s$ and $K_0$, we have systematically calculated the fixation probability as $\nu$ varies from $10^{-3}$ (proxy for $\nu\ll s$)
to $1$ (proxy for $\nu\gg s$) for fixed
$(\gamma, \delta)$, with  $\gamma\gtrsim 0.7$, and $\delta \gtrsim 0.2$. For each pair $(\gamma,\delta)$ we have then found the value of $\nu$ for which it attains its maximum and store it in a matrix. In practice, the diagonal part of the boundary is then found by keeping $\gamma$ fixed and increasing $\delta$ until we find the first entry of the matrix for which $10^{-3}<\nu^*_r < 1$. This determines $\delta_c\left( \gamma,s\right)$. The left hand side of the boundary, $\gamma_c(s)$ is found by finding the largest value of $\gamma$ such that $\nu^*_r = 1$ or $\nu^*_r =10^{-3}$ for all $\delta$.
 Predictions of Eq.~(\ref{eq:Ppdmp-basedSM}) are in good agreement with simulation results which confirm
 that $\phi_{\alpha}(\nu)$ has a nontrivial maximum at $\nu_{\alpha}^*$ when $\gamma>\gamma_c(s)$ and
 $\delta>\delta_c(\gamma, s)$,  see Figs.~3(d) and \ref{fig:FigS1}(c).
  As shown in Fig.~\ref{fig:FigS1}~(a),  $\gamma_c(s)$ is an increasing function of $s$ while
$\delta_c(\gamma, s)$ changes little when $\gamma$ and $s$ are increased. As a result, when the selection intensity $s$
is increased (at $K_0$ fixed), the triangular-like region of the parameter space in which
$\phi_{\alpha}(\nu)$ has a nontrivial maximum is ``squeezed out'', as shown in  Fig.~\ref{fig:FigS1}(a), whereas we have verified
that the optimal fixation probability remains unaltered when $s$ changes but $K_0 s$ is kept fixed.

 A similar analysis can be carried out for the periodic switching. In this case,
 the fixation probability is approximated
 by substituting the PPP approximation (\ref{PPPapprox}) with rescaled switching rate into the PSD in Eq.~(3), yielding
\begin{eqnarray}
\label{eq:Pppp-basedSM}
\phi_p(\nu)\simeq
\int_{N_{\rm min}}^{N_{\rm max}}~P_{\nu/s}^{{\rm PPP}}(N)~\phi(x_0)|_{N}~dN.
\end{eqnarray}
This expression gives a sound approximation of $\phi_p$, and correctly predicts the same qualitative behavior as under random switching, as shown in Figs.~3(d) and \ref{fig:FigS1}(d,e).
We have used Eq.~(\ref{eq:Pppp-basedSM}) to obtain the heatmap of $\nu_p^*$ of Fig.~\ref{fig:FigS1}~(e)
giving the switching rate $\nu_p^*$ for which $\phi_p(\nu)$ is maximal. The comparison with  the heatmap of $\nu_r^*$ of Fig.~3(e) in the main text for the same parameters $s, K_0, x_0$,
shows that both $\phi_p(\nu)$ and $\phi_r(\nu)$ are non-monotonic in qualitatively similar triangular-like regions of the $\gamma-\delta$ parameter space (triangular-like region of  Fig.~\ref{fig:FigS1}~(e) and Fig.~3(e)
are of similar size). We notice that $\nu_p^*\lesssim \nu_r^*$ for given  $\gamma$ and $\delta$, which translates  in the
triangular-like region of Fig.~\ref{fig:FigS1}~(e)
to be  overall more ``reddish'' than the corresponding region of  Fig.~3(e).

\subsection{Critical selection intensity}
\label{app:crit-sel}
As explained in the main text, it is useful to determine the critical selection intensity $s_c$
such that $\phi^{(0)}=\phi^{(\infty)}$. Here, using the diffusion approximation we have
$\phi^{(0)}=[(1-\delta)\phi(x_0)|_{K_-}+(1+\delta)\phi(x_0)|_{K_+}]/2$
and $\phi^{(\infty)}=\phi(x_0)|_{{\cal K}}$. By introducing $z=\exp(-sK_-)$,
$a_1 = (1+\gamma)/(1-\gamma)$ and $a_2 =
(1+\gamma)/(1-\delta \gamma)$, which yields $K_+=a_1 K_-$ and ${\cal K}=a_2 K_-$, $s_c$ is obtained by solving
$\phi^{(0)}=\phi^{(\infty)}$, i.e., it is the solution of the
transcendental equation
\begin{eqnarray}
\label{eq:sc}
	(1-\delta)\left(\frac{z^{-x_0}-1}{1-z}\right)+(1+\delta)z^{a_1 -1}\left(\frac{z^{-a_1 x_0}-1}{1-z^{a_1}}\right)
	-2z^{a_2 -1}\left(\frac{z^{-a_2 x_0}-1}{1-z^{a_2}}\right) =0.\nonumber
\end{eqnarray}
The numerical solutions of this equation, for $s_c=s_c(\gamma,\delta)$,
are reported in Fig.~\ref{fig:FigS1}(b), where we find that $s_c$ decreases with $\delta$, and increases with $\gamma$.
When $s<s_c$, $\phi^{(0)}<\phi^{(\infty)}$ and $\phi^{(0)}>\phi^{(\infty)}$ if $s>s_c$.
This allows us to determine the monotonic behavior of $\phi_{\alpha}(\nu)$ under weak switching asymmetry
($|\delta|<\delta_c$):  When $s<s_c$, $\phi_{\alpha}$ is an increasing function of $\nu$,
while it decreases with $\nu$ if $s>s_c$ (at given $s,\gamma, \delta$).
In the examples of Fig.~\ref{fig:FigS1}(d) we find that $s_c\approx 0.06$  for
$(\gamma,\delta)=(0.9, 0.5)$,
$s_c\approx 0.03$ for
$(\gamma,\delta)=(0.8, 0.6)$, and $s_c\approx 0.095$ for
$(\gamma,\delta)=(0.9, -0.5)$. When $s=0.05$, this corresponds to $\phi_{\alpha}(\nu)$ being an increasing function of $\nu$
for $(\gamma,\delta)=(0.9, 0.5)$ and $(\gamma,\delta)=(0.9, -0.5)$, and decreasing with $\nu$ in the case
$(\gamma,\delta)=(0.8, 0.6)$, which is in  accord with simulation results of Fig.~\ref{fig:FigS1}(d).  It is worth noting that when $0<s\ll 1$ (weak selection) and $K_0\gg 1$, as considered in this work, the generic case is $s>s_c$ and $\phi_{\alpha}(\nu)$ therefore generally decreases with $\nu$ as in Fig.~3(a).

In the regime $\gamma>\gamma_c(s)$ and
 $\delta>\delta_c(\gamma, s)$ where $\phi_{\alpha}(\nu)$ is non-monotonic, we can determine that
 $\phi_{\alpha}(\nu)$ increases steeper at slow/intermediate switching if $s<s_c$
 while it is the opposite when $s>s_c$.
 As confirmed by the results reported in Fig.~\ref{fig:FigS1}(d), where
  $\phi_{r}$ and  $\phi_{p}$ exhibit the same $\nu$-dependence,
 these results hold for both random and periodic
 switching, since in both cases
  $\phi_{\alpha}(\nu)$ essentially coincides with
 $\phi^{(0)}$ and $\phi^{(\infty)}$ for slow/fast switching, respectively.

\subsection{Effective selection intensity under fast switching}
As discussed in the main text, see also Sec.~\ref{sec:Saddle}, under fast random and periodic switching
$\phi_{\alpha}(\nu)\simeq \phi^{(\infty)}=\phi^{(\infty)}|_{{\cal K}}$, with
$\phi^{(\infty)}=e^{m/2}$ and
$m\equiv 2{\cal K}(1-x_0)\ln{(1-s)}=2\left[(1-\gamma^2 )/(1-\gamma\delta)\right]K_0(1-x_0)\ln{(1-s)}$, to leading order in $1/\nu$.
 When $1/K_0\ll s\ll 1/\sqrt{K_0}$ and $\gamma={\cal O}(1)$, the above expression simplifies:
  $\phi^{(\infty)}\simeq e^{-{\cal K}s(1-x_0)}=e^{-s[(1-\gamma^2 )/(1-\gamma\delta)]K_0(1-x_0)}$.
  Hence, in this regime, under fast random and periodic switching,
  the $S$ fixation probability is the same
  as in a population subject to a {\it constant} carrying capacity $K_0$ under
 a rescaled selection intensity $s\to s'=s(1-\gamma^2 )/(1-\gamma\delta)$.
 This result yields the following remarkably simple and enlightening interpretation:
 in the above regime, the effect of  environmental variability, when $\delta<\gamma$,
  is to effectively reduce the selection intensity with respect to the static
  environment, yielding   $\phi_{\alpha}>\phi(x_0)|_{K_0}$ under a selection intensity $s$. Similarly, there is
  an effective increase of selection intensity ($s'>s$)
  when $\delta>\gamma$  resulting in
  $\phi_{\alpha}<\phi(x_0)|_{K_0}$.
\subsection{Duty cycle and general effect of $\delta$ on $\phi_{\alpha}$}
\label{app:duty}
The parameter $\delta$ measures the asymmetry in the switching rate, and
can be used to define the ``duty cycle'' as $(1+\delta)/2$ in the case of periodic switching between $K_+$ and $K_-$
with period $T=(1/\nu_-)+(1/\nu_+)$. The duty cycle gives the fraction $(1/\nu_+)/T$ of one period spent in
the environmental state $\xi=1$. Clearly, when
$\delta>0$, the population spends more time in the environmental state $\xi=+1$ (with $K=K_+$) than in the state
$\xi=-1$ (with $K=K_-$). Since $s>0$, species $S$ has a selective disadvantage with respect to strain $F$
and $\phi_{\alpha}$ is therefore a decreasing function of $\delta$ when all the other parameters are fixed,
see Fig.~\ref{fig:FigS1}(c,d).

\section{Mean fixation time and average number of switches}
\label{AppendixC}

\begin{figure}[t!]
\includegraphics[width=0.31\linewidth,,height=0.25\linewidth]{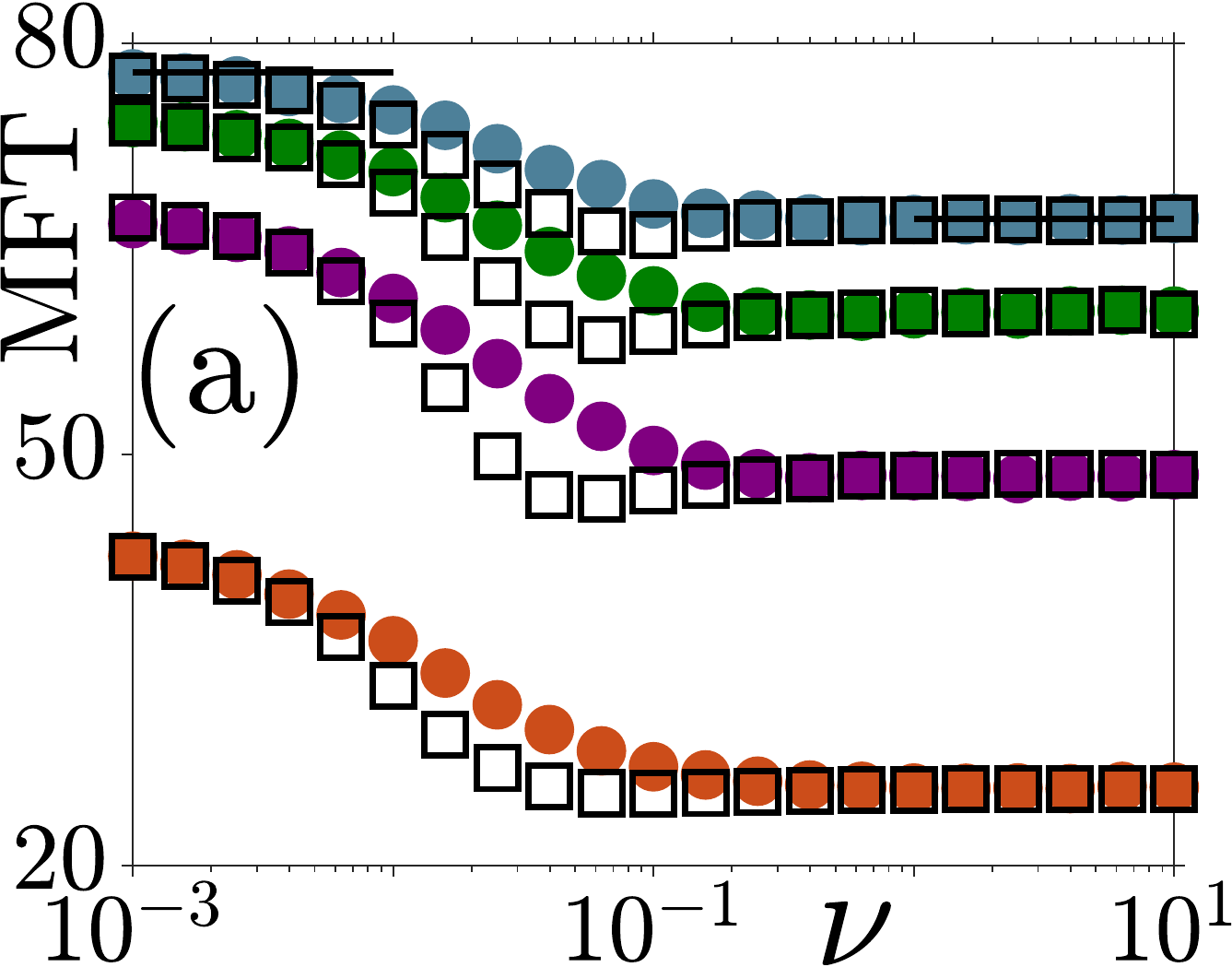}
\includegraphics[width=0.32\linewidth]{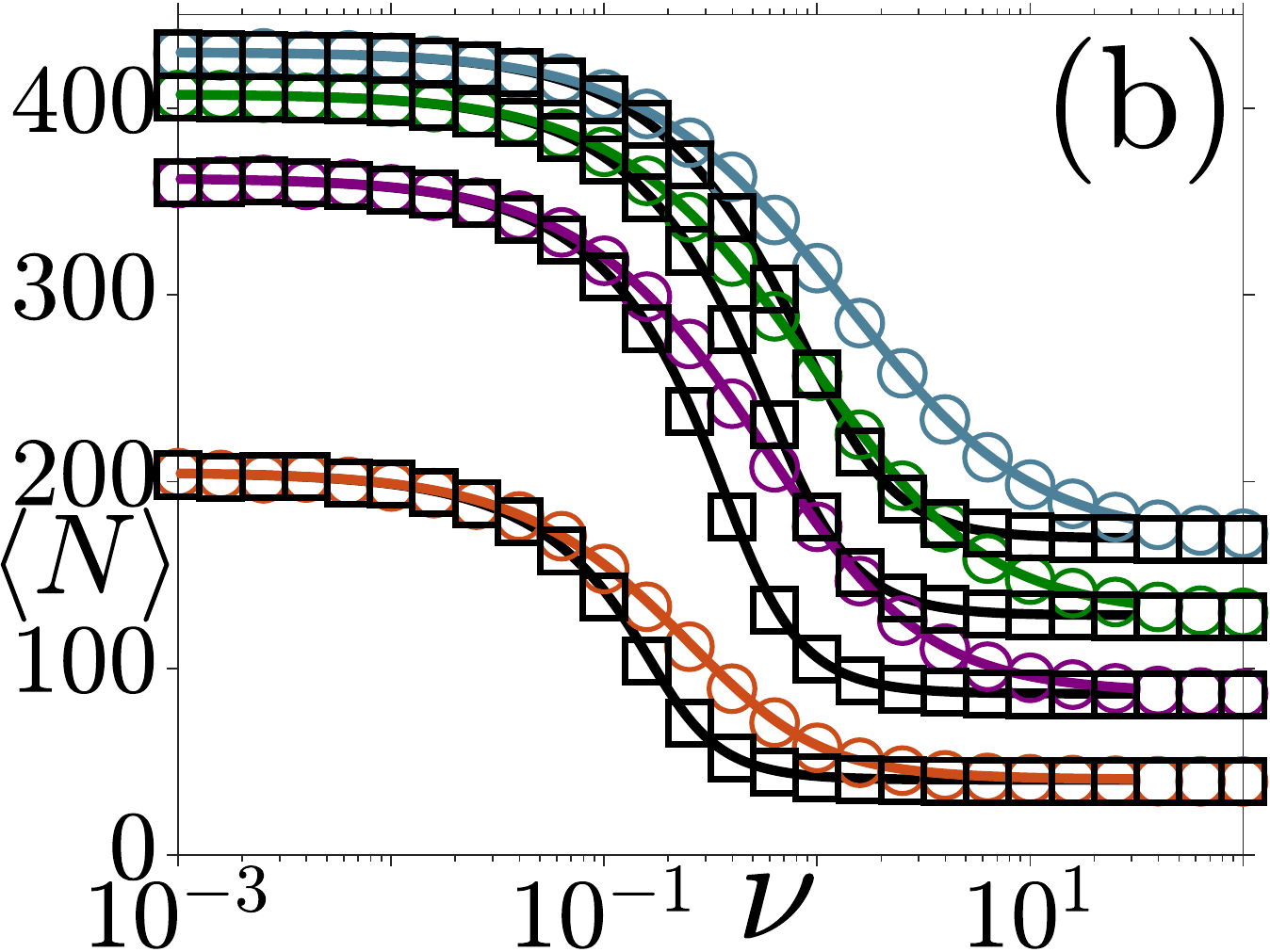}
\includegraphics[width=0.33\linewidth]{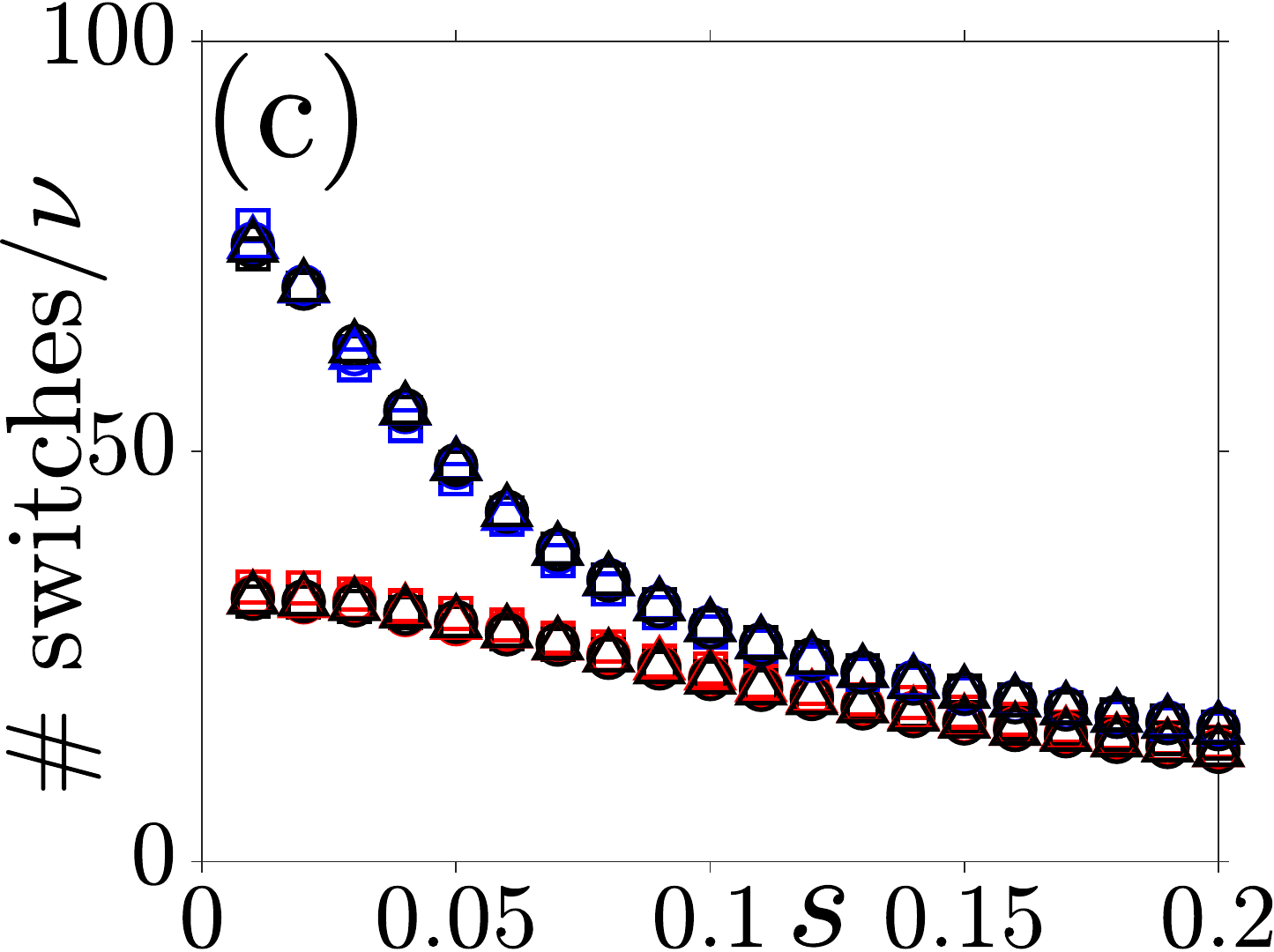}
\caption{(a) Unconditional MFT  under random (circles) and periodic (squares) switching versus $\nu$ for $(K_0,s, x_0) =(250,0.05,0.6)$,  and $(\gamma,\delta) =(0.9, 0.5)$ (purple), $(0.9,-0.2)$ (orange), $(0.9,0.7)$ (green) and
$(0.9,0.8)$ (teal). The MFT scales as $1/s$; the effect of random/periodic switching is to reduce
the subleading corrections due to the decreasing average population size $\langle N \rangle$, see text.
(b) $\langle N \rangle$ versus $\nu$
for random (circles) and periodic (squares) switching, for the same parameters as in (a). Solid colored lines are  given by $\langle N\rangle_{{\rm PDMP}}$ and solid black lines show  $\langle N\rangle_{{\rm PPP}}$, see Eq.~(\ref{means}).
(c) Average number
of switches divided by $\nu$ prior to fixation versus $s$, for $\delta = 0.5, -0.5$ (blue/top, red/bottom) and $\nu = 0.1, 1,
10$ (squares, circles, triangles) are shown to be ${\cal O}(1/s)$ with
data for different values of $\nu$ collapsing together. Other parameters are: $(K_0,\gamma,x_0)=(250, 0.8,0.6)$. Here colored/black symbols are from simulations with random/periodic switching.
}
\label{fig:FigS3}
\end{figure}

In addition to the fixation probability, we have also computed the MFT, $T^{(\alpha)}(x_0)$ -- the unconditional mean
time  until the  fixation  of either species $S$ or $F$, starting from aמ initial
 fraction $x_0$ of individuals of type $S$. As in the case
$\delta=0$, $T^{(\alpha)}(x_0)$
is obtained by averaging the unconditional MFT, $T(x_0)|_{N}$,
obtained in a population of constant size $ N $
over $ P^{(\alpha)}_{\nu/s}(N) $ with a rescaled switching rate $\nu \to \nu/s$~\cite{KEM1,KEM2}.  In the limits of
slow and fast switching, we have $T^{(\alpha)}(x_0)=[(1+\delta)T(x_0)|_{K_+}+
(1-\delta)T(x_0)|_{K_-}]/2$ when $\nu/s\ll 1$ and $T^{(\alpha)}(x_0)=T(x_0)|_{{\cal K}}$ when $\nu/s\gg 1$, see Fig.~\ref{fig:FigS3}(a).
When $1/K_0 \ll s \ll 1$,  $T(x_0)|_{N}\sim {\cal O}(1/s)$~\cite{Ewens,Blythe07},  and  the MFT under random switching also scales as $1/s$, i.e., $T^{(\alpha)}(x_0)={\cal O}(1/s)$; this result is evident since $x$ deterministically relaxes on a time scale ${\cal O}(1/s)$.
As the average population size $\langle N\rangle$  decreases with $\nu$,
see Fig.~\ref{fig:FigS3}(b),
environmental variability reduces the subleading prefactor of
$T^{(\alpha)}(x_0)$~\cite{KEM2}.
 As a consequence, on average the population experiences ${\cal O}(\nu/s)$ switches prior to fixation  when   $1/ K_0 \ll s \ll 1$. Fig.~\ref{fig:FigS3}(c)
confirms that in this regime the average number of switches prior to fixation scales as $1/s$
and increases linearly with $\nu$
to leading order.
Since the PSD greatly varies when $\nu$ and $\delta$ change, see Figs.~2 and \ref{fig:FigS4},
the fact that the average number of switches increases linearly with $\nu$
shows that it is essentially independent of  the population size and supports the
rescaling $\nu \to \nu/s$ in the approximations of Eqs.~(3) and (\ref{eq:Ppdmp-basedSM}).

\section{Eco-evolutionary dynamics \& fixation probability in a public good scenario}
\label{AppendixB}
\begin{figure}[t!]
\centering
\includegraphics[width=0.3\linewidth,height=0.245\linewidth]{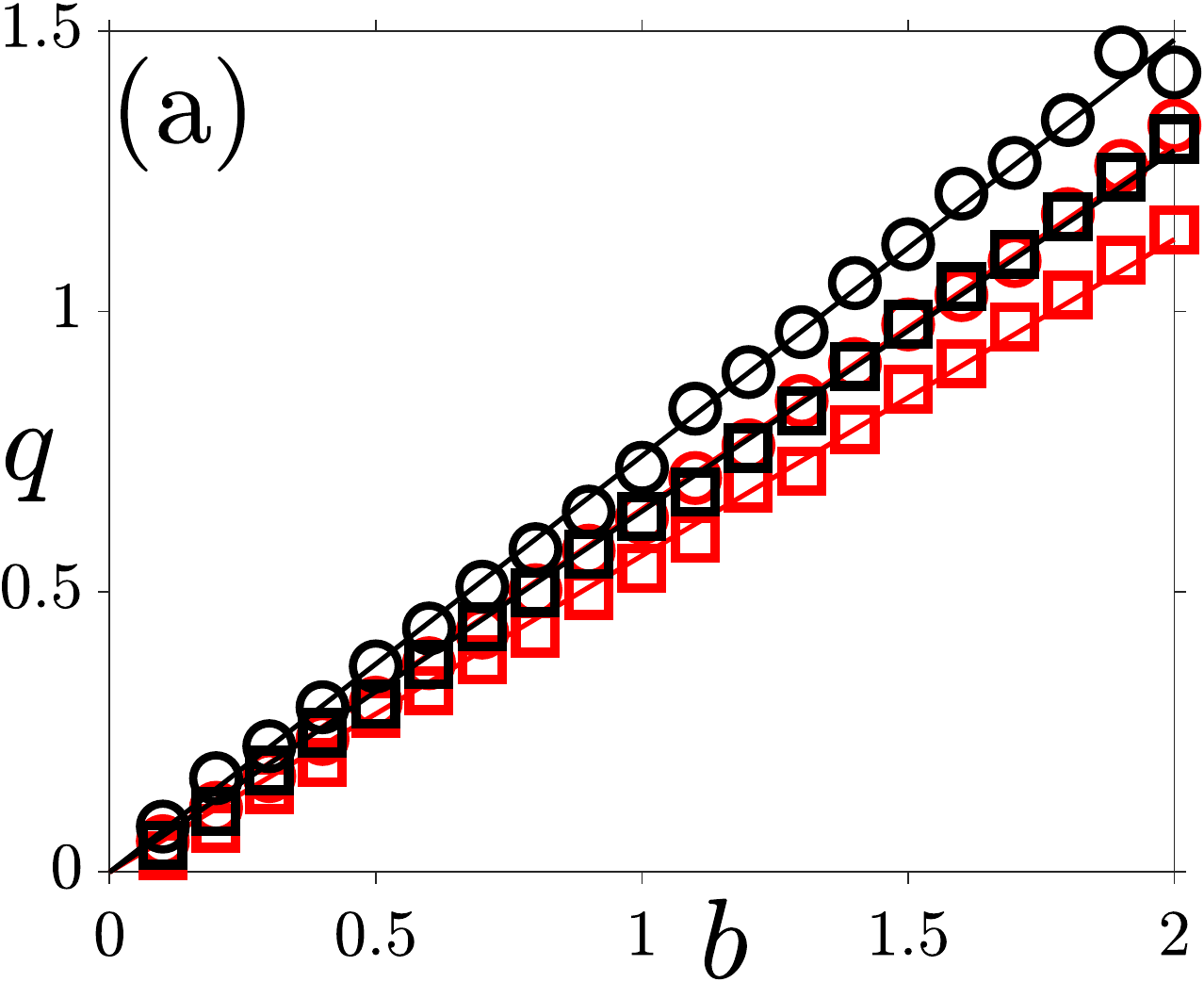}
\includegraphics[width=0.32\linewidth]{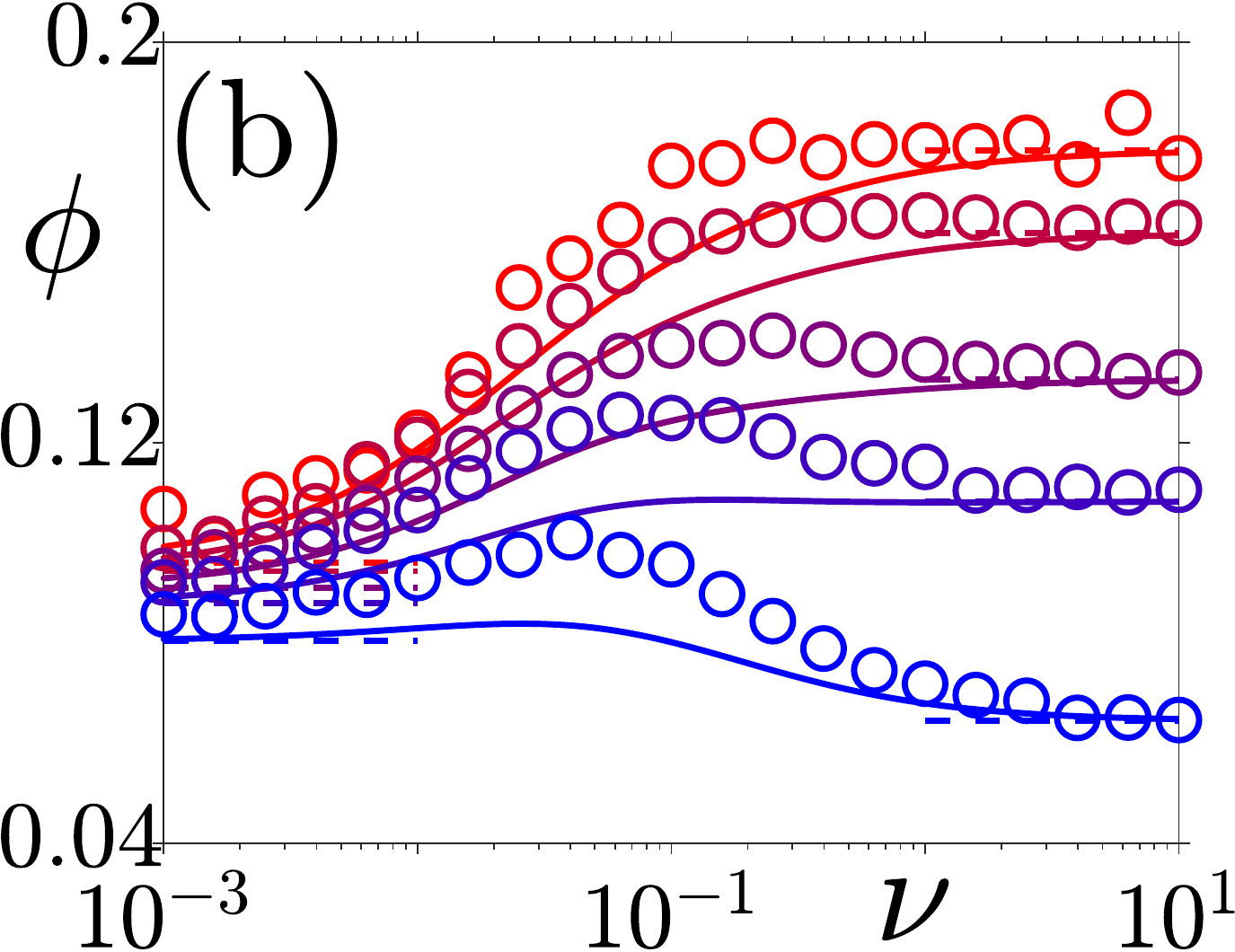}
\includegraphics[width=0.32\linewidth]{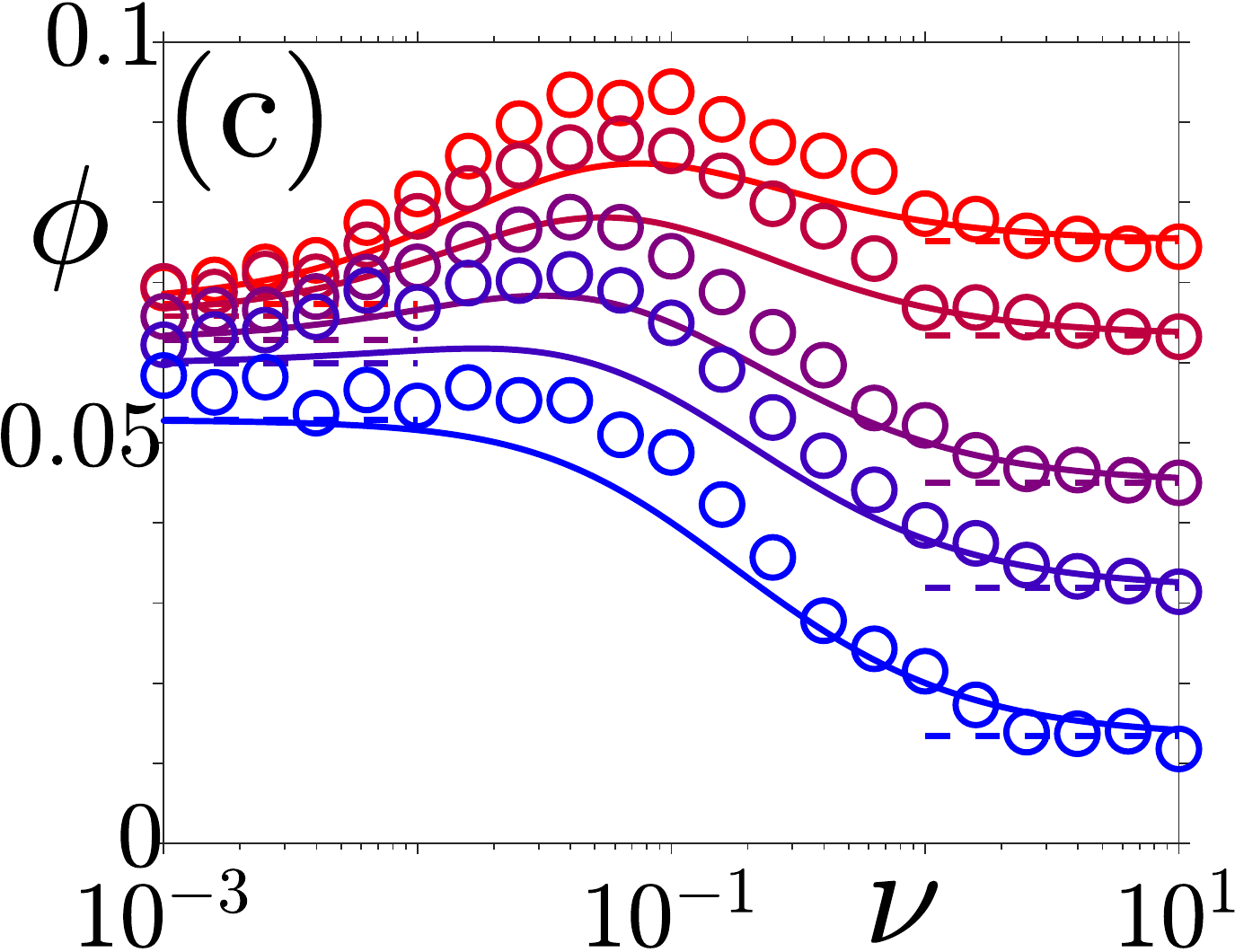}\\
\includegraphics[width=0.32\linewidth]{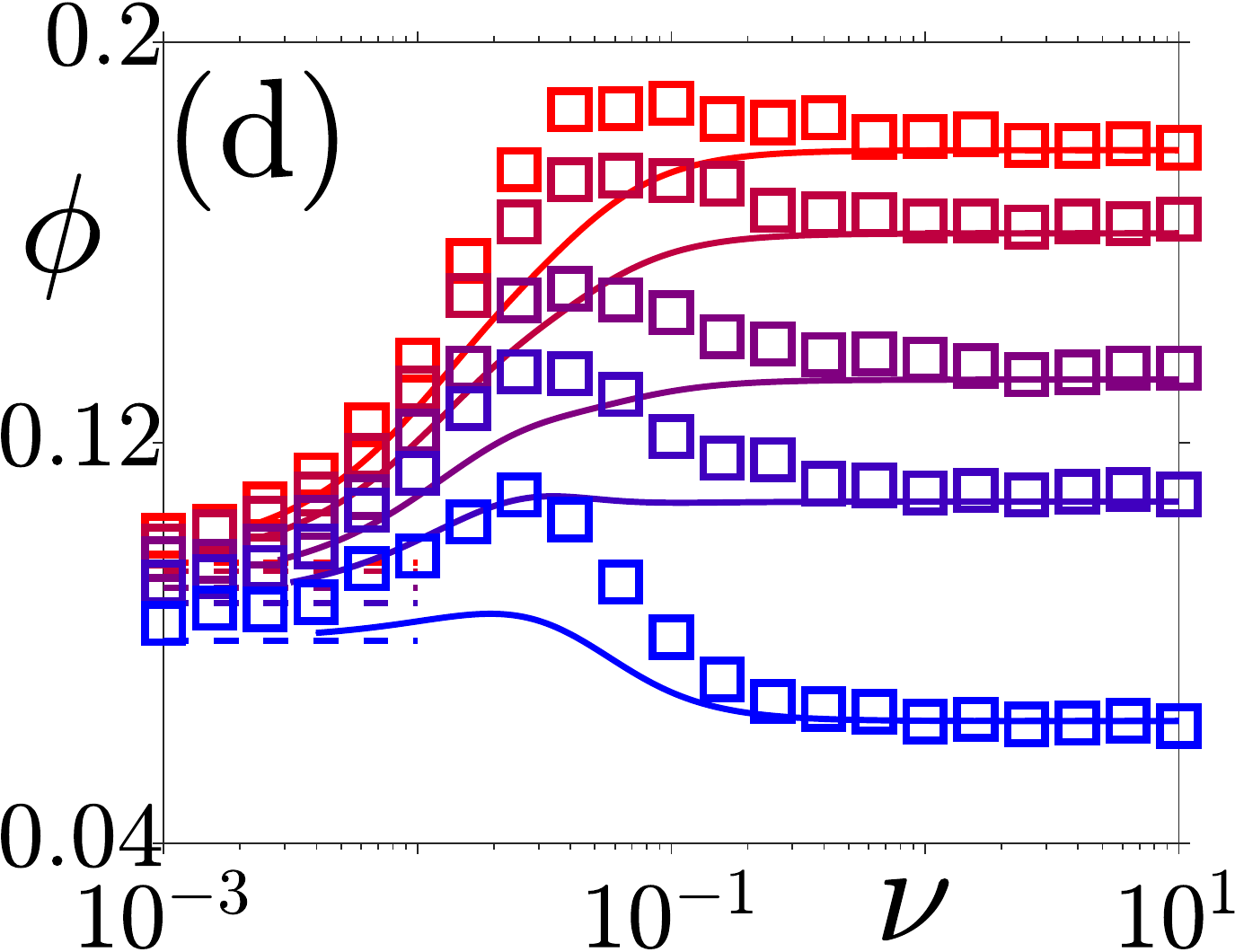}
\includegraphics[width=0.32\linewidth]{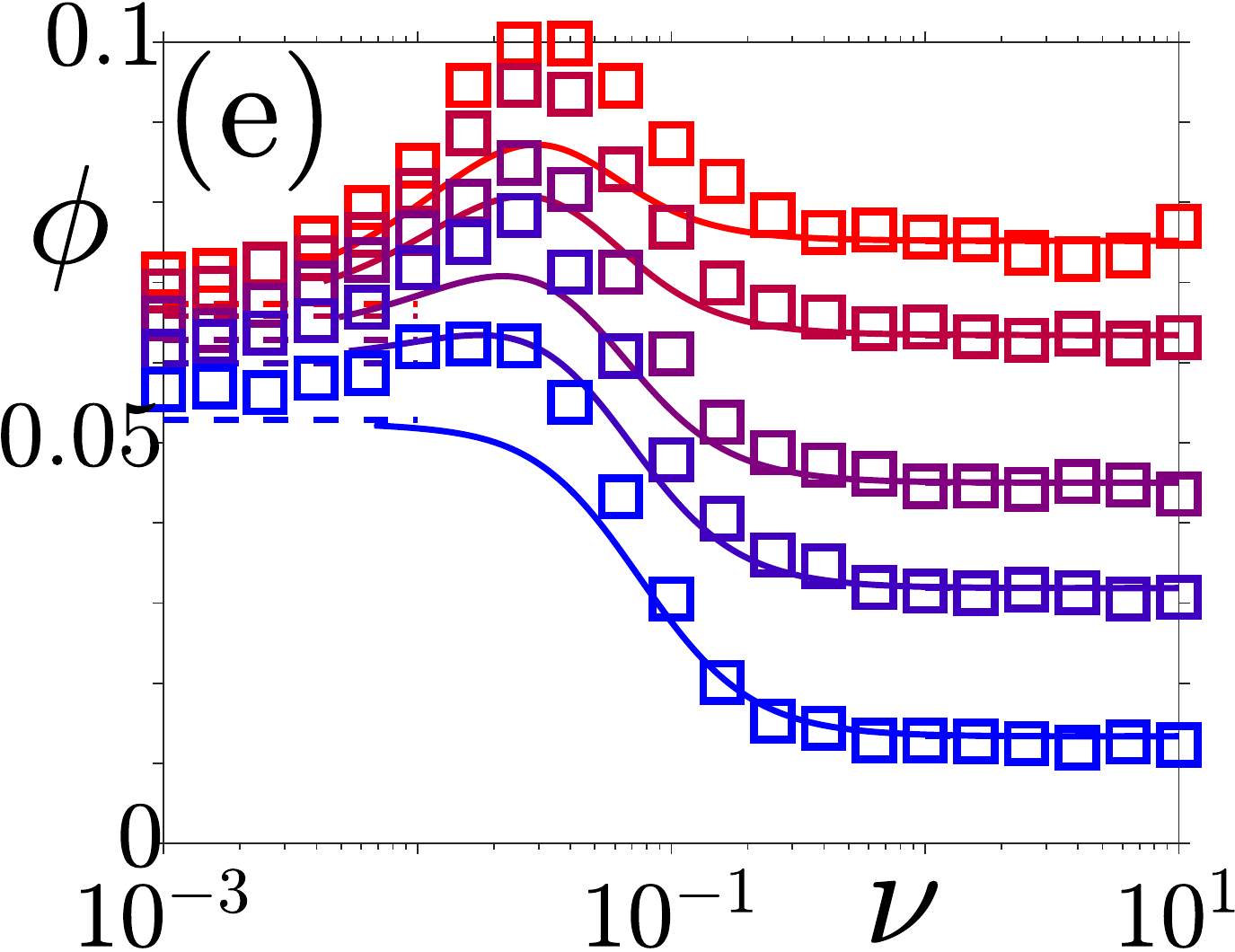}
\caption{(a) Effective parameter $q$ versus $s$ for $\delta = -0.5, 0.5$ (black, red)
	and $s=0.02, 0.05$ (squares, circles). Dependence of $q$  on $b$ is  approximately linear
	while $q$ depends
	weakly on $\delta$ and $s$ (solid lines are eyeguides).
	 (b,c,d,e) $\phi_{\alpha}$ versus $\nu$ for $(K_0,\gamma,s,\delta, x_0) = (250,0.9,0.04,0.6, 0.6)$  in
	(b,d) and $(250,0.9,0.05,0.7, 0.6)$  in (c,e). Here (b,c) and (d,e) show results for random ($\alpha=r$) and periodic ($\alpha=p$) switching, respectively, with the same parameters. In (b,c,d,e) $b = (0,0.1,0.3,0.5,1)$ from red to blue (top to bottom), open circles/squares are simulation results under random/periodic switching. Solid lines are $\phi_{r}^{{\rm  PG}}(\nu)$ from Eq.~\eqref{eq:phi_pg} in (b,c) and $\phi_{p}^{{\rm  PG}}(\nu)$ from Eq.~\eqref{eq:phi_pg_periodic} in (d,e).
	 In (b,d), $\phi_{\alpha}^{{\rm  PG}}(\nu)$
	is an increasing function of $\nu$ at low values of $b$,
	and varies nonmonotonically with $\nu$ for intermediate $b$'s.
	In (c,e), $\phi_{\alpha}^{{\rm  PG}}(\nu)$ is a nonmonotonic function of
	$\nu$ at low $b$'s and becomes a decreasing function of $\nu$ as $b$ increases.}
\label{fig:FigS2}
\end{figure}

The model studied in the main text describes the competition for resources
of the slow and fast growing strains $S$ and $F$ without assuming any explicit interactions between them.
Yet, as discussed in Sec.~S1.1 of this SM, the model can be generalized
to describe the situation where strain $S$ is a
public good (PG) producer. Here, we consider the situation where
$S$ produces a PG that benefits the entire population that
is subject to a time-varying carrying capacity.

A simple way to describe a PG scenario in the general framework outlined in Section S1.1 is to multiply the
per capita birth rate by the global term $g(x)=1+bx$, where
$b\geq 0$~\cite{KEM1,KEM2,Melbinger2010,Cremer2011,Melbinger2015a,Cremer19}.
This PG generalization of the model is thus defined by the continuous-time birth-death
process $N_{S/F}  \xrightarrow{T_{S/F}^{-}}  N_{S/F}- 1$, and $N_{S/F}  \xrightarrow{T_{S/F}^{+}}  N_{S/F}+ 1$,
with the modified transition rates $T_{S}^{+}= g(x)\frac{1-s}{\bar{f}}N_{S},\; T_{F}^{+}= \frac{g(x)}{\bar{f}} N_{F}$, and  $T_{S}^{-}= \frac{N}{K(t)} N_{S}, \; T_{F}^{-}= \frac{N}{K(t)} N_{F}$,
where $b={\cal O}(1)$ while $K(t)$ is given by Eq.~(1) of the main text.
To discuss how the properties of this model can be studied by extending the analysis carried out in the main text, it is convenient to first
consider specifically the case of random switching ($\alpha=r$). When demographic noise is neglected and the only source of randomness stems from the randomly switching  $K(t)$,
the population's mean-field  dynamics obeys~\cite{KEM1,KEM2} (see also Sec.~S1.2)
\begin{eqnarray}
\label{eq:PDMPcoupled}
		\frac{dx }{dt} = -\frac{sg(x)x(1-x)}{1-sx} \quad \text{and} \quad 	
		\frac{dN }{dt} =N\left[g(x)-\frac{N}{{\cal K}}\left(\frac{1-\gamma \xi_r(t)}{1-\gamma\delta}\right)\right],
\end{eqnarray}
with
${\cal K}\equiv  K_0(1-\gamma^2)/(1-\gamma\delta)$.
 $N$ and $x$ are thus explicitly coupled, which breaks the time separation and   yields an explicit form of eco-evolutionary dynamics.
Analytical progress can be made by using the effective theory devised in Refs.~\cite{KEM1,KEM2}.
Since the model's dynamics under a {\it constant} carrying capacity is well described in terms of
a population of an effective size, as in the case $\delta=0$~\cite{KEM1,KEM2}, we introduce a  suitable parameter
$q$ (with $0\leq q\leq b$) and replace $g(x)$ by  $1+q$ in~(\ref{eq:PDMPcoupled}).
This decouples $N$ and $x$, and one can thus perform a similar PDMP-based approximation as before, yielding
\begin{equation}
\label{eq:PDMPqpdf}
P^{{\rm PDMP}}_{\nu,q}(N) \propto \frac{1}{N^2}
\left[\left(\frac{(1+q)K_+}{N}-1\right)^{\frac{\nu_+}{(1+q)} -1}
\left(1-\frac{(1+q)K_-}{N}\right)^{\frac{\nu_-}{(1+q)} - 1}\right],
\end{equation}
where we have omitted the normalization constant.
As in Refs.~\cite{KEM1,KEM2}, the parameter $q$ is obtained by matching the simulation results for the $S$
fixation probability
in the fast switching limit (i.e., when $\nu/s \gg 1$)  with $\phi(x_0)|_{(1+q){\cal K}}$. Results reported in
Fig.~\ref{fig:FigS2}(a), obtained using the diffusion approximation [see below and Eq.~(\ref{eq:Ppdmp-basedSM})], show that $q=q(b)$ increases almost linearly with $b$, and depends only weakly on $s$ and $\delta$,
with $q(0)=0$ when $b=0$.
From $P^{{\rm PDMP}}_{\nu,q}$ it is clear that the effect of increasing $b$,
and therefore the effective parameter  $q(b)$, results in effectively
increasing the carrying capacity $K_{\pm} \to (1+q(b))K_{\pm}$ and reducing the switching rates $\nu_{\pm} \to \nu_{\pm}/(1+q)=\nu(1\mp \delta)/(1+q)$.
Proceeding as in the case $b=q=0$ and $\delta=0$~\cite{KEM1,KEM2},
the fixation probability is obtained by averaging
$\phi(s,x_0)|_N$ over the PSD in Eq.~(\ref{eq:PDMPqpdf}) with  $\nu \to \nu/s$~\cite{footnote8}.
 Furthermore, by changing the variable of integration to $N' = N/(1+q)$ we find that this is equivalent to rescaling the selection strength to $s_{\text{eff}} = \left(1+q\right)s$ in the model without a PG
\begin{eqnarray}
\label{eq:phi_pg}
	\phi^{\text{PG}}_{r}(\nu,q)
	=\int_{(1+q)K_-}^{(1+q)K_+} \phi(s,x_0)|_N~
	P^{{\rm PDMP}}_{\nu/s,q}(N)~  dN =\int_{K_-}^{K_+} \phi(s_{\text{eff}},x_0)|_{N'}~
	P^{{\rm PDMP}}_{\nu/s_{\text{eff}},0}(N')~  dN',
\end{eqnarray}
where $N'=N/(1+q)$ and we have used $\phi(s,x_0)|_N \simeq  (e^{-Ns(1-x_0)}-e^{-Ns})/(1-e^{-Ns})$. 

According to Eq.~(\ref{eq:phi_pg}), the effect of increasing $b$ results in raising the value of the corresponding value of $q$, see Fig.~\ref{fig:FigS2}(a), which in turn results in a  carrying capacity switching between $(1+q(b))K_{\pm}$. Thus, as in the case $\delta=0$, one can transform the expression of the $S$ fixation probability, $\phi^{\text{PG}}_{r}(\nu,q)$, to the (approximate)
fixation probability in the absence of PG but under an effective (increased) selection intensity $s_{{\rm eff}} = \left[1+q(b)\right]s$.  This results in $\phi^{\text{PG}}_{r}$ decaying  approximately exponentially with $b$~\cite{KEM2}.

Equation~(\ref{eq:phi_pg}) is an approximation of the actual fixation probability $\phi_r$ that is valid over a broad range of frequencies $\nu$ and gives an accurate description of $\phi_r$ when $\delta=0$~\cite{KEM1,KEM2} and $|\delta|\ll 1$ (small switching asymmetry); its accuracy deteriorates as $|\delta|$ and $b$ increase. Here, we are chiefly interested in the {\it qualitative} dependence of the fixation probability on $\nu$ when $b$ changes  and $\gamma={\cal O}(1)$,
$\delta={\cal O}(1)$ (see Figs.~3(e) and \ref{fig:FigS1}(e)). With Fig.~\ref{fig:FigS1}(a) in mind,  we can understand  how
raising $b$ changes the diagram of Fig.~\ref{fig:FigS1}(a):
As $b$ is increased,
the triangular-like region is squashed since $\gamma_c$ increases under
the effect of $s\to s_{{\rm eff}}=(1+q(b))s$. This allows us to qualitatively explain how the fixation probability
$\phi_r^{\text{PG}}$ varies with $\nu$
under intermediate switching at $\gamma, \delta, s$ fixed. In the case of Fig.~\ref{fig:FigS2}(b),
$\delta<\delta_c$ at low $b$ and therefore $\phi_r^{\text{PG}}(\nu)$ increases monotonically;
then as $\gamma_c$ increases together with $b$, $(\gamma, \delta)$
 enter the triangular-shaped region (i.e., $\gamma>\gamma_c,\delta>\delta_c$) of Fig.~\ref{fig:FigS1}(a) where $\phi_r^{\text{PG}}(\nu)$ varies non-monotonically with $\nu$. In the example of Fig.~\ref{fig:FigS2}~(c),
$\gamma>\gamma_c$ and $\delta>\delta_c$ at low $b$ implying that $\phi_r^{\text{PG}}(\nu)$
is a nonmonotonic function of $\nu$; then
$\gamma_c$ increases along with $b$ and attains a value such that
$\gamma<\gamma_c$ with $\delta>\delta_c$, and in this case $\phi_r^{\text{PG}}(\nu)$
decreases monotonically with $\nu$. Hence, while Eq.~(\ref{eq:phi_pg})
cannot accurately predict the full $\nu$ dependence of $\phi_r$, it qualitatively  captures the emergence of a peak in  \ref{fig:FigS2}(b) at some nontrivial intermediate switching rate, and the disappearance of the peak in  \ref{fig:FigS2}(c),  when $b$ is increased.
These are examples
of the rich and complex behavior that eco-evolutionary loops can generate.

The results of this section have so far focused on the case of random switching, but we have again obtained a similar qualitative behavior with periodic switching, as shown in Fig.~\ref{fig:FigS2}(d,e). This can be explained in terms of a PPP-based approximation in the realm of an effective theory as in the random switching case. In fact, we have verified that the effective parameter $q(b)$ allows us to obtain
a suitable approximation of the fixation probability under fast periodic switching,
i.e.  $\phi_p \simeq \phi|_{(1+q){\cal K}}$. This suggests to use the
PPP-based approximation $P_{\nu,q}^{{\rm PPP}}$ as an effective approximate PSD
in the PG scenario with periodic switching, where $P_{\nu/s,q}^{{\rm PPP}}$
is obtained from Eqs.~(\ref{PPPapprox})-(\ref{C}) by rescaling $\nu_{\pm}\to \nu_{\pm}/[(1+q)s]$
and $K_{\pm} \to (1+q)K_{\pm}$. This rescaling of $\nu_{\pm}$ and $K_{\pm}$ results in
a support of  $P_{\nu/s,q}^{{\rm PPP}}$ that is now denoted by $[N_{\rm min}(q), N_{\rm max}(q)]$. In the same vein as in the random switching case, we thus write
\begin{eqnarray}
\label{eq:phi_pg_periodic}
	\phi^{\text{PG}}_{p}(\nu,q)
	=\int_{N_{\rm min}(q)}^{N_{\rm max}(q)} \phi(s,x_0)|_N~
	P^{{\rm PPP}}_{\nu/s,q}(N)~  dN,
\end{eqnarray}
which is expected to be  a suitable approximation of $\phi_{p}$ when
$|\delta|\ll 1$, and to qualitatively capture the $\nu$ dependence of $\phi_{p}$ when $\delta={\cal O}(1)$.
The results of Fig.~\ref{fig:FigS2}(d,e) indeed show that $\phi^{\text{PG}}_{p}$ provides the same qualitative description of the fixation probability  as Eq.~(\ref{eq:phi_pg})
in the random switching case. In particular, Eq.~(\ref{eq:phi_pg_periodic}) qualitatively
reproduces the emergence of a peak
at a nontrivial frequency in  \ref{fig:FigS2}(d), and the disappearance of the peak in
\ref{fig:FigS2}(e), as $b$ is increased.


\begin{thebibliography}{99}
\bibitem{rect}
A rectangular function is defined as follows: ${\rm rect}(x)=1$ if $|x|<1/2$, ${\rm rect}(x)=0$ if $|x|>1/2$, while ${\rm rect}(\pm 1/2)=0$.
\bibitem{meta}
After the exinction of one species, the other  has a logistic-like dynamics, see text, with a mean time to extinction scaling exponentially with its carrying capacity~\cite{Assaf10,Assaf17}.
\bibitem{IC}
When  $N$ and $x$ are not coupled and $s\ll 1$, the  initial condition $N(0)$ is here irrelevant. We  set $N(0)=K_0$
or $N(0)=K_0(1-\gamma^2)/(1-\gamma\delta)={\cal K}$ and confirmed that our results were independent of  $N(0)$.
\bibitem{monotone} This  occurs for $s>s_c$ (generic case), when $\phi_\alpha (\nu)$ is a decreasing monotonic function. On the other hand,  $\phi_{r}\lesssim \phi_{p}$  when $s < s_c$ and  $\gamma<\gamma_c$ and $\delta<\delta_c$, see below and Sec.~S5.2 in \cite{SM}.
\bibitem{footnote8} When $b>0$, the average number of switches prior to fixation is ${\cal O}(\nu/s)$
	as in the cases $\delta=0$~\cite{KEM2} and $b=q=0$, see Sec.~\ref{AppendixC} and Fig.~\ref{fig:FigS3}(c).
\end{thebibliography}
\end{document}